\documentclass[sigconf]{acmart}
\AtBeginDocument{%
  }

\setcopyright{acmlicensed}
\copyrightyear{2018}
\acmYear{2018}
\acmDOI{XXXXXXX.XXXXXXX}
\acmConference[Conference 'XX]{Make sure to enter the correct
  conference title from your rights confirmation email}{June 03--05,
  2018}{Woodstock, NY}
\acmISBN{978-1-4503-XXXX-X/2018/06}

\definecolor{hlcolorCQ}{RGB}{255, 224, 204}   
\definecolor{hlcolorA}{RGB}{255, 255, 179}    
\definecolor{hlcolorB}{RGB}{204, 255, 204}    
\definecolor{hlcolorC}{RGB}{204, 255, 255}    
\definecolor{hlcolorD}{RGB}{204, 229, 255}    
\definecolor{hlcolorE}{RGB}{229, 204, 255}    
\definecolor{hlcolorF}{RGB}{255, 204, 229}    

\definecolor{hlcolorG}{RGB}{0, 150, 0}
\definecolor{hlcolorH}{RGB}{175, 82, 222}

\newcommand{\frameworkName}{FlashANNS}

\usepackage{bookmark}
\usepackage{mdframed}
\usepackage{needspace}
\usepackage{soulutf8}

\usepackage{graphicx}       
\usepackage{subcaption}     
\usepackage{multirow}
\usepackage{array}
\usepackage{mathtools}
\usepackage[export]{adjustbox} 

\usepackage{tikz}
\newcommand*\cycle[1]{\tikz[baseline=(char.base)]{
            \node[shape=circle,fill,color=black,text=white,inner sep=0.05pt](char){#1};}}
\usepackage[normalem]{ulem}

\usepackage{xcolor,framed} 
\definecolor{shadecolor}{gray}{0.93}

\begin{document}
\settopmatter{authorsperrow=4}




\onecolumn
\setcounter{figure}{0}

\title{\frameworkName{}: GPU-Driven I/O Pipelining for Eliminating Storage-Compute Bottlenecks in Billion-Scale Similarity Search}



\author{Yang Xiao}
\affiliation{%
  \institution{Zhejiang University}
  \city{Hangzhou}
  \country{China}
}
\email{12221061@zju.edu.cn}

\author{Mo Sun}
\affiliation{%
  \institution{Zhejiang University}
  \city{Hangzhou}
  \country{China}
}
\email{sunmo@zju.edu.cn}

\author{Ziyu Song}
\affiliation{%
  \institution{Zhejiang University}
  \city{Hangzhou}
  \country{China}
}
\email{songziyu@zju.edu.cn}

\author{Bing Tian}
\affiliation{%
  \institution{Huazhong University of Science and Technology}
  \city{Wuhan}
  \country{China}
}
\email{tbing@hust.edu.cn}

\author{Jie Sun}
\affiliation{%
  \institution{Zhejiang University}
  \city{Hangzhou}
  \country{China}
}
\email{jiesun@zju.edu.cn}

\author{Jie Zhang}
\affiliation{%
  \institution{Zhejiang University}
  \city{Hangzhou}
  \country{China}
}
\email{carlzhang4@zju.edu.cn}

\author{Zeke Wang}
\affiliation{%
  \institution{Zhejiang University}
  \city{Hangzhou}
  \country{China}
}
\email{wangzeke@zju.edu.cn}

\author{Zonghui Wang}
\affiliation{%
  \institution{Zhejiang University}
  \city{Hangzhou}
  \country{China}
}
\email{zhwang@zju.edu.cn}

\author{Wenzhi Chen}
\affiliation{%
  \institution{Zhejiang University}
  \city{Hangzhou}
  \country{China}
}
\email{chenwz@zju.edu.cn}

\author{Fei Wu}
\affiliation{%
  \institution{Zhejiang University}
  \city{Hangzhou}
  \country{China}
}
\email{wufei@zju.edu.cn}

\renewcommand{\shortauthors}{Trovato et al.}

\begin{abstract}
Approximate Nearest Neighbor Search (ANNS) enables efficient similarity retrieval in high-dimensional vector spaces, and becomes a fundamental component of upper-layer workloads ranging from recommendation systems to retrieval-augmented generation (RAG). Modern ANNS systems integrate SSDs to support terabyte-scale vector datasets, primarily employing cluster-indexing and graph-indexing. However, cluster-indexing ANNS systems suffer from suboptimal query throughput because of the coarse-grained vector indexing, while graph-indexing systems suffer from suboptimal performance due to two inherent limitations: 1) failing to overlap SSD accesses with distance computation processes and 2) poor I/O performance due to long tail latency. 

To address these challenges, we present \frameworkName{}, a GPU-accelerated out-of-core graph-based ANNS system through I/O-compute overlapping. Our core insight lies in the careful orchestration of I/O and computation through three key innovations: 1) Dependency-relaxed asynchronous pipeline with rigorous theoretical convergence guarantee: \frameworkName{} decouples I/O-computation dependencies to fully overlap between GPU distance calculations and SSD data transfers;  2) Query-grained concurrent SSD access: \frameworkName{} implement a lock-free I/O stack with query-grained concurrency control, to avoid I/O performance degradation due to long tail latency; and 3) Computation-I/O balanced graph degree selection, which ensures optimal balance between computational load and storage access latency across different hardware characteristics, different datasets, and different query requirements. 

We implement \frameworkName{} and compare it with three state-of-the-art out-of-core ANNS systems (SPANN, DiskANN, and FusionANNS). Experimental results demonstrate that at the same $\geq$95\% recall@10 accuracy, our method achieves 2.7–5.9× higher query throughput compared to existing SOTA methods with a single SSD, and further achieves 3.9–12.2× query throughput improvement in the multi-SSD configurations. 

\end{abstract}





\maketitle

\section{Introduction}
\label{introduction}
Approximate Nearest Neighbor Search (ANNS) refers to a set of methods for finding the top-k vectors most similar to a given query vector in a high-dimensional vector dataset. 
Compared with exact search, ANNS sacrifices a little precision for much less retrieval time.
ANNS is widely applied in various domains, including information retrieval~\cite{image1,information_retrieval1,information_retrieval2,pattern_recognition,Data_mine1}, recommendation systems~\cite{recommend1,recommend2,recommend3}, and large language models~\cite{rag1,llm1,llm2,llm3,llm4}.
Especially, the rise of generative AI and large-scale recommendation systems has driven the demand for billion-scale ANNS, with modern datasets (often exceeding 1000M vectors) and their indices (up to 6$\times$ dataset size) overwhelming traditional in-memory frameworks due to prohibitive memory and computation requirements. 

\begin{figure}
	\centering
	{\includegraphics[width=2.3in]{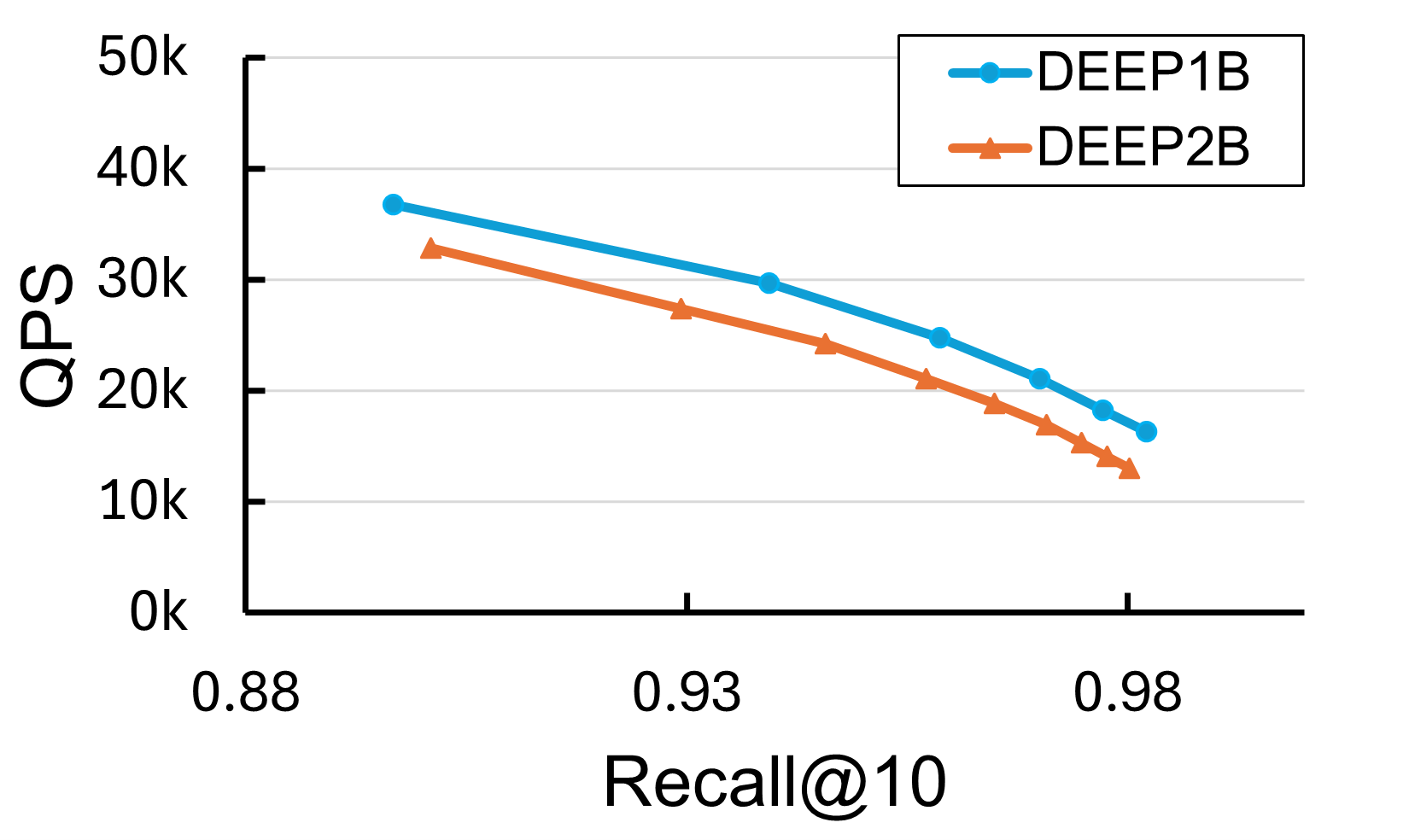} } 
	\caption{Comparison of query QPS performance between dataset DEEP1B and DEEP2B} 
	\label{deep_1B2B}
\end{figure}  

A straightforward scale-out solution is to employ a distributed in-memory solution~\cite{Milvus, li2018design} that utilizes the host memory of multiple servers for index storage. Vearch~\cite{li2018design} partitions a size-N index into n shards of size N/n, each shard is stored in the DRAM of a server. During query execution, all nodes process the same query concurrently on their local shards, and each node computes its local top-K results from its assigned shard. Finally, global results are merged by sorting and truncating the $n \times K$ local results to yield the final K-nearest neighbors. However, bringing $n\times$ computation resource can not increase query throughput by $n\times$, mainly because the query time does not scale linearly with the dataset shard size. As shown in Figure~\ref{deep_1B2B}, halving the dataset size can only increase query throughput by 10.8\%-19.5\%.

{Considering the suboptimal performance of scale-out solutions, recent ANNS systems~\cite{fusion,diskann,spann,ni2023diskann++,xu2023spfresh} are equipped with Solid State Drives (SSDs) to fit such huge datasets/indices (up to tens of Terabytes), to further extend single node capacity. According to the kind of indexing method used, these systems can be categorized into cluster-based and graph-based systems. We identify that both of these systems suffer from low system resource utilization.}

\noindent{\bf Cluster-Based Indexing. } 
Cluster-indexing methods such as SPANN~\cite{spann} partition the dataset into cluster-based inverted lists, storing original vectors on disk while maintaining centroid indices in memory. However, its coarse-grained clustering leads to a large candidate search space, requiring extensive vector comparisons in high-precision scenarios.

FusionANNS~\cite{fusion} mitigates this issue via GPU-accelerated pre-filtering to reduce I/O overhead by selectively fetching vectors from SSDs. Despite this improvement, FusionANNS still incurs high SSD accesses due to the inherent limitations of coarse-grained clustering. Furthermore, its computational pipeline does not fully utilize the GPU’s processing capability due to high overhead from the CPU-managed task synchronization.



\noindent{\bf Graph-Based Indexing. } 
Fine-grained graph indexing such as DiskANN~\cite{diskann} can achieve the target accuracy with a smaller candidate size, thereby reducing the number of SSD fetches and the overall I/O cost. However, it increases per-fetch computational work (e.g., distance evaluations and neighbor management). As a result, a larger share of the end-to-end latency is attributable to computation, indicating the potential for compute-side optimization.

{While graph-based indexing reduces I/O, it increases compute intensity, making computation a larger share of latency. This motivates a GPU-offloading direction for compute-side optimization. However, directly integrating GPUs introduces three key challenges: (1) serialized or coarse-grained execution that fails to fully exploit GPU compute capacity; (2) fine-grained synchronization that amplifies I/O latency; and (3) small random SSD accesses that are IOPS-bound, leaving SSD bandwidth underutilized.}



To this end, we present \frameworkName{}, a GPU-accelerated, SSD-resident graph indexing framework that balances computation and I/O for billion-scale similarity search. Our approach centers on three key designs:


\begin{itemize}
\item \textbf{Dependency-Relaxed Asynchronous I/O Pipeline. }
To address the low GPU/SSD utilization problem caused by the serialized execution of computation and I/O operations, we propose a dependency-relaxed asynchronous I/O pipeline, and provide a rigorous theoretical convergence guarantee to ensure that \frameworkName{} can achieve the same recall.

\item \textbf{Query-Grained Concurrent SSD Access. }
To address the prolonged I/O latency due to synchronization overhead in the kernel-grained GPU I/O stack, we propose a query-grained concurrent SSD access architecture optimized for ANNS workloads, eliminating the GPU kernel-grained global synchronization inherent to conventional CAM-based I/O stacks. Under this architecture, each query-issued SSD access can complete and resume execution independently within the warp without kernel-wide global stalls.

\item \textbf{Computation-I/O Balanced Graph Degree Selection. }
To address the I/O amplification due to SSD page realignment, we propose a hardware-aware graph degree selection mechanism. This approach quantifies SSD I/O bandwidth and GPU computational capacity through sampling-based profiling, enabling adaptive selection of graph degree parameters in the index structure. This maintains computation-I/O equilibrium throughout pipeline operations.

\end{itemize}

We implement \frameworkName{} and evaluate it on widely adopted billion-scale vector datasets (SIFT1B, SPACEV1B, DEEP1B), benchmarking against state-of-the-art out-of-core ANNS systems (SPANN, DiskANN) and a GPU-accelerated out-of-core baseline (FusionANNS). Experimental results show that at $\geq$95\% recall@10 accuracy, \frameworkName{} achieves 2.7–5.9$\times$ higher QPS than existing SOTA methods with a single SSD configuration, and scales to up to 12.2$\times$ QPS improvements in multi-SSD setups. We will make \frameworkName{} open-sourced at GitHub to benefit our community once this paper gets accepted.

\section{Background}
\begin{figure}[t]
    \centering
    \begin{subfigure}[b]{0.48\textwidth}
        \includegraphics[width=2.8in]{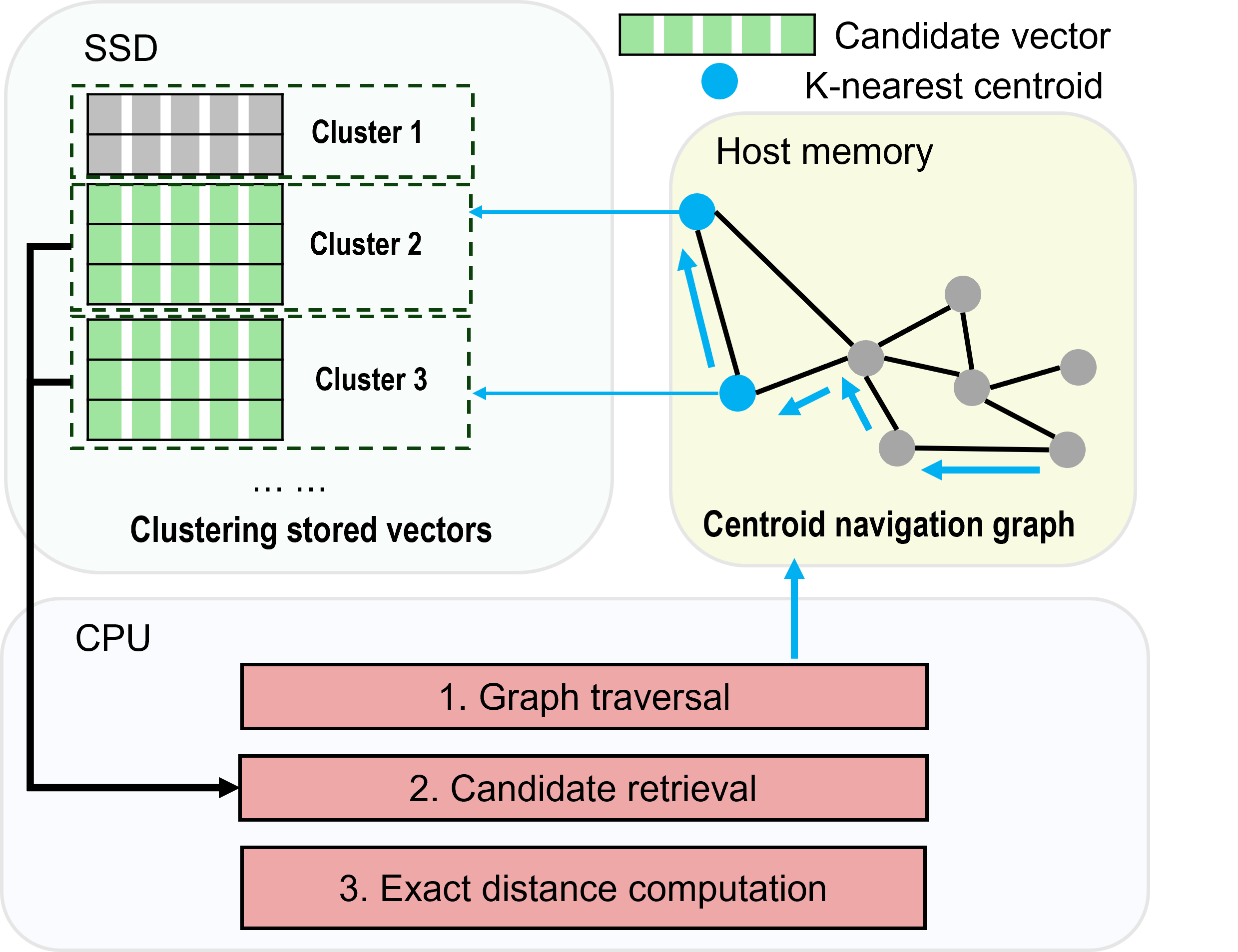}
        \caption{Cluster-Based indexing}
        \label{cluster-based_overview}
    \end{subfigure}
    \begin{subfigure}[b]{0.48\textwidth}
        \includegraphics[width=3.2in]{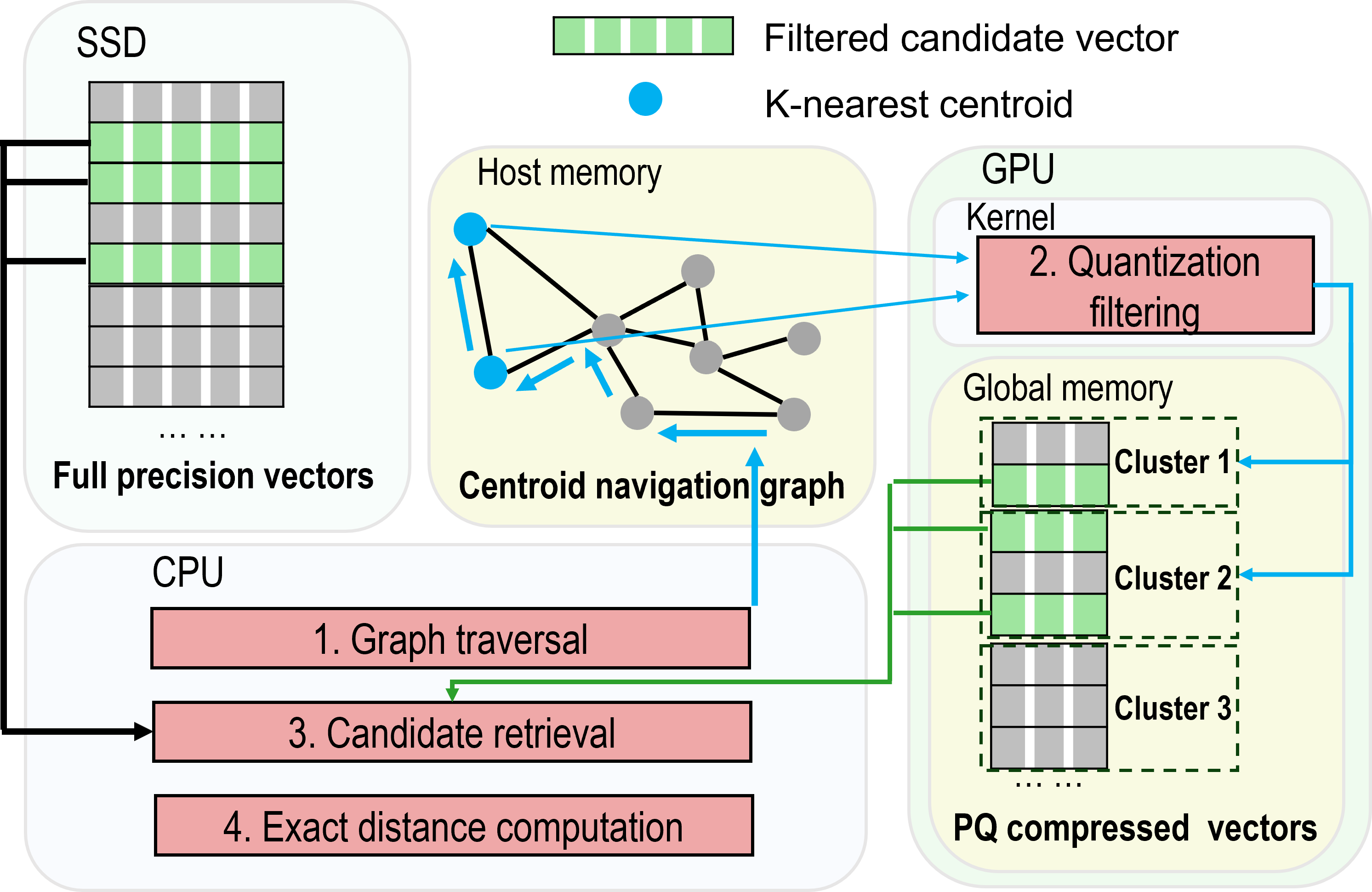}
        \caption{Filtered Cluster-Based indexing}
        \label{filter-cluster-based_overview}
    \end{subfigure}
    \begin{subfigure}[b]{0.48\textwidth}
        \includegraphics[width=3.2in]{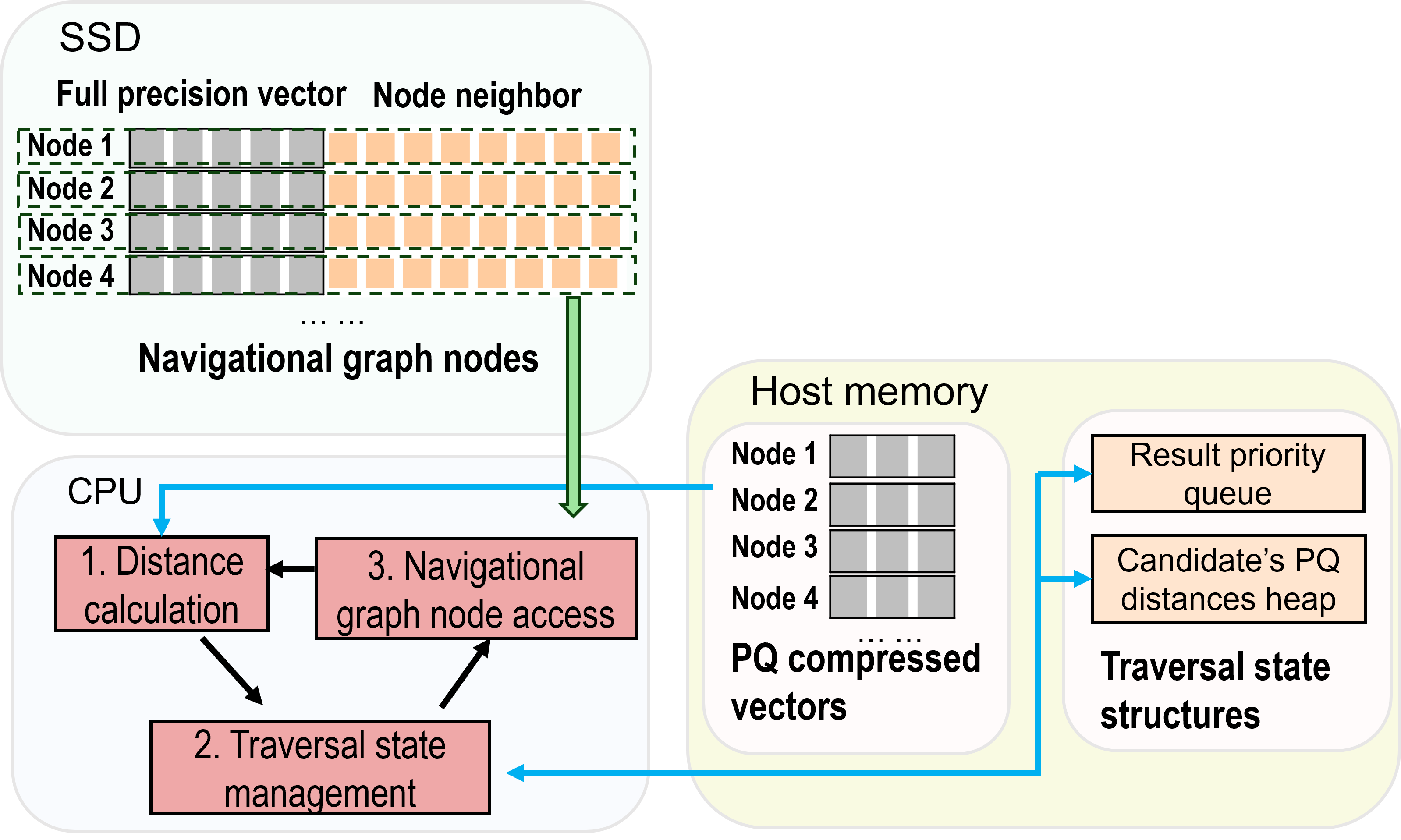}
        \caption{CPU-Based graph indexing}
        \label{graph-based_overview}
    \end{subfigure}
    \caption{Overall comparison of indexing architectures}
\end{figure}

The Approximate Nearest Neighbor Search (ANNS) has emerged as a fundamental operation in modern data processing systems, enabling efficient similarity retrieval in high-dimensional vector spaces. As a fundamental component of upper-layer workloads ranging from recommendation systems to retrieval-augmented generation (RAG), ANNS evolves rapidly as the scale of these workloads becomes larger and larger. Vector indexing serves as a critical ANNS component, organizing vector data and determining search methodologies. Substantial index storage overheads consume up to 86\% of the query execution storage footprint. Modern ANNS systems~\cite{spann,fusion,diskann} integrate SSDs to support terabyte-scale vector datasets required by contemporary applications.



\subsection{Cluster-Based Vector Indexing}
\begin{figure}
	\centering
	{\includegraphics[width=2.3in]{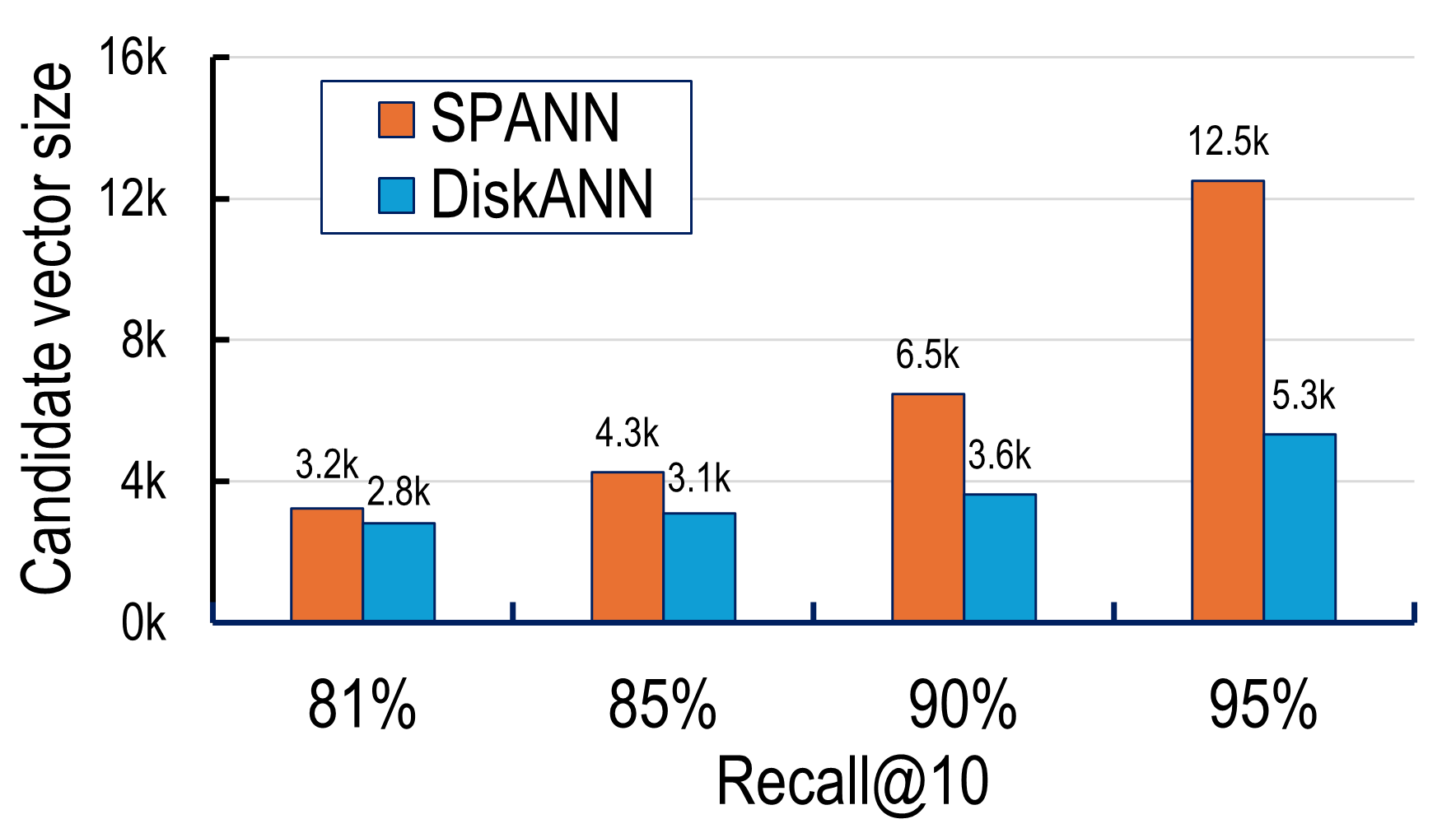} } 
	\caption{Candidate vector size during searching progress of cluster-indexing (SPANN) and graph-indexing (DiskANN) under different recall accuracy} 
	\label{accessed_candidates}
\end{figure}  

\begin{figure}
	\centering
	{\includegraphics[width=2.3in]{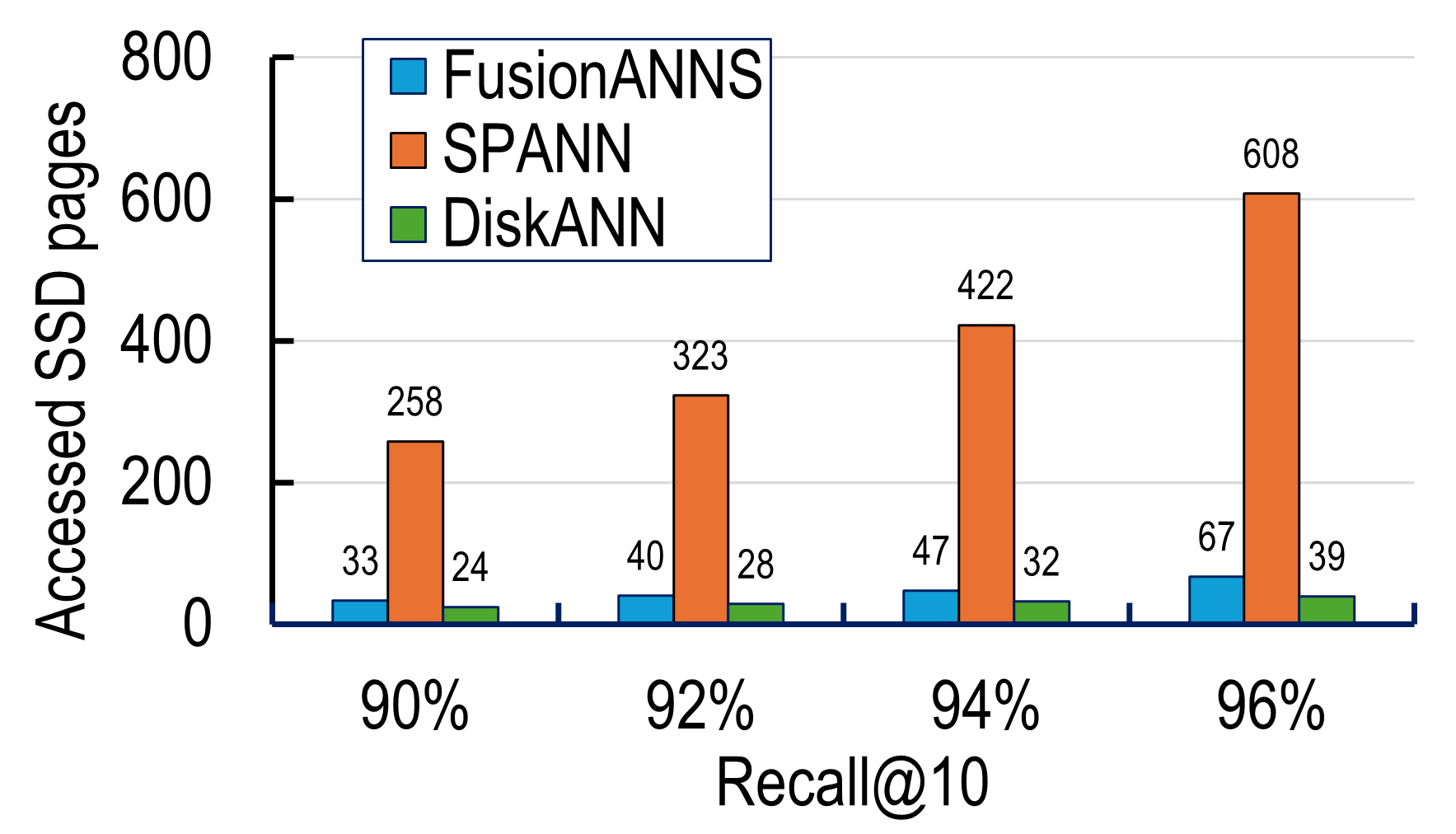} } 
	\caption{SSD page read requirements of FussionANNS, SPANN and DiskANN under different recall accuracy} 
	\label{accessed_pages_3}
\end{figure} 
Cluster-based vector indexing is widely used due to its low query latency. These systems mainly maintain two data structures: 1) SSD-resident clustered vectors and 2) an in-memory navigation graph of cluster centroids. During the index construction, the system scans the entire vector dataset and partitions vectors into several clusters. The vectors are stored in the SSD by clusters, while the centroid of each cluster forms a graph and is stored in the CPU memory. Figure~\ref{cluster-based_overview} demonstrates the three-phase search workflow of cluster-based vector indexing. 
\begin{itemize}
\item \textbf{1. Graph Traversal Phase.} Within the in-memory centroid navigation graph, the search process traverses the graph to identify the K-nearest centroids to the query vector.
\item \textbf{2. Candidate Retrieval Phase.} The retrieval phase fetches clusters corresponding to the selected K centroids from SSD-resident clustered vectors.
\item \textbf{3. Distance Calculation Phase.} The search process computes precise distances between all vectors in retrieved clusters and the query vector, then ranks vectors by distance to select the final result. 
\end{itemize}

However, cluster-indexing ANNS systems suffer from suboptimal query throughput, because these systems use coarse-grained vector indexing, and graph traversal phase (\cycle{1})'s computation can be very light-weight, while there would be numerous candidates to be accessed in candidate retrieval phase (\cycle{2}) and corresponding distance calculation in distance calculation phase (\cycle{3}).


The fundamental limitation stems from the graph traversal phase (\cycle{1})'s assumption that centroids within posting lists adequately represent associated cluster vectors. In practice, this representation proves to be inaccurate, manifesting two distinct deficiencies:

\textbf{1. False Inclusion.} Centroids proximate to the query may correspond to distant vectors, forcing candidate retrieval phase (\cycle{2}) and distance calculation phase (\cycle{3}) to process massive volumes of irrelevant candidates.

\textbf{2. False Exclusion.} During candidate retrieval (\cycle{2}), the search process may erroneously exclude distant centroids containing neighboring vectors, misclassifying the truly nearest neighbors within them as distant. This initial error leads to their absence in the candidate pool, severely degrading query accuracy.

Figure~\ref{accessed_candidates} demonstrates the candidate size disparity between the cluster-indexing implementation (SPANN) and the graph-indexing implementation (DiskANN) on SIFT1B. Under 81\%-95\% recall@10, to achieve equivalent recall, cluster-indexing implementation requires 1.14-2.34$\times$ more candidates than graph-indexing. And this gap is widening with the increasing accuracy. This imposes prohibitive computational and I/O pressure during queries. Consequently, DiskANN achieves 1.23$\times$ to 1.88$\times$ higher QPS than SPANN.

\noindent{\bf GPU-Accelerated Cluster-Based Vector Indexing.}
To mitigate excessive computational and SSD access demands, FusionANNS~\cite{fusion} leverages a GPU and stores quantified vectors in GPU memory. These vectors reduce dimension and are quantified as uint8 values via product quantization (PQ). As Figure~\ref{filter-cluster-based_overview} demonstrates, compared to the cluster-indexing approach's three-phase search workflow, FusionANNS introduces a GPU-based quantization filtering phase (\cycle{2}) between graph traversal (\cycle{1}) and candidate retrieval (\cycle{3}).

In quantization filtering phase (\cycle{2}), the GPU calculates quantization distances for vectors within clusters selected during graph traversal (\cycle{1}) using PQ-coded vectors. It screens vectors with distances below a predetermined threshold, directing subsequent phases to retrieve and process only these filtered candidates from SSD storage for exact distance calculations.





Despite efficiency gains from quantization filtering, FusionANNS can only partially mitigate severe SSD I/O pressures because its filtering stage relies solely on PQ-compressed distances for candidate selection, lacking the fine-grained navigation guidance formed by exact distances. This necessitates additional SSD accesses to obtain exact distances for compensating quantization errors. As shown in Figure~\ref{accessed_pages_3}, FusionANNS still incurs 1.36–1.70$\times$ more SSD page accesses than DiskANN when evaluating on the SIFT1B dataset.



\subsection{Graph-Based Vector Indexing}

\begin{figure}
	\centering
	{\includegraphics[width=2.3in]{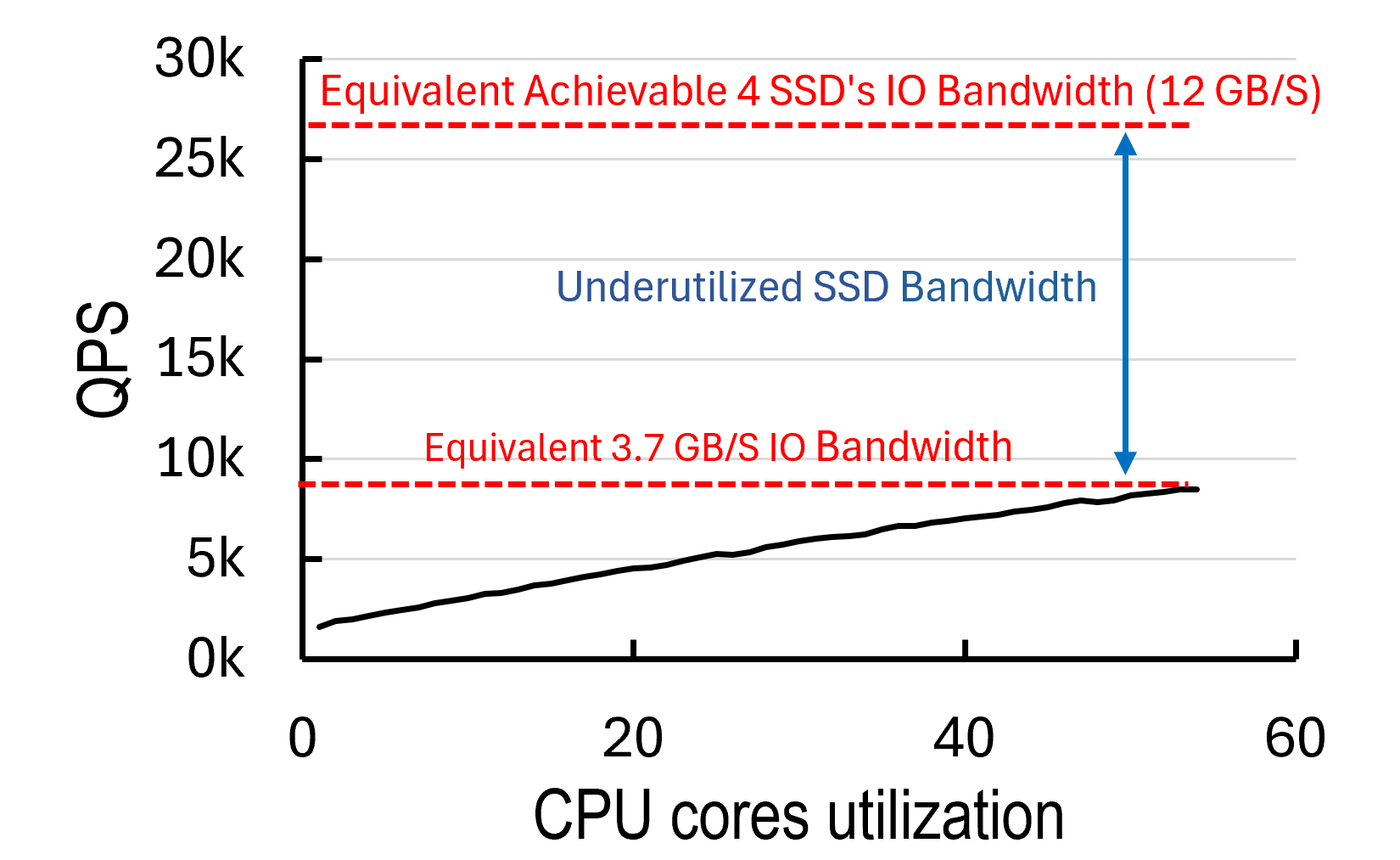} } 
	\caption{DiskANN query throughput variation with increasing CPU cores} 
	\label{disk_ann_core}
\end{figure}  

\begin{figure}
	\centering
	{\includegraphics[width=2.3in]{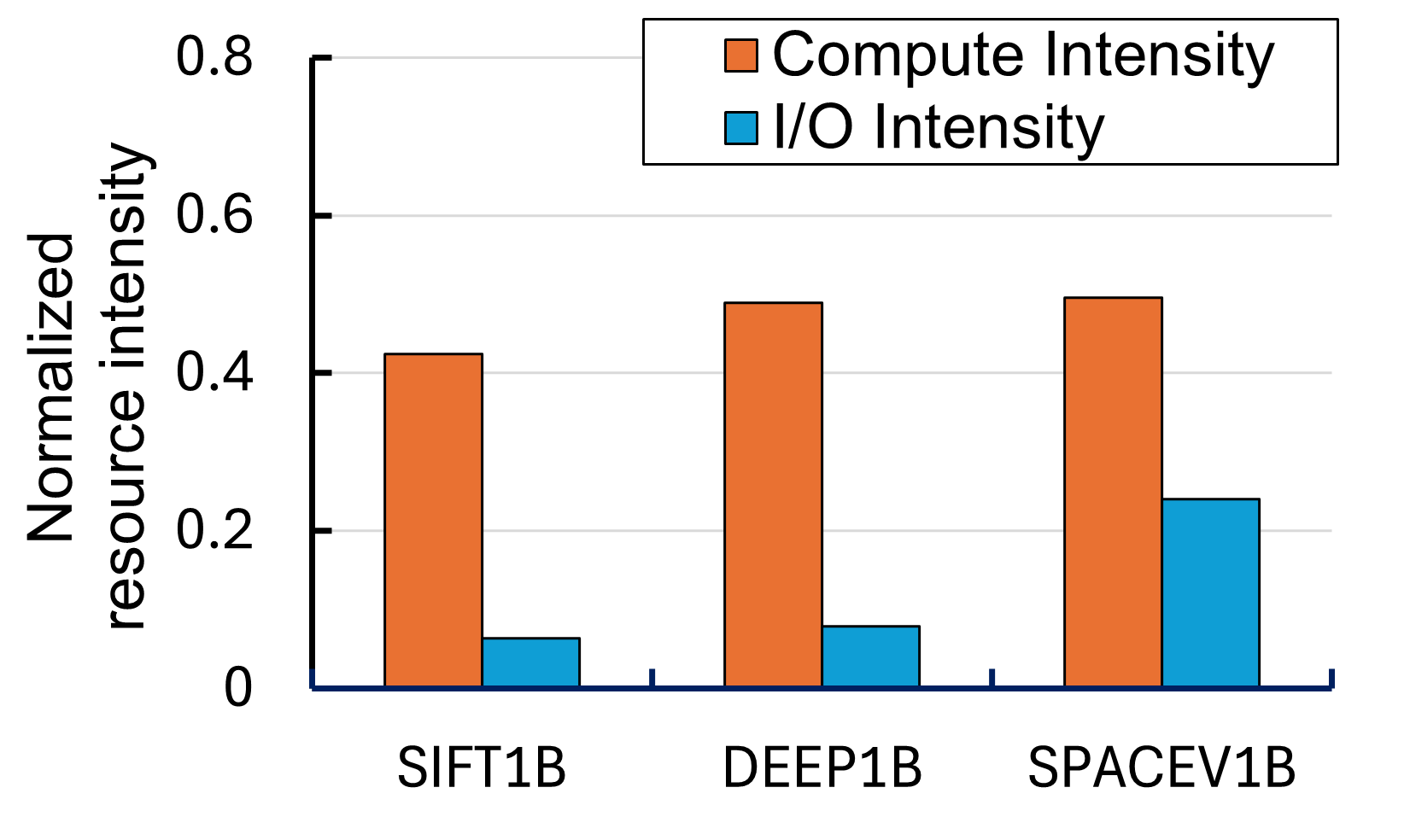} } 
	\caption{Normalized I/O and compute demands of graph indexing across datasets (vs. cluster indexing)} 
	\label{resource_intensity}
\end{figure}  

To minimize SSD accesses and the corresponding computation, emerging ANNS systems (e.g., DiskANN~\cite{diskann}, DiskANN++~\cite{ni2023diskann++}) use a fine-grained vector graph instead of a coarse-grained cluster centroid graph. These systems mainly maintain two data structures: 1) in-memory quantified vectors, and 2) SSD-resident adjacency list of the vector graph and full-precision vectors. Figure~\ref{graph-based_overview} demonstrates how graph-indexing ANNS systems retrieve results iteratively.


\begin{itemize}
\item \textbf{1. Distance Calculation Phase.} The system computes exact query-to-node distances and calculates PQ distances for neighbors using in-memory quantized vectors.
\item \textbf{2. Traversal State Management Phase.} The system maintains exact distances in a result heap while managing PQ distances via a min-heap. The next traversal node is dynamically selected from the min-heap by minimal PQ distance.
\item \textbf{3. Neighbor Access Phase.} The system accesses selected node's full-precision vectors and its neighbor lists from SSDs. 
\end{itemize}




By employing high-precision vector indexing, this approach reduces the candidate size and minimizes distance computations. As shown in Figure~\ref{accessed_candidates} and Figure~\ref{accessed_pages_3}, DiskANN achieves significantly smaller candidate size and lower SSD access requirements compared to SPANN and FusionANNS. 

\section{Motivation}


As shown in Figure~\ref{resource_intensity}, graph-indexing reduces many more SSD accesses than calculation, which makes the system easily bottlenecked by the computation. 
Figure~\ref{disk_ann_core} shows how query throughput changes with varying CPU core counts. We observe that query throughput scales nearly linearly with core count. In particular, when all 52 cores are used, the consumed SSD throughput is only 3.7 GB/s, which is slightly higher than the I/O bandwidth of a PCIe 4.0x4 SSD (Intel P5510~\cite{solidigm_p5510}). When the system scales to multiple SSDs, the SSD I/O throughput is underutilized.

Inspired by FusionANNS, which adopts GPU to address the computation bottleneck of preranking, it is straightforward to consider introducing GPUs to alleviate the computational pressure from distance calculation. 
The overall design seems straightforward, but faces three severe challenges.


\textbf{C1: Low GPU Utilization due to Serialized Execution of Computation and I/O.} Graph-indexing ANNS algorithms inherently suffer from a fundamental computational dependency: the distance calculation phase requires neighbor vector data stored on storage devices, and the neighbor access phase fetches data depending on preliminary distance results. This cyclic dependency creates unavoidable computation stalls. With GPUs, accelerated distance calculations make severe pipeline imbalance - computation completes quicker, leading to a larger proportion of time waiting for I/O. In our experiments (\S~\ref{sec:pipe}), under the same recall@10 condition, the QPS of serial execution demonstrates 67.6\%--73.1\% of our asynchronous execution.

\textbf{C2: Prolonged I/O Latency due to Synchronization in the Kernel-grained GPU I/O Stack.}
Kernel-grained GPU I/O stacks optimize I/O throughput via batched SSD accesses, but any single long-tail latency event within a batch propagates delays to all co-batched requests. In the GPU-accelerated ANNS systems, massively increased concurrent SSD access traffic critically intensifies this latency amplification. In our experiments (\S~\ref{sec:warp-level}), under the same recall@10 condition, the QPS of kernel-grained execution demonstrates 55.3\%--67.4\% of query-grained execution.

\textbf{C3: I/O Amplification due to SSD Page Misalignment.}
In the graph-indexing ANNS access patterns, SSD reads typically use a 4KB minimum granularity since IOPS remain consistent when the access size is smaller than 4 KB~\cite{diskann}. During graph traversal, each query step accesses one graph node. However, standard graph nodes storing vector data and neighbor indices are often significantly smaller than 4KB, causing severe I/O amplification. For example, a 64-degree graph node occupies merely 384B (9.37\% of 4KB), resulting in 90.63\% bandwidth waste per access.


 
\section{Design of \frameworkName{}}
\label{design}

\begin{figure}
	\centering
	{\includegraphics[width=3.4in]{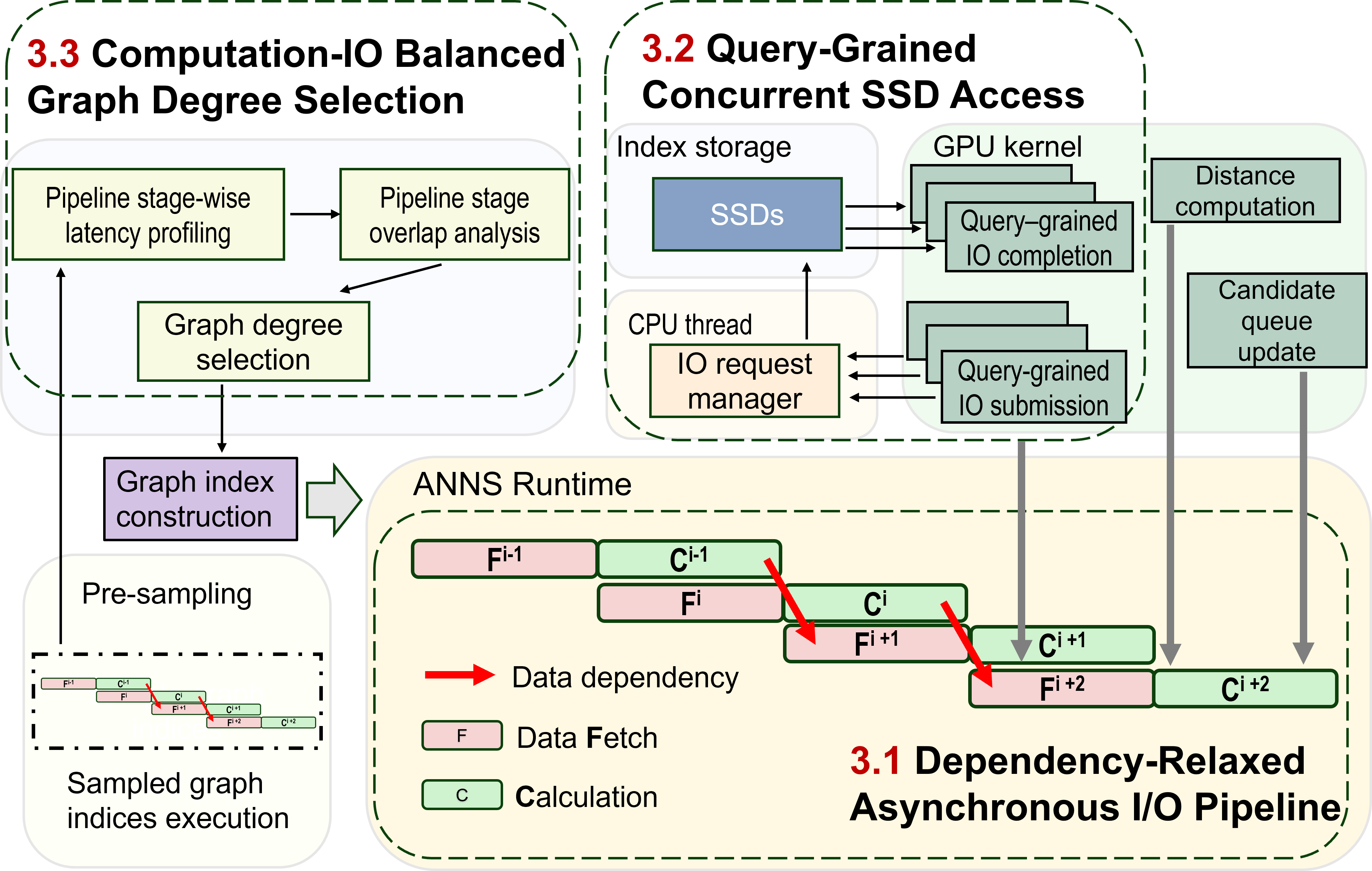} } 
	\caption{\frameworkName{}~overview} 
	\label{Flash-architecture}
\end{figure}

To address these issues, we propose \frameworkName{}, a GPU-accelerated out-of-core graph-indexing ANNS system.
Figure~\ref{Flash-architecture} shows the overall architecture of \frameworkName{}. \frameworkName{} consists of three novel designs. 1) Pipelined I/O-compute processing. It parallelizes GPU computation and I/O accesses by loosening data dependency. 2) query-grained GPU-SSD direct I/O stack. It employs query-grained synchronization to reduce tail latency interference during SSD data retrieval. 3) Sampling-based graph degree selector. Using sample indices to characterize pipeline performance across varying graph degrees under current hardware profiles, thereby selecting optimal degrees that maximize pipeline overlap for distinct SSD quantities. 

\subsection{Dependency-Relaxed Asynchronous I/O Pipeline}
\label{Dependency-Relaxed}


\begin{figure}
	\centering
	{\includegraphics[width=3.4in]{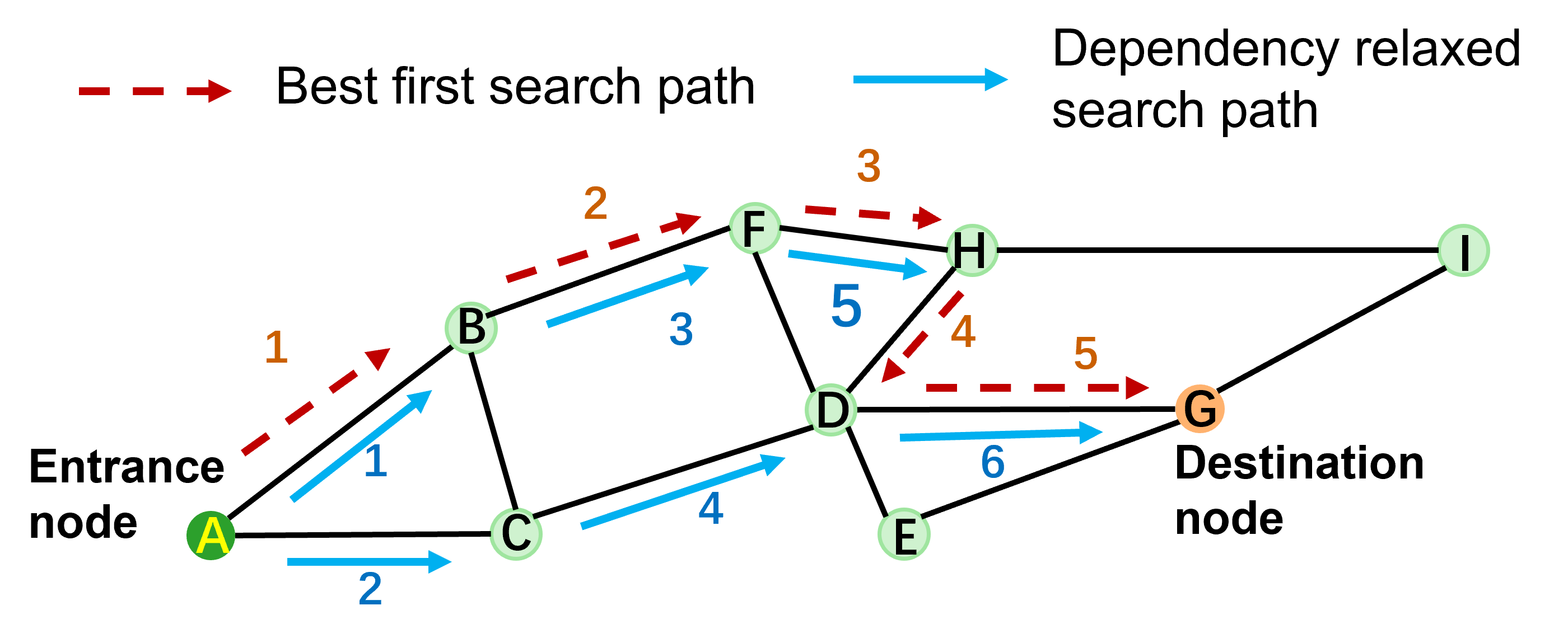} } 
	\caption{Exampled search paths of best-first search vs. relaxed dependency search} 
	\label{fig_relaxed_search_path}
\end{figure} 

\begin{figure}[t!]
    \centering
    \begin{subfigure}[t]{0.48\textwidth}
        \includegraphics[width=\textwidth]{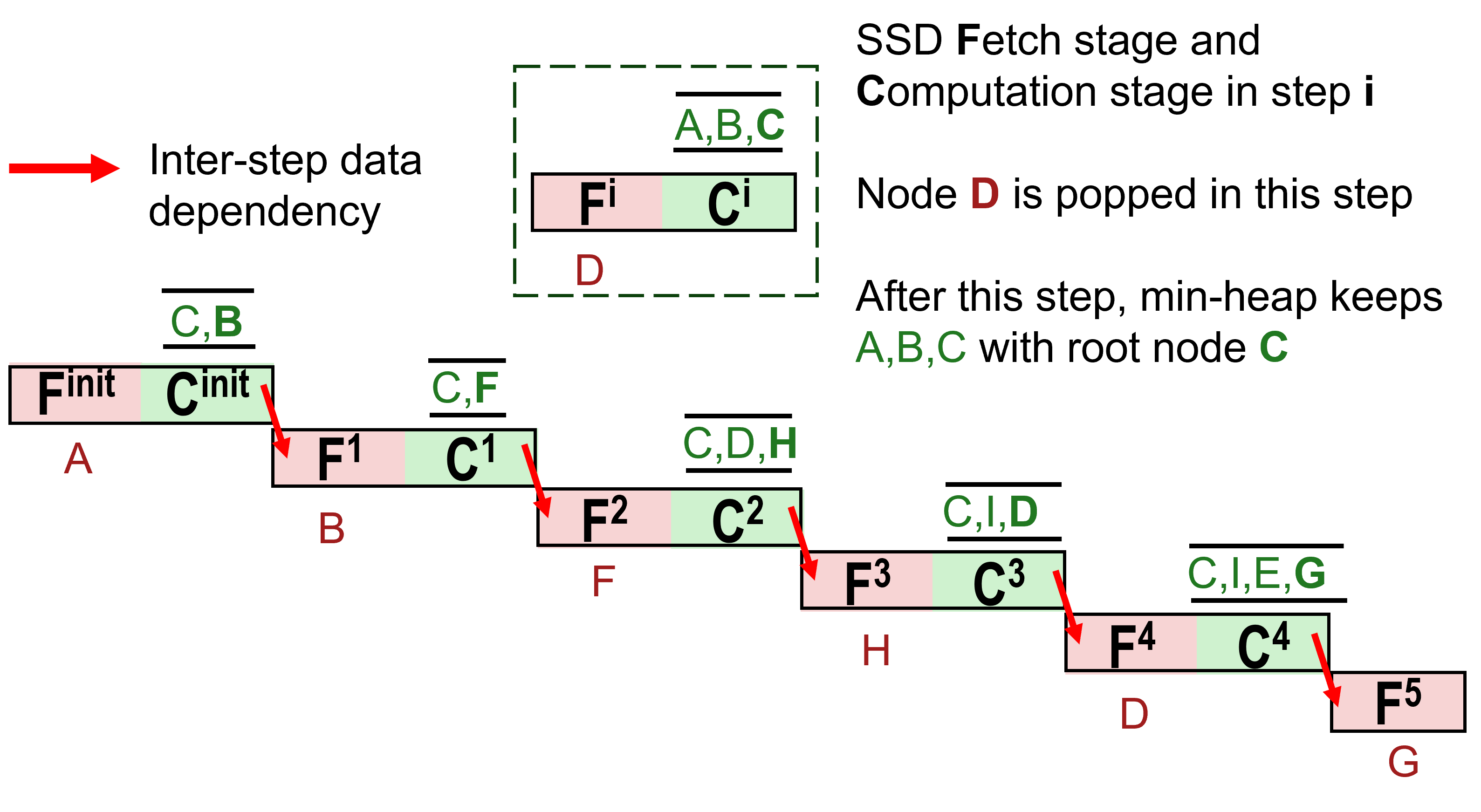}
        \caption{Best-first search pipeline}
        \label{fig:pipeline_a}
    \end{subfigure}
    \\[\baselineskip]  
    \begin{subfigure}[t]{0.46\textwidth}
        \includegraphics[width=\textwidth]{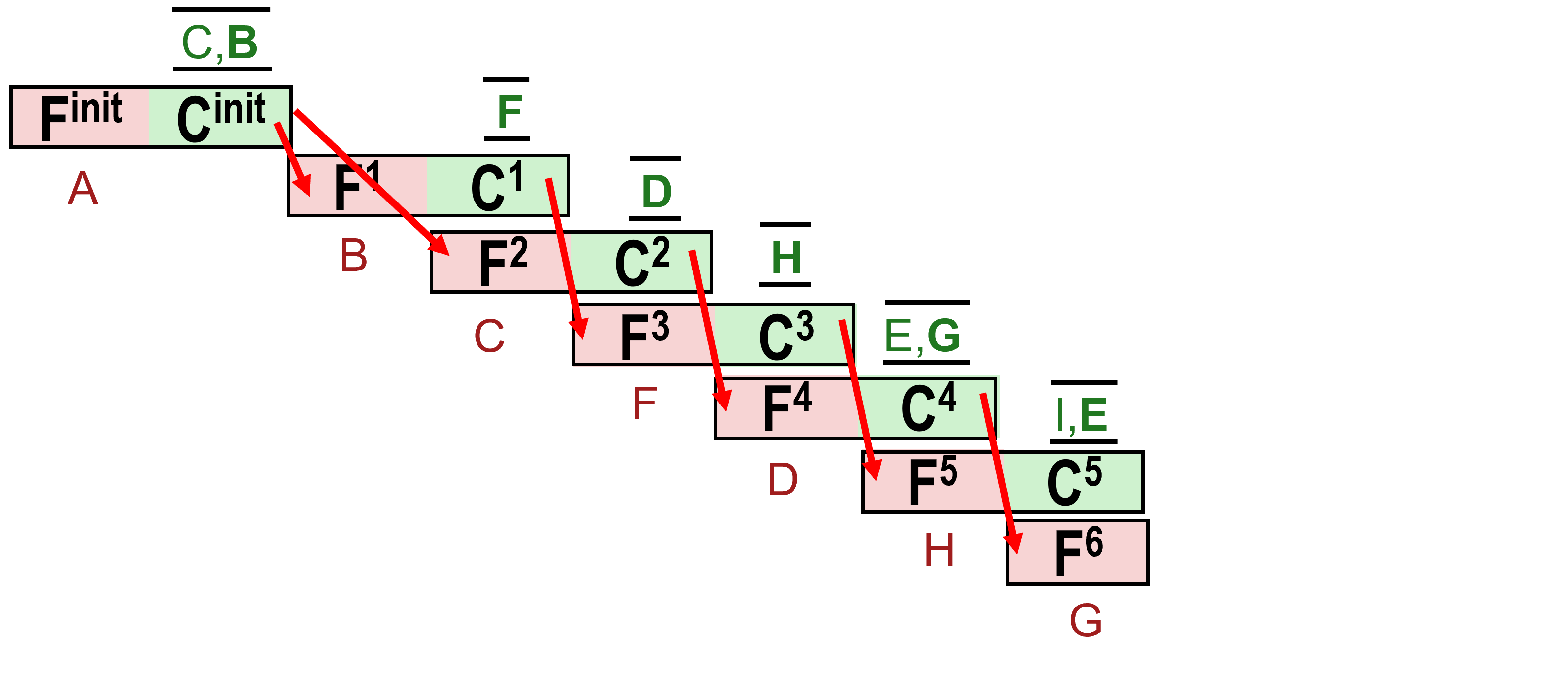}
        \caption{Relaxed dependency search pipeline}
        \label{fig:pipeline_b}
    \end{subfigure}
    
    \caption{Execution pipeline of best-first search and relaxed dependency search}
    \label{fig_dependency_pipeline}
\end{figure}



To address C1, we propose a dependency-relaxed I/O pipeline. Instead of enforcing strict step-by-step dependencies, our design permits SSD accesses to proceed before the preceding compute stage has finished. In this section, we first identify where strict sequential dependencies arise and then present a staleness-aware pipeline that relaxes them. {Finally, we establish convergence guarantees for the dependency-relaxed search and}{ provide an explanation supporting the choice of staleness step.} 


\subsubsection{Data-Dependency in Traditional Best-First Search}

{As the search path shown in Figure~\ref{fig_relaxed_search_path}, traditional best-first search in ANNS graph indices begins from an entrance node and iteratively visits neighboring nodes. In this process, we define one \emph{search step} as a pop–expand iteration: at step $i$, the algorithm pops the current closest node $v_i$ from the candidate min-heap, issues an SSD read to fetch all neighbors of $v_i$, computes their distances to the query, and inserts them into the candidate min-heap.

Figure~\ref{fig:pipeline_a} shows that the algorithm computes neighboring nodes' distances to the query and maintains a candidate min-heap that is ranked by nodes' distances to the query vector. 
Each "min-heap" in figure~\ref{fig_dependency_pipeline} is a snapshot of the candidate min-heap at the \emph{end} of a search step, and the node at the head of the queue is the heap root that will be popped and expanded. Once a node has been popped and its neighbors have been processed, the node is removed from the heap and therefore does not appear in the later 'min-heap'. Each step has two time-consuming stages: 1) an SSD access stage (reading neighbors of the popped node), and 2) a distance computation stage (calculating distances of these neighbor nodes to the query vector and updating the candidate min-heap). 

These two stages exhibit two types of dependencies: a) \textbf{intra-step dependency} that requires distance computation and uses the results of SSD access from the same step, and b) \textbf{inter-step dependency} that requires SSD access in the current step, uses the candidate min-heap updated by the computation results of the previous step. The dependencies prevent the overlapping of the two stages.}

\subsubsection{Dependency Relaxed Strategy}

\begin{figure}
	\centering
	{\includegraphics[width=2.3in]{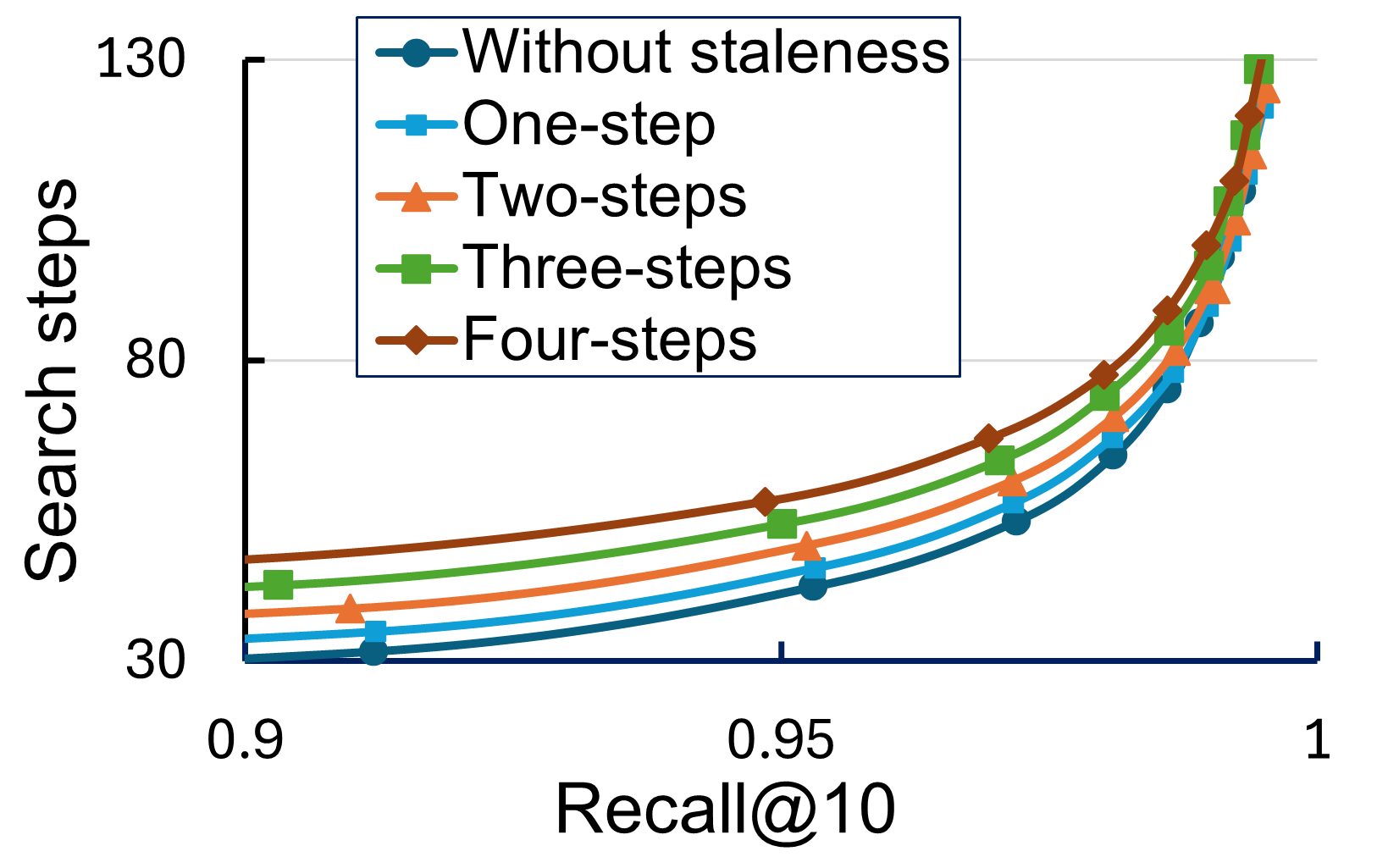} } 
	\caption{Search step count under different staleness steps} 
	\label{Staleness_step}
\end{figure}

{To mitigate low computational resource utilization, we propose relaxing inter-step dependencies to overlap the computation and I/O stages. This approach is grounded in a key insight: the graph search process exhibits a natural tolerance to staleness (using candidate sets from up to k steps prior). This insight is supported by two observations. First, our analysis on the SIFT1B dataset reveals that only 24.3\% of search steps directly depend on the immediately up-to-date candidate min-heap. Second, even with the stale candidate min-heap, the search direction often remains valid.  Consequently, the total number of additional steps incurred by staleness is limited and slight. As shown in Figure~\ref{Staleness_step}, the step count rises by just 2.4\% to 9.8\% per additional staleness step.

Upon the inherent tolerance of graph search to staleness, we implement a dependency-relaxed pipeline to maximize resource utilization. As Figure~\ref{fig:pipeline_b} shows, we break the inter-step dependency by allowing the SSD access stage of the current step \textbf{i} to start without waiting for the completion of the distance calculation stage in the previous step \textbf{i-1}. The node selected for access in the step \textbf{i} is chosen from the candidate min-heap updated based on the distance calculation results from the step \textbf{i-2}. As illustrated in Figure~\ref{fig_relaxed_search_path}, in this case, the step \textbf{i} proceeds without incorporating the distance calculation results from the step \textbf{i-1}, which may overlook a closer node that should be chosen in the step \textbf{i} in the traditional best-first search.  As such, the staleness mechanism may require slightly more steps to search the vector.

However, as shown in the comparison of time consumption between Figure~\ref{fig:pipeline_a} and Figure~\ref{fig:pipeline_b}, the staleness algorithm allows the overlapping of the SSD access stage and distance calculation stage, and the overall execution time can be greatly reduced even with slightly more steps.}


\subsubsection{Convergence Analysis}

We provide the convergence guarantee of the search algorithm under the staleness mechanism with bounded search steps.   

Define $\Delta_t$ as the maximal positional deviation between paths $\mathcal{P}_{\text{relax}}$ and $\mathcal{P}_{\text{strict}}$ at the step $t$, constrained by at most $k$-step directional variances. The deviation satisfies: 
\begin{equation}
    \Delta_t \leq \Delta_{t-1} + k 
    \quad \text{for} \quad t \geq 1
\end{equation}
with initial condition at $t=0$:
\begin{equation}
\Delta_0 = 0 
\end{equation}
since no staleness exists at initialization.

By mathematical induction over steps:

\begin{itemize}
    \item \textbf{Base Case ($t=0$).} Trivially $\Delta_0 = 0 \leq k \cdot 0$
    \item \textbf{Inductive Step.} Assume $\Delta_{t-1} \leq k(t-1)$. Then
\end{itemize}
\begin{equation}
\Delta_t \leq \Delta_{t-1} + k \leq 
k(t-1) + k = kt
\end{equation}

After $T$ steps, the maximal cumulative detour depth $d$ satisfies:
\begin{equation}
d \leq \Delta_T \leq kT
\end{equation}
Consequently, the total traversal steps of the relaxed path are bounded by:

\begin{equation}
    |\mathcal{P}_{\text{relax}}| \leq \underbrace{T}_{\text{strict path steps}} + \underbrace{k \times T}_{\text{max detour steps}} = (k+1) \times T \quad 
\end{equation}
This guarantees that the relaxed search converges within a bounded number of additional steps determined by $k$.

\begin{figure}
	\centering
	{\includegraphics[width=2.3in]{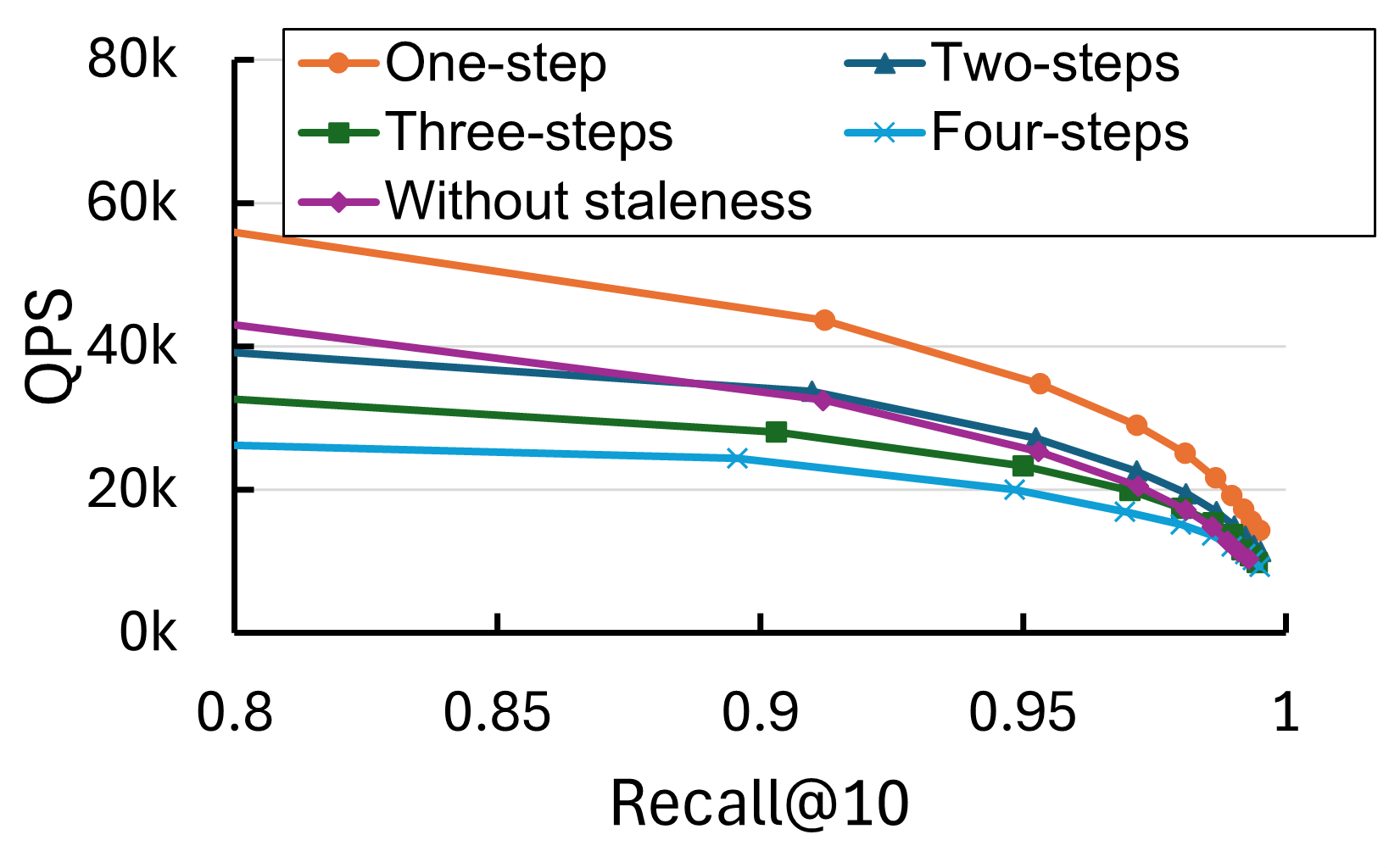} } 
	\caption{End-to-end QPS performance under different staleness steps} 
	\label{Staleness_qps}
\end{figure} 

\subsubsection{Staleness Step Selection}
\label{step_select}

{In \frameworkName{}, we fix the staleness step at k = 1  as the default optimal configuration. This decision is grounded in a co-design principle under our graph degree selector (Section~\ref{sec:degree_sel}). The degree selector tunes the graph degree before index construction to explicitly balance the per-step GPU computation time $T_{c}$ and SSD I/O time $T_{f}$, making them approximately equal. Under this balanced condition, a staleness of $k = 1$  is sufficient to achieve full overlap between the computation and I/O stages, thereby maximizing pipeline efficiency without introducing unnecessary latency, as illustrated in Figure~\ref{fig:pipeline_b}.}

{Increasing the staleness to k = 2 or higher is not beneficial. For example, increasing the staleness to k=2 does not reduce the intrinsic durations of $T_{f}$ and $T_{c}$. While higher staleness can initiate data fetches earlier, the computation unit cannot finish its work ahead of time. As a result, as shown in Figure~\ref{fig:pipeline_b}, the pipeline cycle time remains unchanged compared to k=1. Furthermore, as shown in Figure~\ref{Staleness_step}, a larger k generally increases the total number of search steps due to the use of a more stale candidate, ultimately resulting in higher overall query latency. The same rationale applies to even larger values of k. As shown in Figure~\ref{Staleness_qps}, 
k=1 consistently achieves the highest QPS in FlashANNS's end-to-end evaluation.}


\begin{figure}
	\centering
	{\includegraphics[width=3.1in]{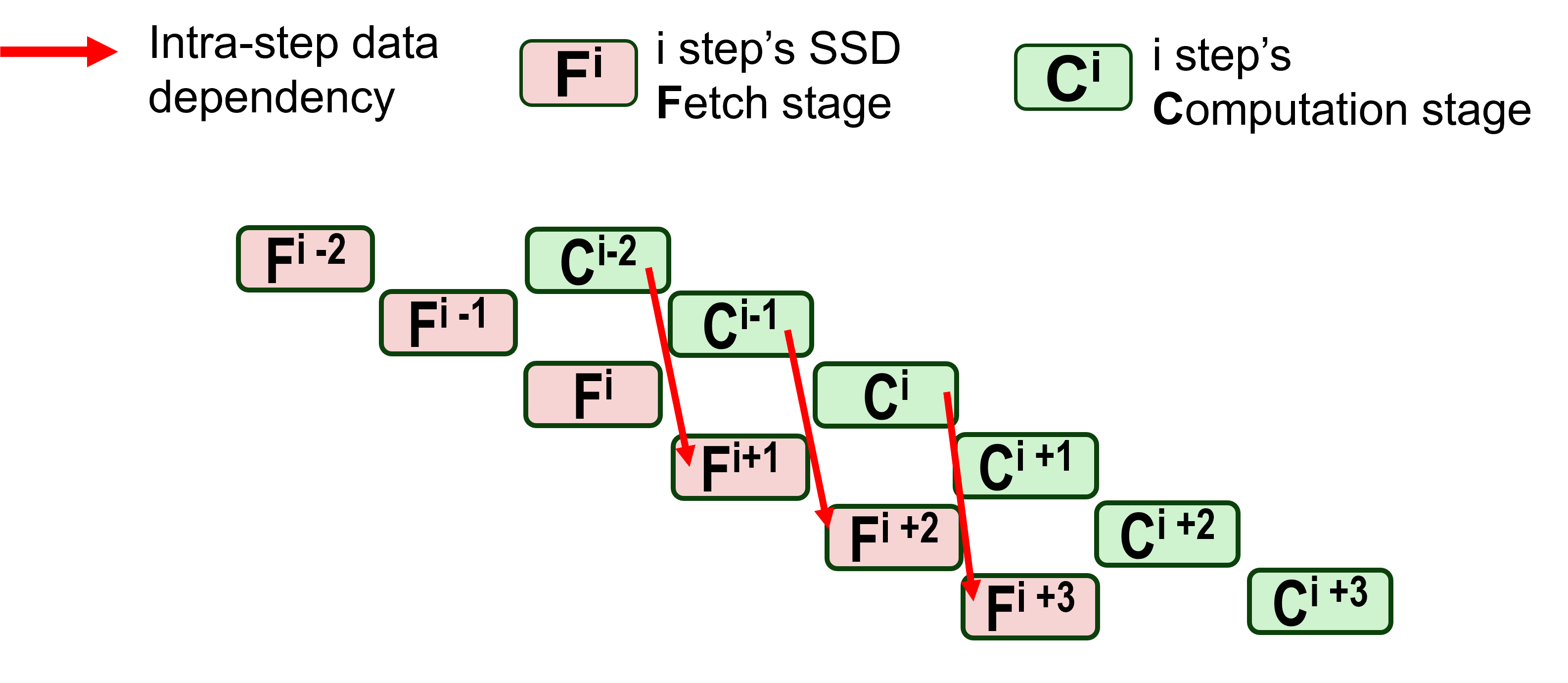} } 
	\caption{Two-steps staleness pipeline} 
	\label{fig:pipeline_ideal_two_steps}
\end{figure}

\subsection{Resource-Efficient Query-Grained Concurrent I/O Stack}
\label{Query-Level}

\begin{figure}
	\centering
	{\includegraphics[width=3.4in]{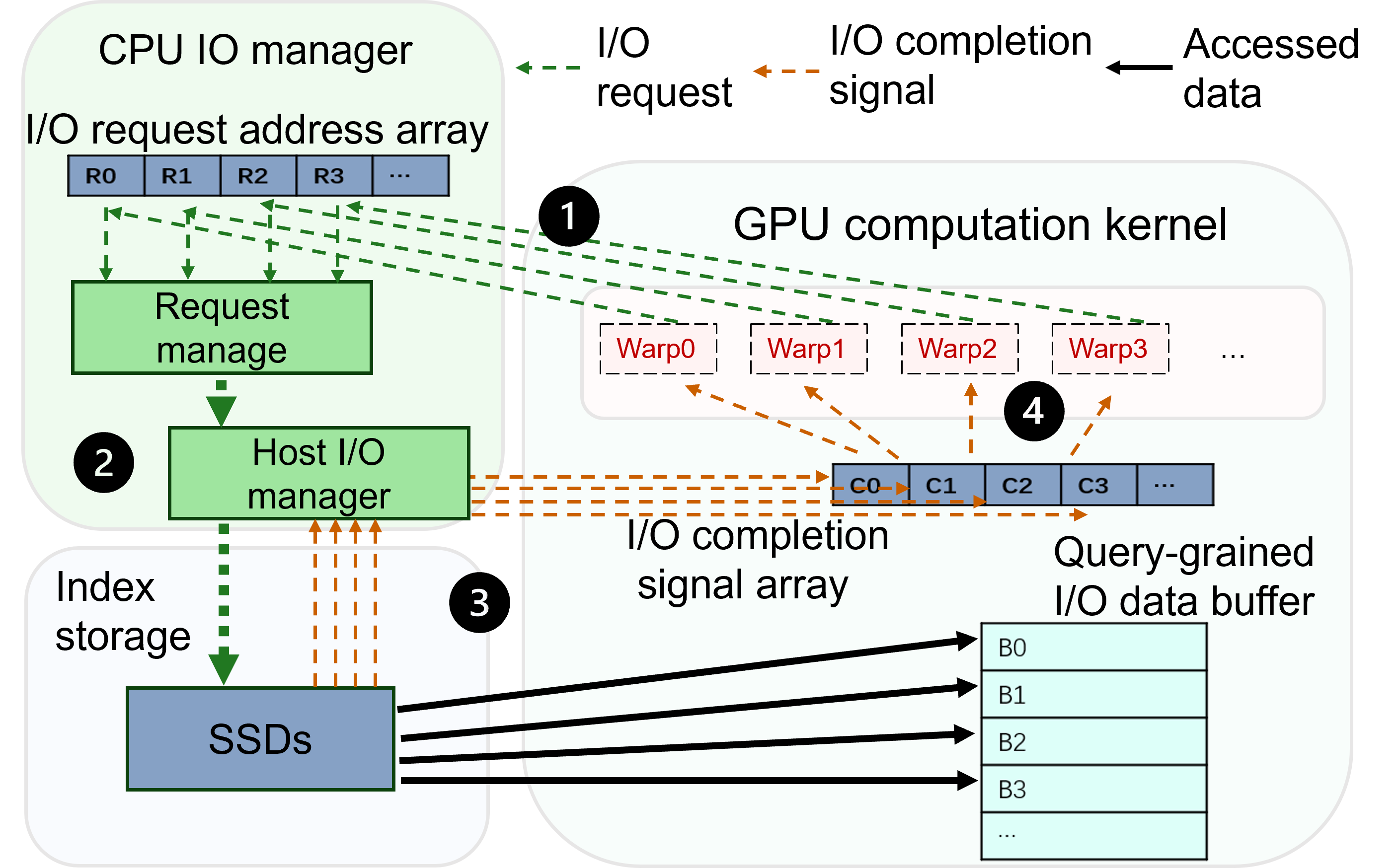} } 
	\caption{I/O process of \frameworkName{}} 
    \vspace{-2ex}
	\label{warp-level_IO_stack}
\end{figure}


\begin{figure}[t]
  \centering
  \begin{subfigure}[t]{\columnwidth}
    \includegraphics[width=\linewidth]{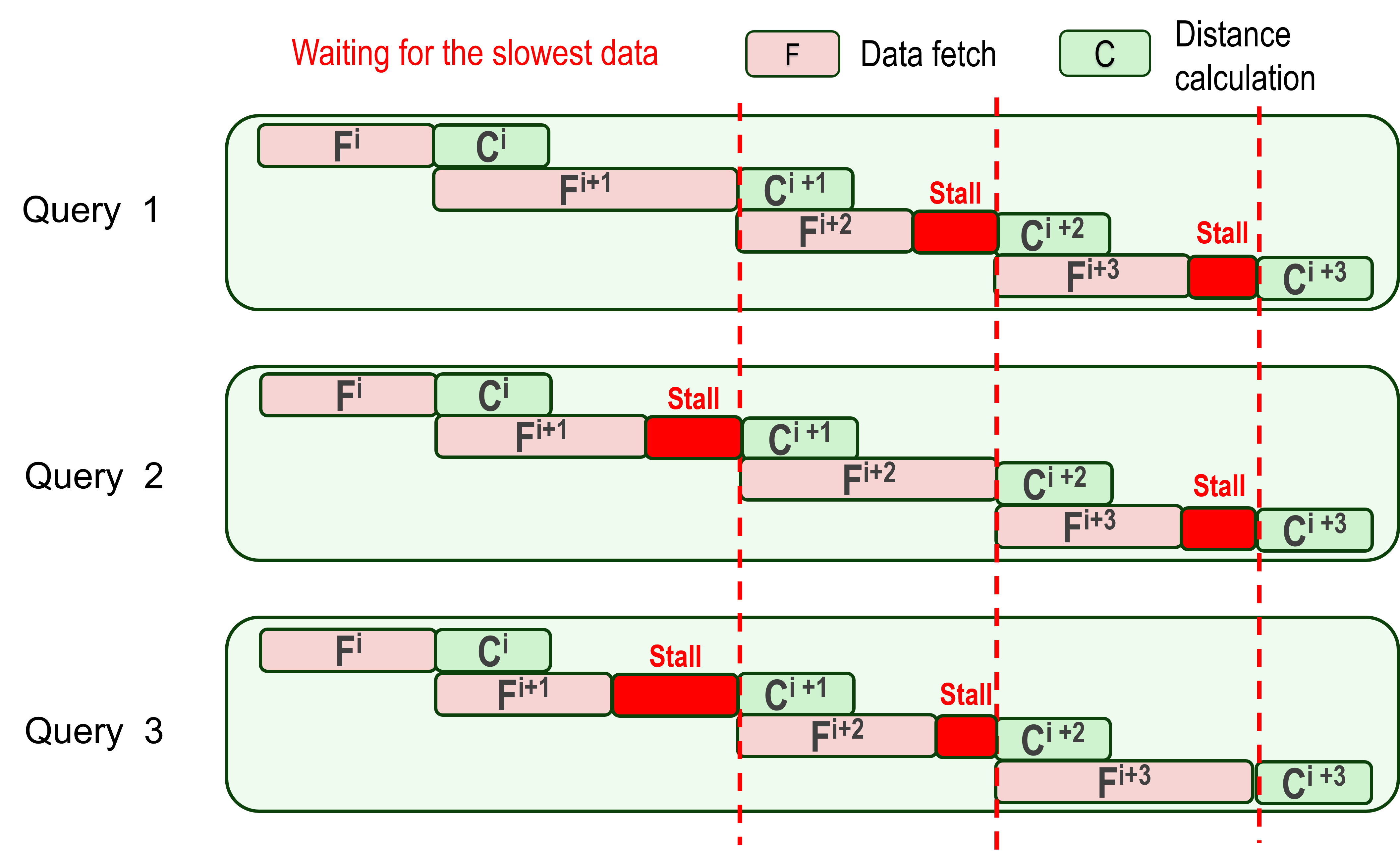}
    \caption{Kernel-Grained SSD access}
    \label{fig:kernal_warp_a}
  \end{subfigure}

  \vspace{0.6em} 

  \begin{subfigure}[t]{\columnwidth}
    \includegraphics[width=\linewidth]{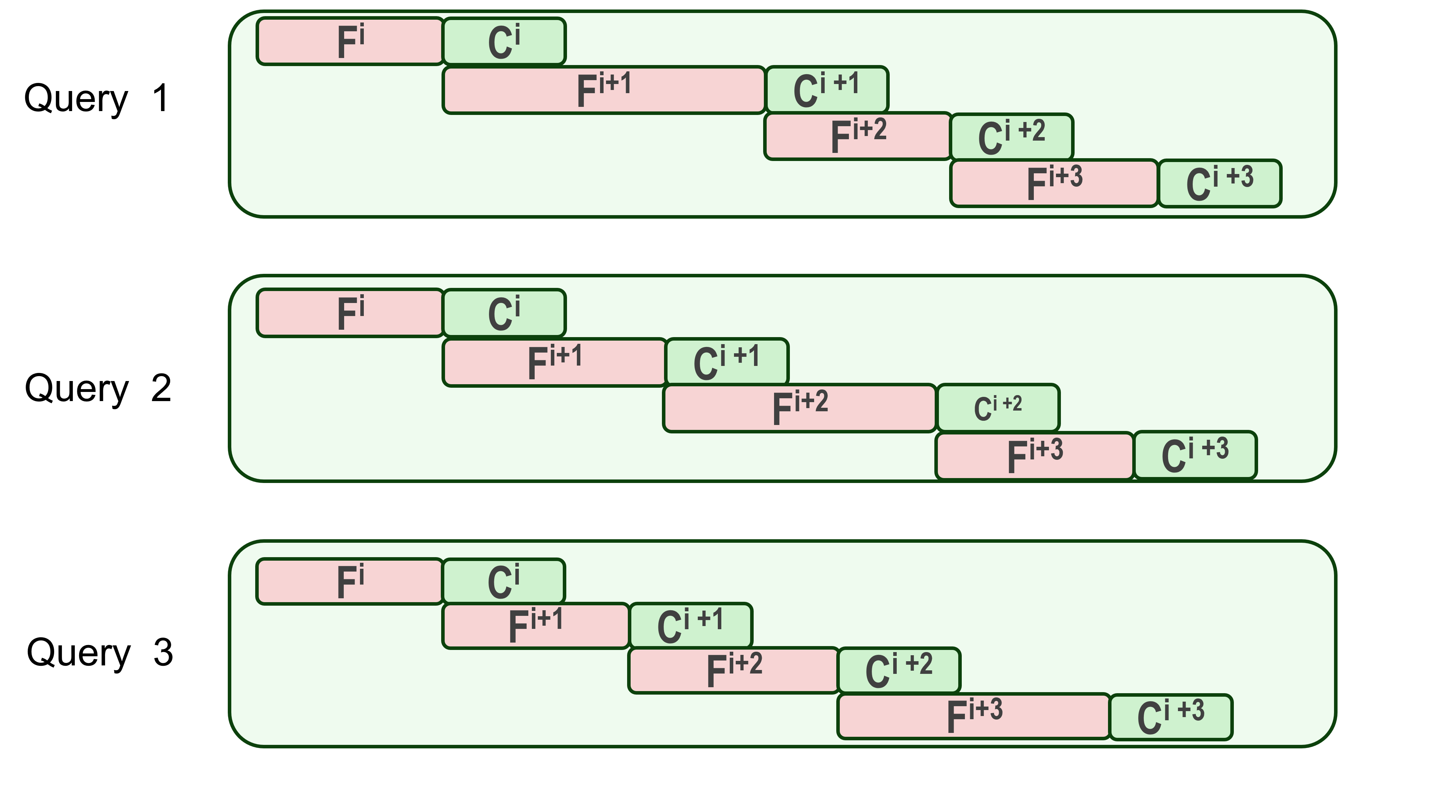}
    \caption{Query-Grained SSD access}
    \label{fig:kernal_warp_b}
  \end{subfigure}

  \caption{Pipeline performance comparison: kernel-grained vs.\ query-grained SSD access}
  \label{fig_kernal_warp}
\end{figure}

\begin{figure}
	\centering
	{\includegraphics[width=2.3in]{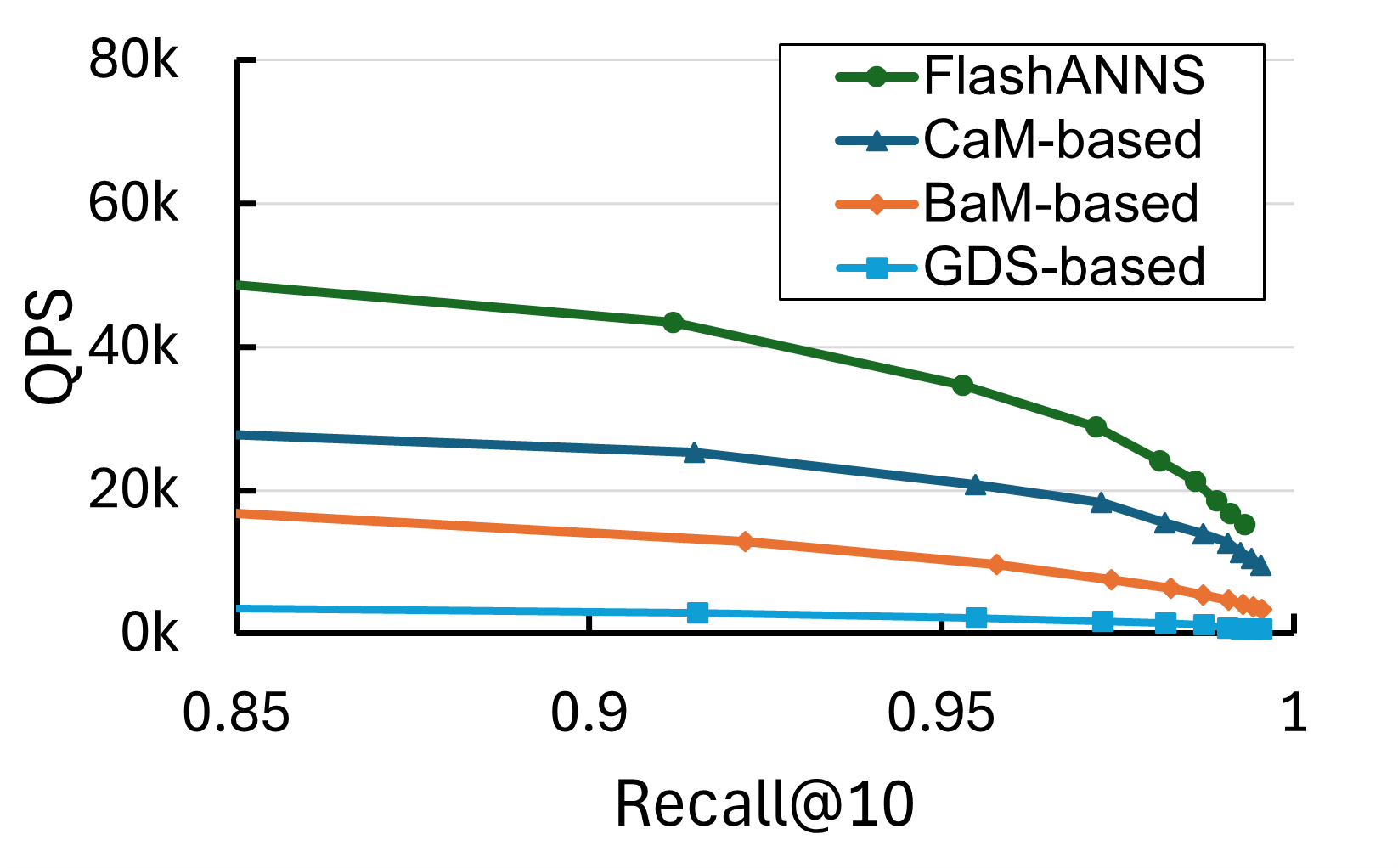} } 
	\caption{End-to-End QPS performance with various I/O stack-based FlashANNS} 
	\label{IO_stack_qps}
\end{figure} 


{To address C2, we introduce a resource-efficient, query-grained concurrent I/O stack tailored to ANNS workloads. It overcomes the limitations of existing I/O stacks when processing the iterative, batched, and fine-grained SSD requests characteristic of ANNS workloads. Our design facilitates direct GPU-SSD data transfer with minimal GPU overhead, entirely bypassing the host OS file system.}


{\noindent\textbf{Design and Execution of the Query-Grained I/O Stack.}
As illustrated in Figure~\ref{warp-level_IO_stack}, \frameworkName{} implements the query-grained I/O stack with three components: (1) A CPU-hosted I/O request array, (2) A GPU-resident completion signal array (3) A GPU data buffer. Each array element corresponds to a single query, and each query is executed by a warp (a group of 32 threads in GPU computing). 


When issuing an SSD read, the pipeline proceeds in four stages that enable computation and I/O overlap: (\cycle{1}) Each warp writes its target SSD block address into a request buffer that is allocated in pinned host memory and mapped into the GPU address space using CUDA’s mapped memory. After submitting the address, the warp is free to resume its computational work, without blocking for the I/O request to complete. (\cycle{2}) A CPU agent polls this array, batches the addresses into read requests, and submits them to the SSD. (\cycle{3}) On completion, the SSD DMA transfers the payload directly into the query-grained data buffer in GPU global memory, and the CPU I/O threads post a completion signal to the GPU-side array. (\cycle{4}) When a warp finishes its current computational tasks and reaches a synchronization point, it checks for the completion signal. If the data is ready, the warp retrieves the data.}

{\textbf{ANNS workload's unique character and challenge}
ANNS workloads are characterized by a large batch of queries, where the processing of each query involves an iterative execution of the SSD access stage and GPU computation stage. This introduces three challenges:
First, ANNS workloads contain a massive number of small, random I/Os. 
Second, our proposed SSD access/GPU computation overlapping requires handling SSD accesses asynchronously.
Third, synchronizing SSD accesses across different queries causes the long-tail latency (stragglers) of individual I/Os to delay the entire batch. In an iterative search process, these delays accumulate significantly, severely degrading overall throughput.
To the best of our knowledge, none of the existing I/O stacks address these challenges simultaneously.}


{\textbf{Limitations of existing GPU–SSD IO stacks.} Existing I/O stacks introduce critical bottlenecks that undermine performance. \textbf{GDS (GPU Direct Storage)}~\cite{cuda2023gpudirect} bypasses the CPU for data transfer but relies on the host OS filesystem for control operations. This reliance necessitates system calls and frequent kernel-user mode transitions, which incur substantial overhead when managing the massive number of small, random I/Os typical in ANNS. \textbf{BaM (GPU-Initiated On-Demand Storage)}~\cite{qureshi2023gpu} employs a GPU-centric I/O control path. BaM can't resolve the second challenge due to its synchronous interface. BaM's threads are forced to wait for their I/O requests to complete instead of performing computations. Thereby, it prevents the overlapping of computation and I/O stages in ANNS workloads. \textbf{CAM (Asynchronous GPU-Initiated, CPU-Managed SSD Management)}~\cite{CAM} attempts a balanced approach with asynchronous, CPU-managed access but enforces a kernel-grained global synchronization model. This means the GPU must wait for all I/O requests in a batch to complete before proceeding. When facing the third challenge of ANNS offload, as shown in Figure~\ref{fig:kernal_warp_a}, CAM's synchronization mechanism waits for the completion of all pending SSD accesses across different queries and thus becomes a source of significant latency inflation.}

{\textbf{How does the query-grained I/O stack overcome the limitations of existing designs?} To overcome GDS's overhead, FlashANNS leverages SPDK to entirely bypass the kernel stack, thereby eliminating filesystem overhead. To improve BaM's synchronous design, \frameworkName{} offloads I/O management to the CPU, and implements an asynchronous data-passing mechanism to make computation and IO process parallelize. To mitigate CAM's long waiting time for batch SSD access, as shown in Figure~\ref{fig:kernal_warp_b}, \frameworkName{} enables each query to independently issue I/O requests and receive completion signals, thereby preventing straggler requests from delaying the entire batch.}

We evaluate FlashANNS with three alternative I/O stacks integrated into its architecture: GDS, BaM, and CAM. As shown in Figure~\ref{IO_stack_qps}, under the 4-SSD SIFT1B setup, our custom I/O stack demonstrates superior performance, achieving 14.5$\times$, 3.9$\times$, and 1.5$\times$ higher QPS than these respective alternatives. These results conclusively demonstrate that FlashANNS' I/O stack delivers superior performance and is better optimized for the access patterns of ANNS workloads compared to existing solutions.

\subsection{Sampling-Based Graph Degree Selector}
\label{sec:degree_sel}

To address C3, we propose a sampling-based graph degree selector. This is a pre–index-construction procedure. It analyzes pipeline behavior on sample indices with different degrees to estimate the relative latencies of I/O and computation, and then selects the degree configuration that maximizes pipeline overlap by making use of the wasted bandwidth caused by I/O amplification.

\subsubsection{Impact of Graph Degree on Computation / IO Bottleneck}

The I/O characteristics of SSDs dictate that when the access granularity is no more than 4KB, IOPS instead of bandwidth becomes the primary performance bottleneck. Graph-index's node sizes are mostly smaller than 4KB. For instance, in DiskANN's~\cite{diskann} default configuration(degree 64) for SIFT1B, a node is 384 bytes. In this case, SSD access is usually performed in 4KB blocks (as noted in DiskANN).  This mismatch between the SSD access block and the graph-index's size node results in significant I/O amplification, as each read retrieves an entire 4KB block but utilizes only a small portion.

The node is composed of an original full-precision vector and indices of its neighbors. While increasing the graph degree to inflate the node size to 4KB seems like a direct solution to eliminate I/O amplification, it introduces a different issue. A higher degree linearly increases the per-step GPU computation time, as more neighbors necessitate more distance calculations. Meanwhile, adding more neighbors yields diminishing marginal returns. Therefore, selecting the optimal graph degree requires a careful balance between I/O and computational load, and must be adapted to the available SSD bandwidth. 


As mentioned in Section~\ref{step_select}, to enable the one-step staleness pipeline, the degree selector chooses to balance the per-step GPU computation time $T_{c}$ and SSD I/O time $T_{f}$, making them approximately equal.

\subsubsection{Workflow of Graph Degree Selector} 
\label{degree_selector_workflow}

Prior to the full index construction, \frameworkName{} uses the lightweight graph degree selector to determine the optimal graph degree. This process operates on a compact data sample (e.g., 100k nodes, index size<200MB) that matches the target dataset's data type and dimensionality. It constructs temporary graph indices for a set of candidate degrees (e.g., 64, 150, 250), where edges are formed using random links rather than true neighbor relationships. It is sufficient to accurately probe the memory and I/O patterns for each degree. Using the same runtime pipeline and a short warm-up of synthetic queries, the selector measures for each candidate \(d\) the data fetch latency $T_{f}(d)$ and per-step computation latency $T_{c}(d)$. The objective is to \textbf{make I/O and computation take the same time per step to maximize pipeline overlap}, so $d$ is calculated as
\begin{equation}
d \;=\; \arg\min_{d}\, \left|\, T_{\mathrm{c}}(d) - T_{\mathrm{f}}(d) \,\right|.
\end{equation}

\subsubsection{Overhead of Graph Degree Select}
By operating on a small sample (e.g., 100k nodes, 0.01\% of a billion-scale dataset) with random links, this profiling avoids the cost of building true neighborhoods, making both graph construction and performance measurement extremely low-cost. The entire process completes in minutes, incurring less than 1\% overhead compared to the multi-hour runtime of a full index construction.



\subsubsection{Hardware Adaptation via the Degree Selector}
\label{hardware_change}
{The degree selector equips \frameworkName{} with the capability to effectively utilize diverse hardware settings. The degree selector guides the user to leverage hardware improvements by re-balancing the computational load $T_c$ and the I/O latency $T_f$, ensuring the pipeline remains efficient.

In case of using SSDs with higher IOPS, the degree selector guides the user to decrease the graph degree. This reduces $T_c$ to realign with the shorter $T_f$, thereby shortening the pipeline cycle and thus accelerating queries. In case of using a faster GPU, the degree selector guides the user to increase the graph degree. This leverages the additional computational capacity to examine more neighbors per step while maintaining the same pipeline cycle time, and ultimately reduces the total number of search steps and thus speeds up overall query execution.}


\section{Evaluation}
\label{sec:experiments}

Our evaluations aim to answer the following questions:
\begin{itemize}
\item How does the performance of \frameworkName{} compare to other SSD-based frameworks on a single SSD and multiple SSDs (\S\ref{sec:end})?
\item How effective is the dependency-relaxed asynchronous I/O pipeline  (\S\ref{sec:pipe})?
\item How effective is the query-grained concurrent SSD access  (\S\ref{sec:warp-level})?
{\item How does throughput scale when increasing the returned top-$k$ (\S\ref{sec: return sets})?}
\item How does the computation-I/O-balanced graph degree selector guide degree selection (\S\ref{sec:degree_select})?
\item How does \frameworkName{} perform on larger-than-memory indices  (\S\ref{sec:30B})?
\end{itemize}

\subsection{Experimental Setting}

\textbf{Experimental Platform. }The experiments are conducted using a single server equipped with dual Intel Xeon Gold 5320 processors operating at 2.20 GHz (52 threads), 768 GB of DDR4 system memory, and an NVIDIA 80 GB-PCIe-A100 GPU. The storage subsystem comprises eight 3.84TB Intel P5510 NVMe SSDs configured in a PCIe 4.0 x16 topology. The platform runs Ubuntu 22.04 LTS.

\noindent \textbf{Datasets. }Our experimental configuration incorporates three canonical billion-scale datasets extensively adopted in high-dimensional similarity search benchmarks:
\begin{itemize}
\item \textbf{SIFT-1B} comprising 1 billion 128-dimensional vectors with unsigned 8-bit integer (uint8) precision, evaluated using 10,000 query instances.
\item \textbf{DEEP-1B} featuring 96-dimensional floating-point vectors (float32) across 1 billion entries, benchmarked with 10,000 queries.
\item \textbf{SPACEV-1B} containing 100-dimensional signed 8-bit integer (int8) vectors at billion-scale, tested with 29,300 queries.

\end{itemize}

\noindent \textbf{Baselines. }\frameworkName{} has three baselines to compare.

\begin{itemize}
\item \textbf{SPANN~\cite{spann}}: A clustering-based SSD-resident framework that stores cluster lists on SSDs while maintaining cluster centroids in CPU memory. Its search mechanism achieves low latency at the cost of computationally intensive per-query operations. 
\item \textbf{DiskANN~\cite{diskann}}: A graph-indexed SSD-resident framework that stores both index graphs and raw vectors on SSDs, complemented by product quantization compressed vectors in CPU memory. It achieves optimized performance through parallel searching. 
\item \textbf{FusionANNS~\cite{fusion}}: A GPU-accelerated clustering-based SSD-resident framework leveraging cooperative CPU-GPU processing to optimize ANNS.
\end{itemize}

\noindent \textbf{Evaluation Metrics. }
We quantify query throughput in queries per second (QPS) and accuracy via recall@10. Recall@10 measures the proportion of true top-10 neighbors retrieved from ground truth among ANNS-returned candidates. Cluster-indexing systems tune recall by adjusting the count of retrieved posting lists during graph traversal, whereas graph-indexing systems control recall by configuring the candidate min-heap size. On QPS-recall@10 tradeoff curves, superior implementations occupy top-right positions, achieving higher QPS with greater accuracy under identical configurations.

\subsection{End-to-End QPS-Recall Tradeoff}
\label{sec:end}

\begin{figure*}[t!]
    \centering
    \renewcommand{\arraystretch}{1.5}
    \begin{tabular}{c c c c}
        & \textbf{SIFT1B} & \textbf{DEEP1B} & \textbf{SPACEV1B} \\
        \rotatebox[origin=c]{90}{\textbf{1 SSD}} & 
        \includegraphics[width=0.3\linewidth,valign=c]{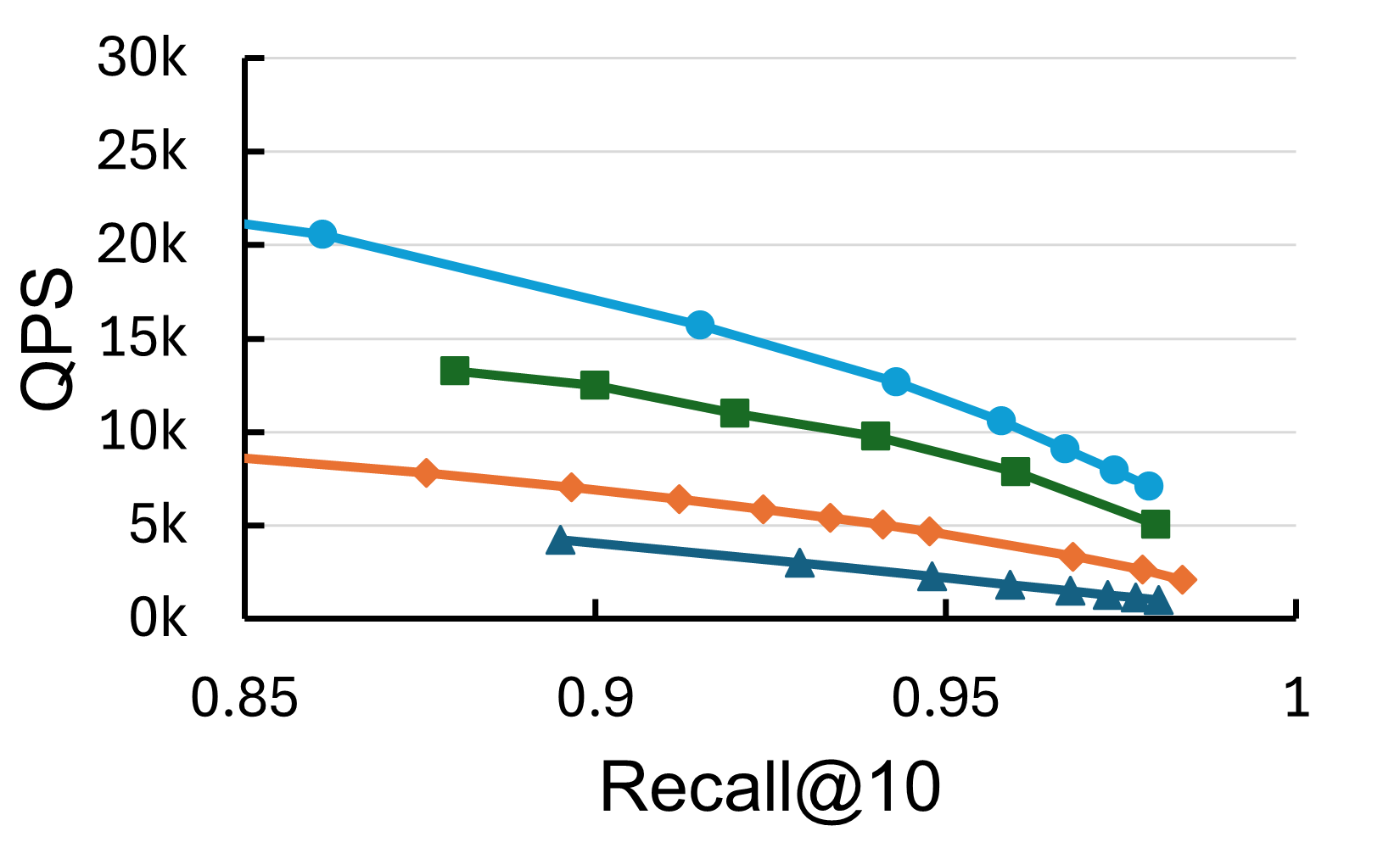} & 
        \includegraphics[width=0.3\linewidth,valign=c]{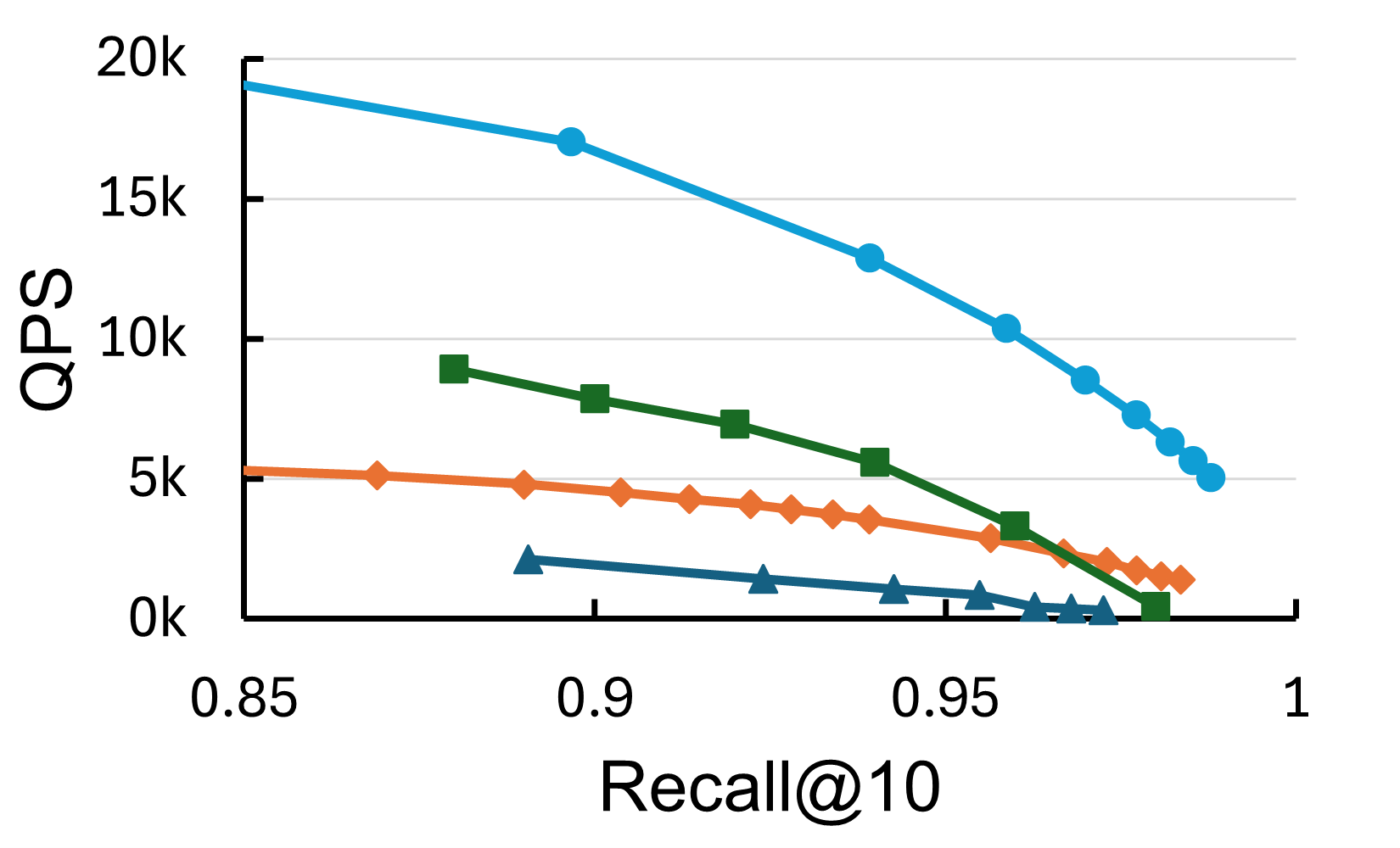} & 
        \includegraphics[width=0.3\linewidth,valign=c]{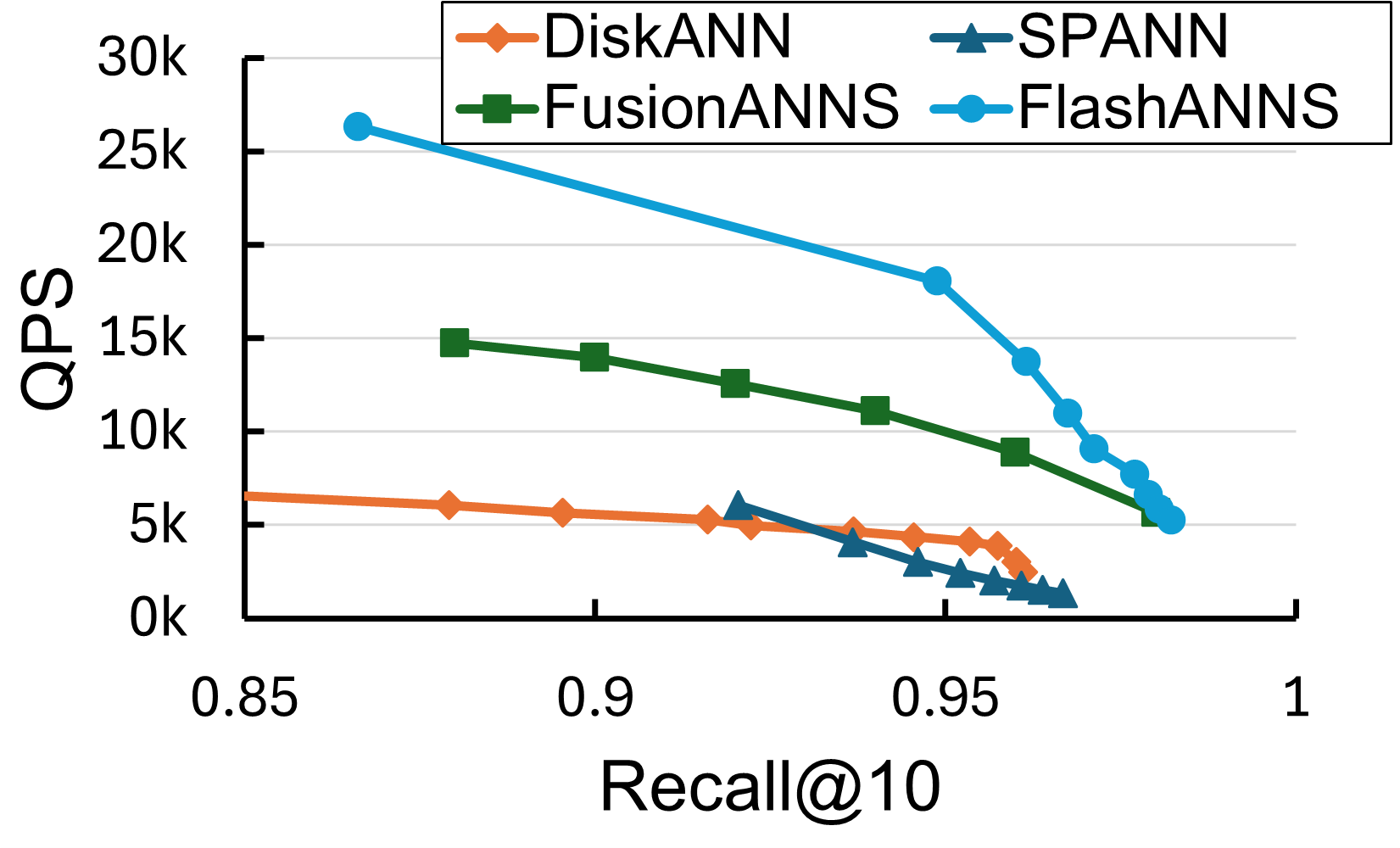} \\
        
        \rotatebox[origin=c]{90}{\textbf{2 SSDs}} & 
        \includegraphics[width=0.3\linewidth,valign=c]{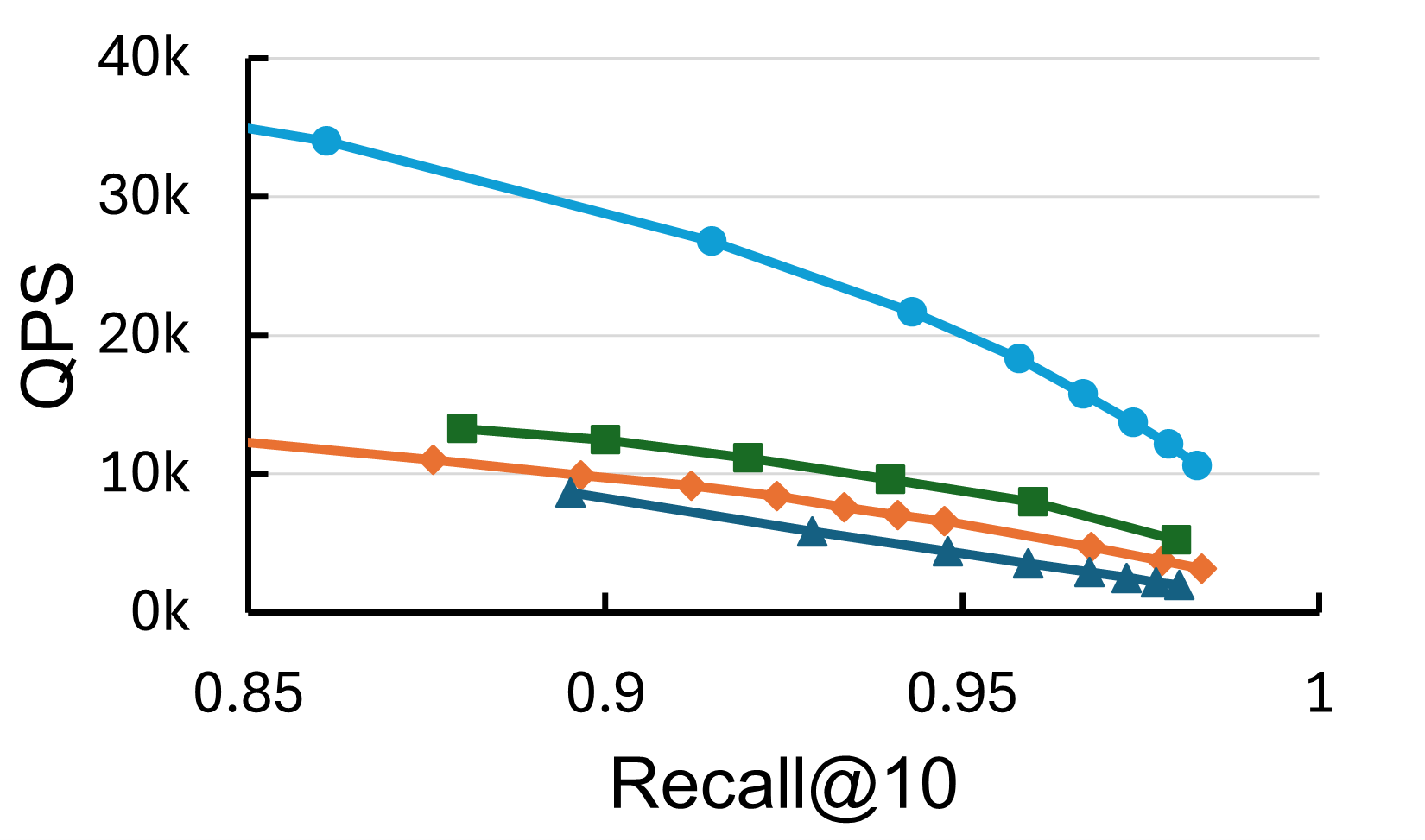} & 
        \includegraphics[width=0.3\linewidth,valign=c]{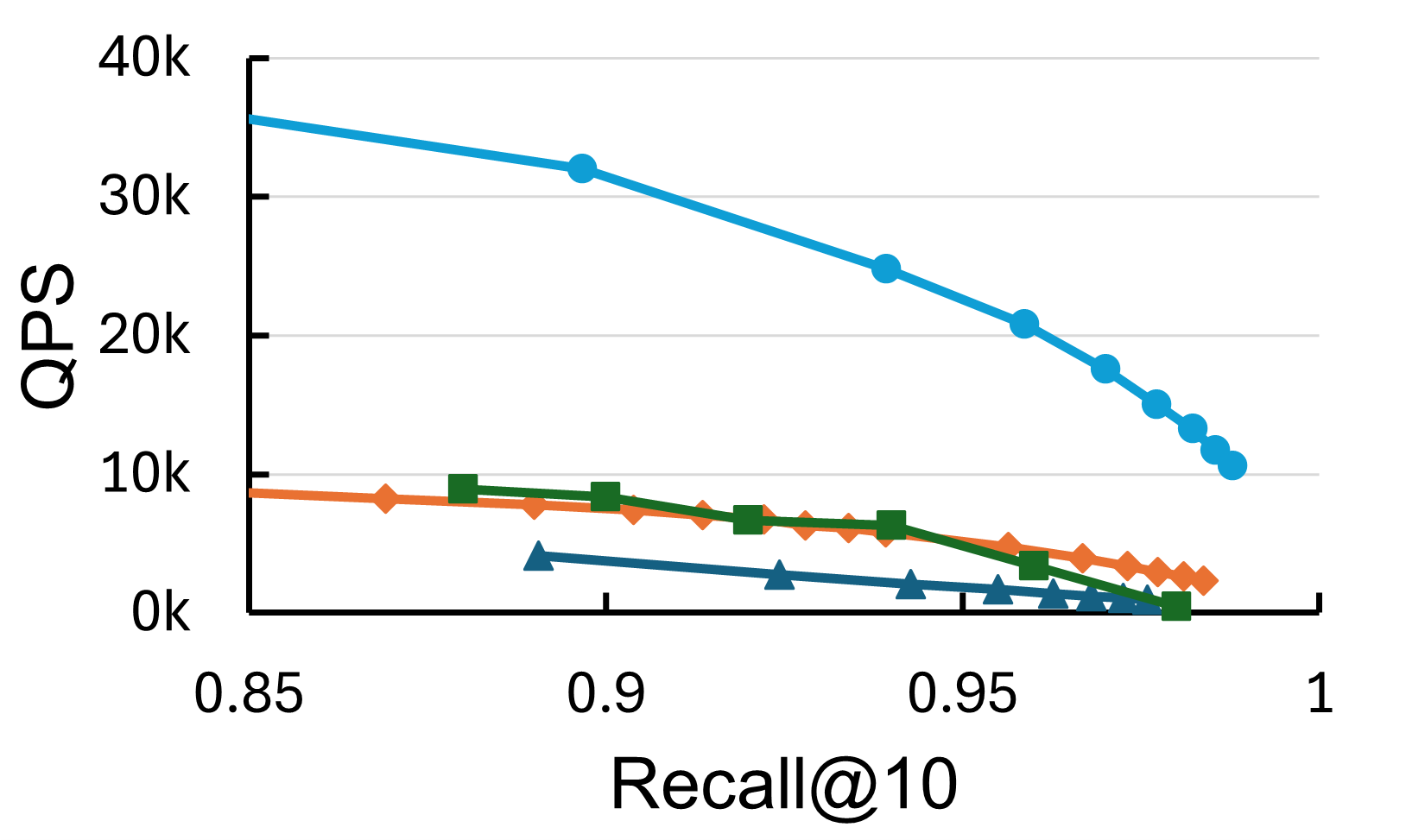} & 
        \includegraphics[width=0.3\linewidth,valign=c]{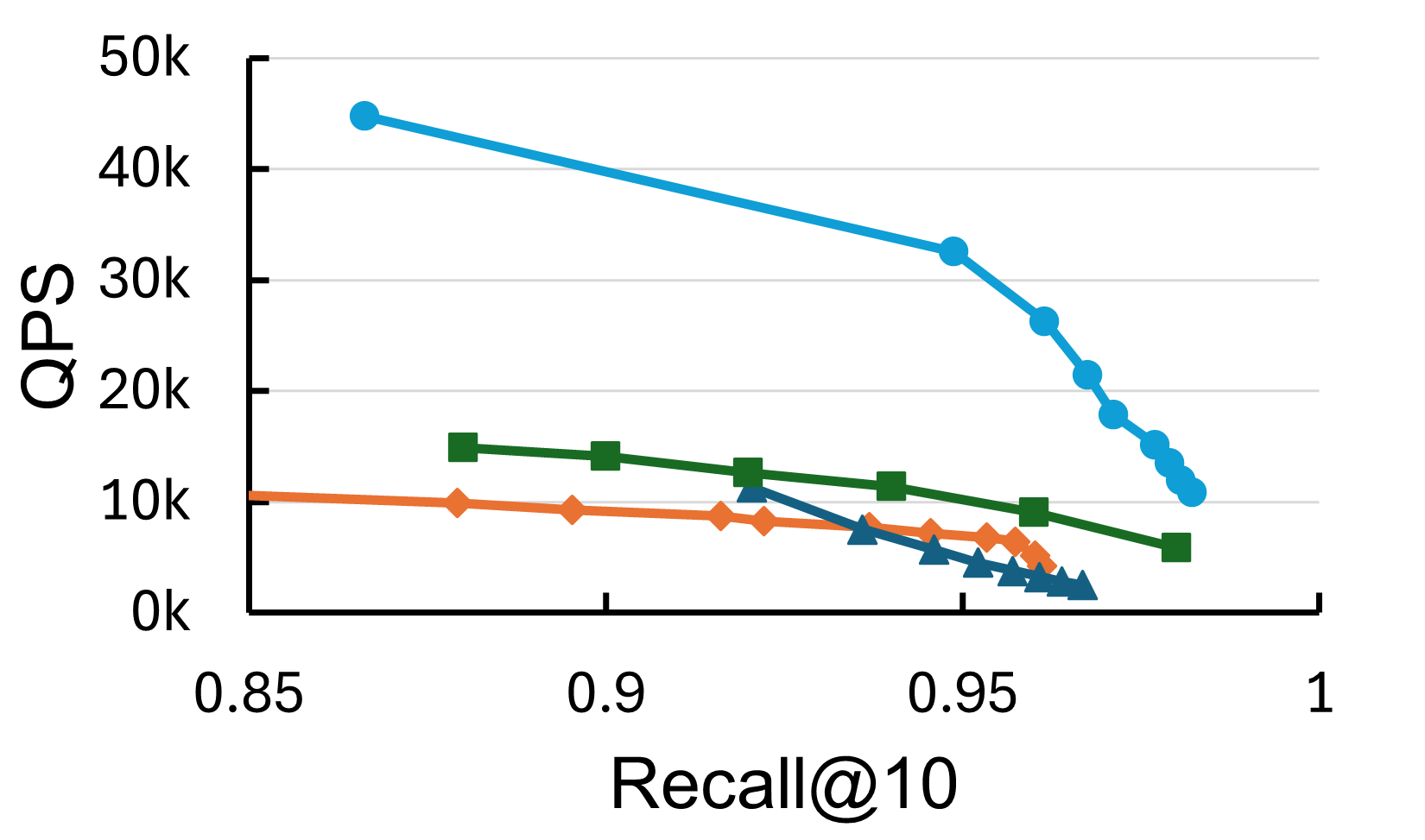} \\
        
        \rotatebox[origin=c]{90}{\textbf{4 SSDs}} & 
        \includegraphics[width=0.3\linewidth,valign=c]{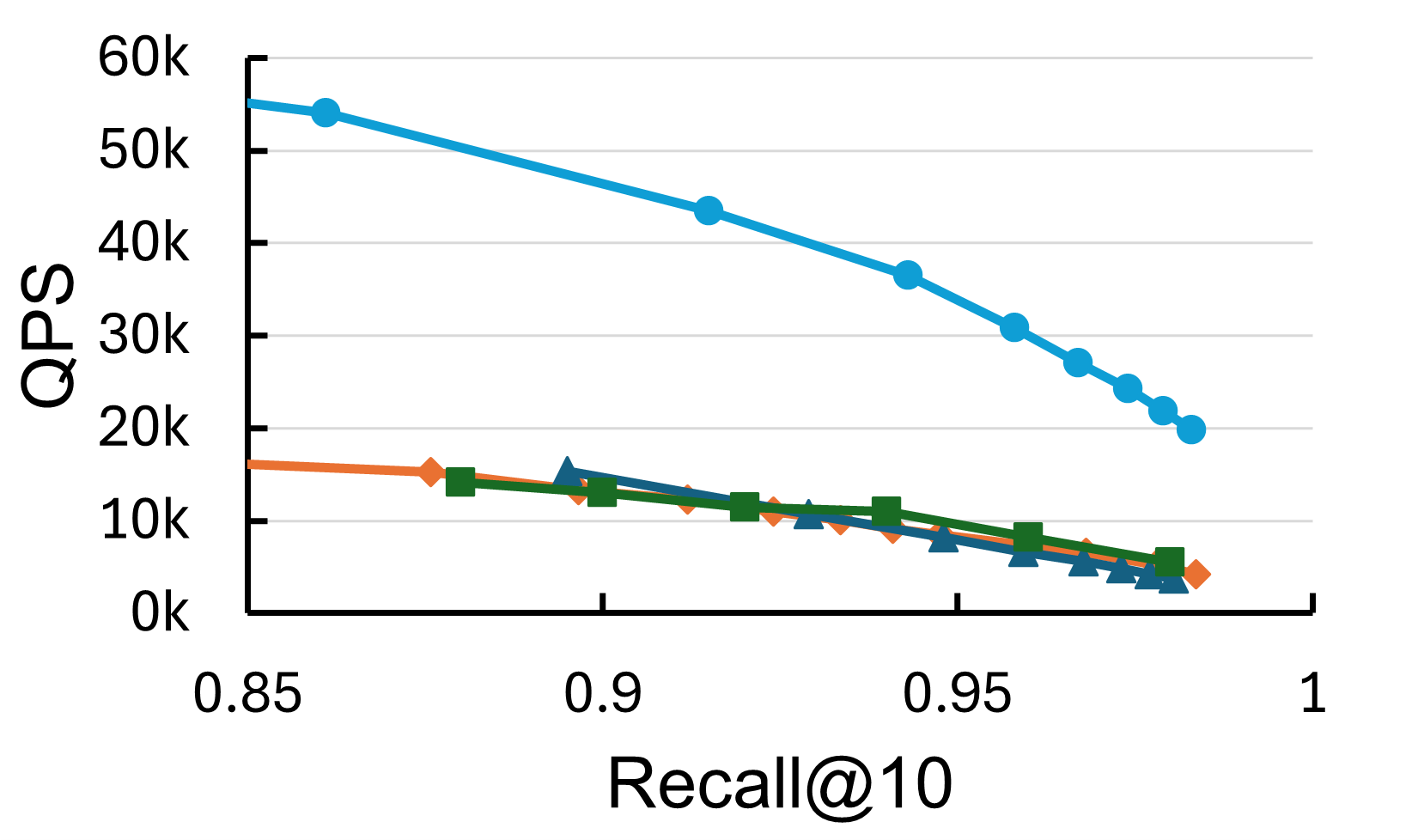} & 
        \includegraphics[width=0.3\linewidth,valign=c]{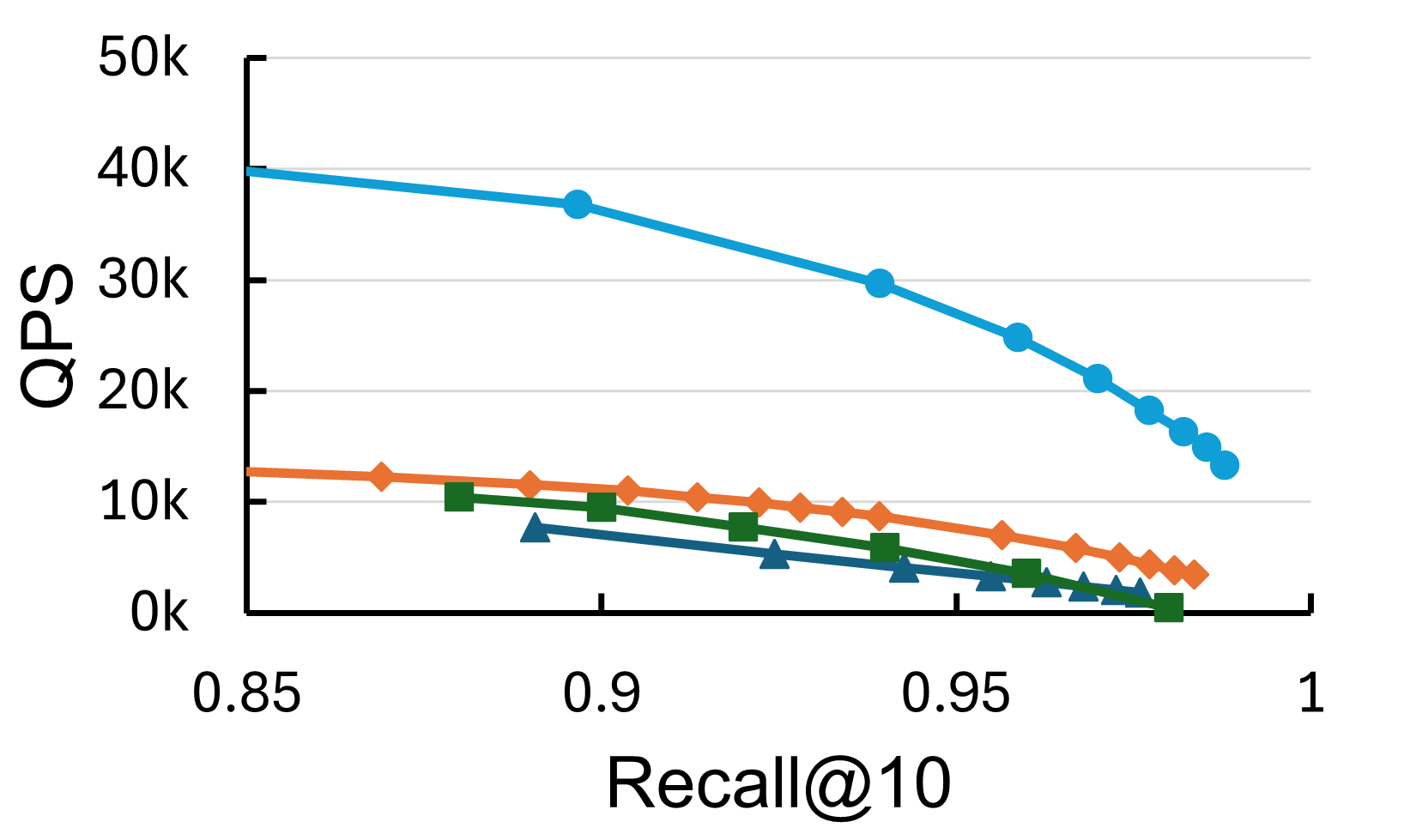} & 
        \includegraphics[width=0.3\linewidth,valign=c]{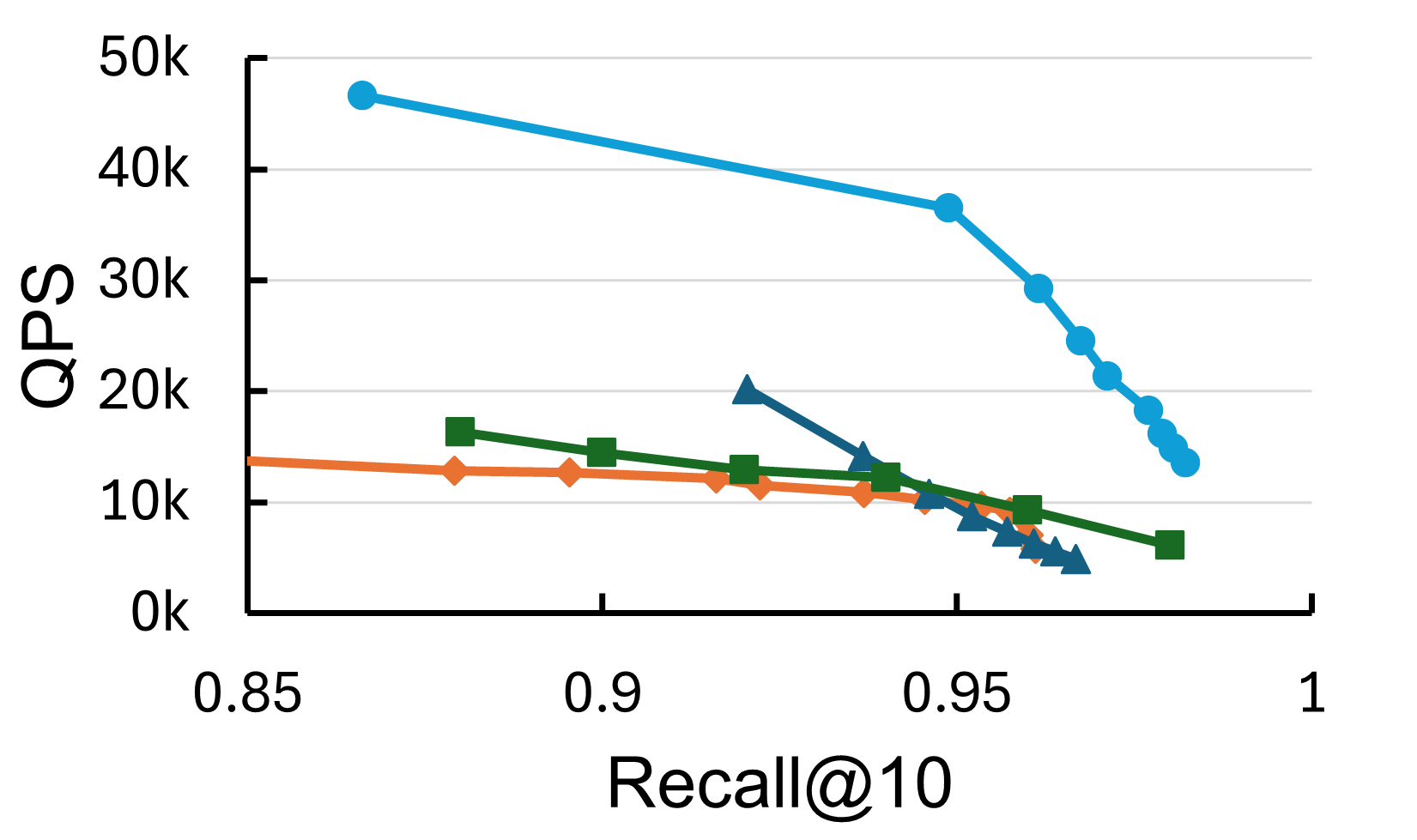} \\
        
        \rotatebox[origin=c]{90}{\textbf{8 SSDs}} & 
        \includegraphics[width=0.3\linewidth,valign=c]{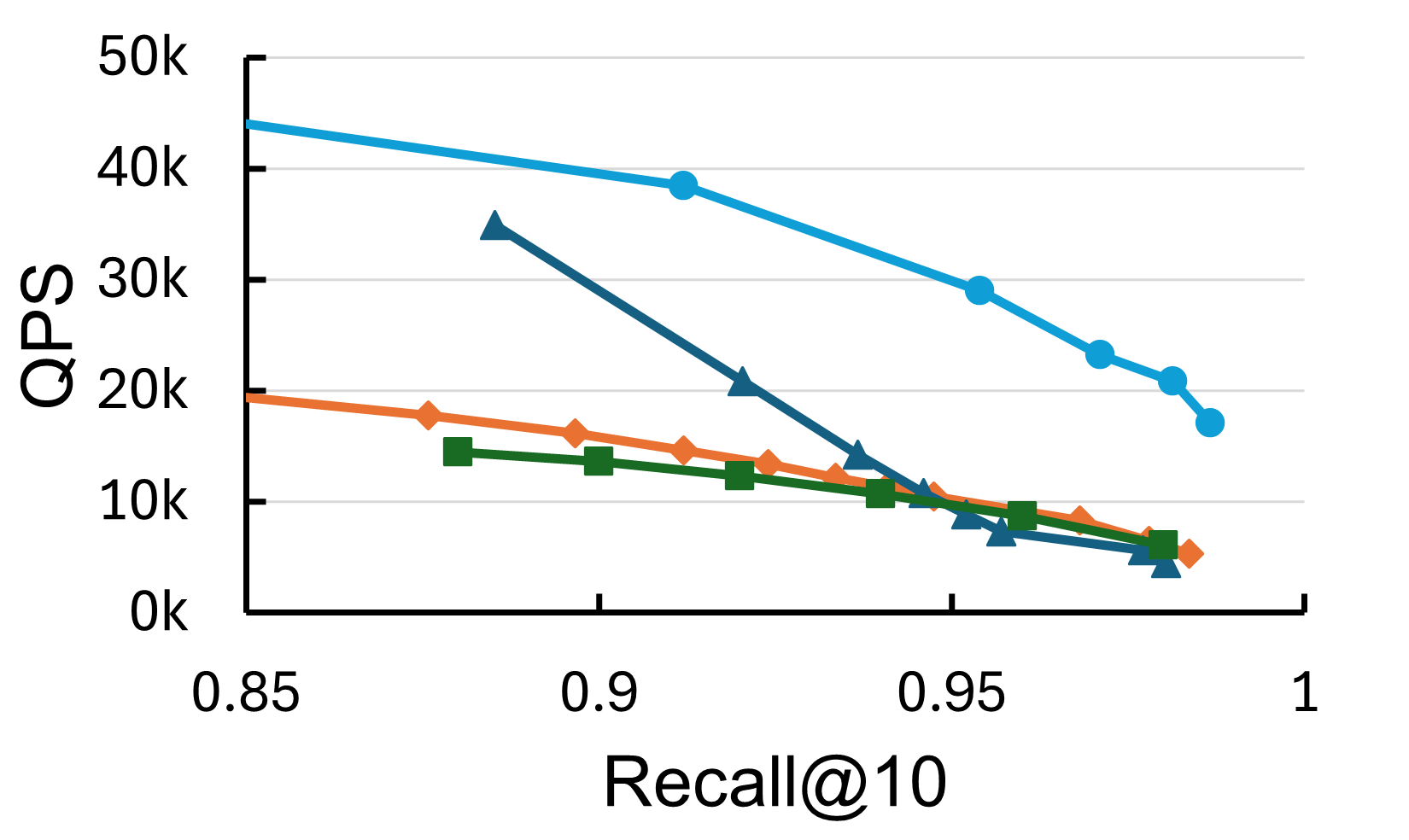} & 
        \includegraphics[width=0.3\linewidth,valign=c]{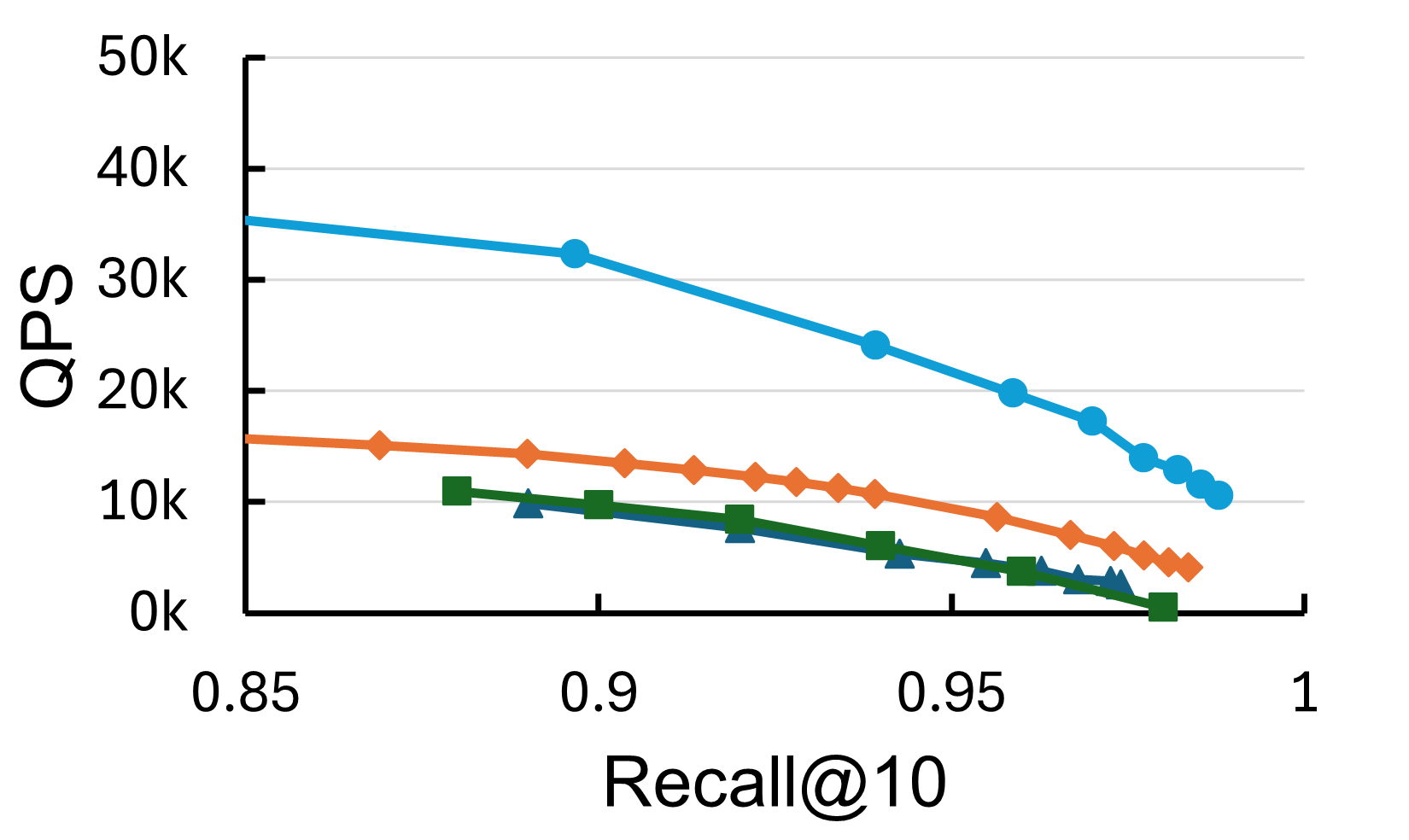} & 
        \includegraphics[width=0.3\linewidth,valign=c]{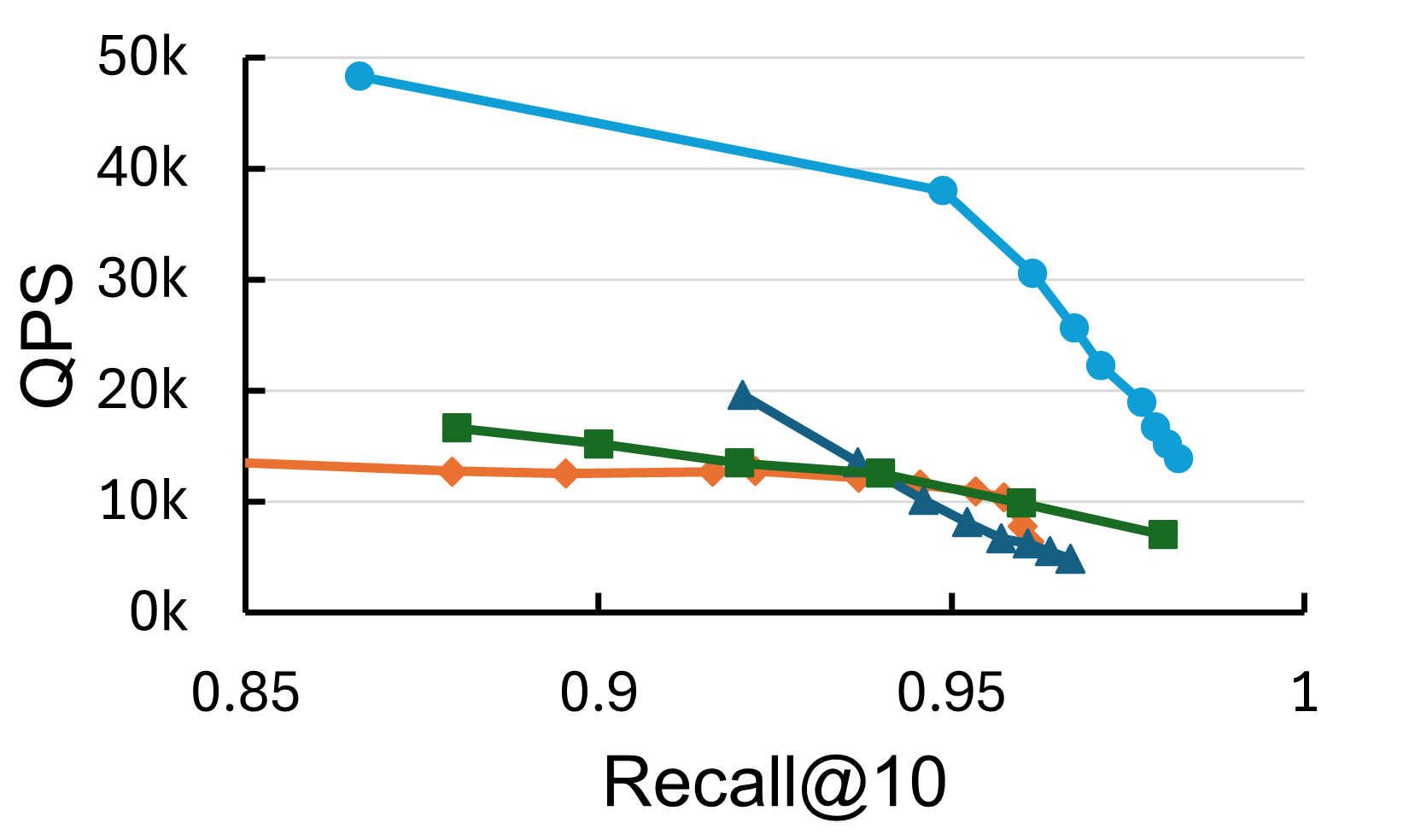} \\
    \end{tabular}
    \caption{The query throughput of different ANNS systems varying with the recall@10 on different SSD counts}
    \label{ssd_comparison}
\end{figure*}

\begin{figure}
	\centering
	{\includegraphics[width=2.3in]{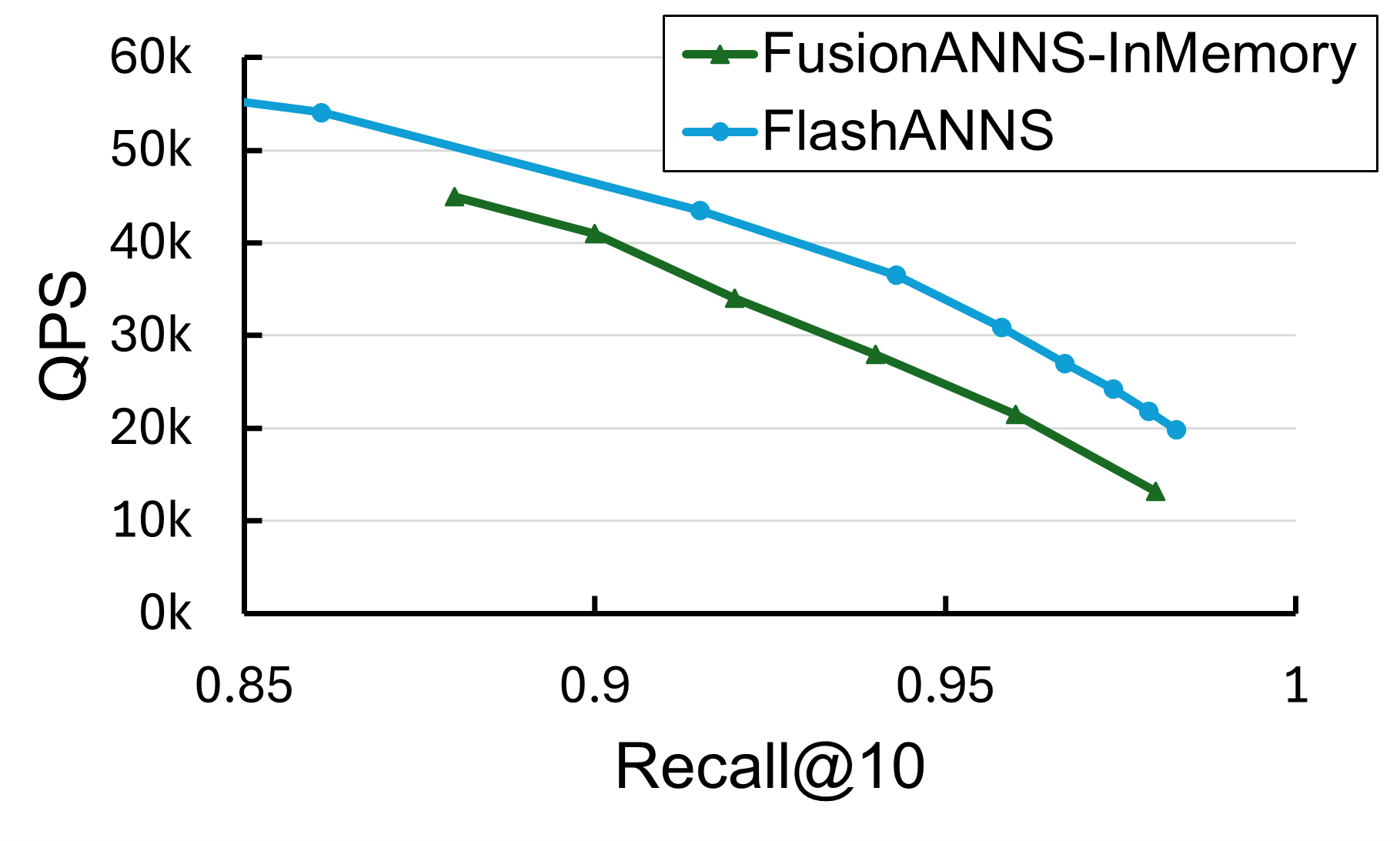} } 
	\caption{\frameworkName{}’s QPS performance compared with in-memory FusionANNS.} 
	\label{sift-inmemory-fusion}
\end{figure}


We evaluate \frameworkName{} against all SSD-based baselines~\cite{spann,fusion,diskann} by maximizing CPU threads to 52. Figure 8 reports query throughput versus recall@10 across three datasets, scaling SSD count from 1 to 8 for index storage.

\noindent \textbf{Comparing under Single SSD Configuration. }
We first examine QPS-recall@10 performance under single-SSD configurations. The top row of Figure~\ref{ssd_comparison} shows \frameworkName{} achieving 2.7-5.9$\times$ query throughput gains over CPU-based baselines (SPANN/DiskANN) at 96\% recall@10. Against GPU-enhanced FusionANNS, \frameworkName{} delivers comparable query throughput (1.03-1.4$\times$) on SIFT1B and SPACEV1B at 98\% recall@10, where constrained I/O bandwidth reduces computational demands and avoids CPU bottlenecks. However, on DEEP1B at 98\% recall@10, \frameworkName{} achieves 14.3$\times$ higher query throughput than FusionANNS. This performance discrepancy originates from DEEP1B's single-precision floating-point (FP32) format, which imposes substantial CPU pressure on FusionANNS during precision distance calculations, revealing CPU-bound limitations. Overall, \frameworkName{} consistently occupies the top-right region of QPS-recall@10 curves, demonstrating superior query throughput through parallel pipeline optimization and I/O stack efficiency under limited I/O bandwidth.

\noindent \textbf{Comparison under an Increasing Number of SSDs.}
As available I/O bandwidth increases with additional SSDs, \frameworkName{} demonstrates disproportionately greater performance improvements than alternatives by navigation graph degree optimizing and I/O-computation latency balancing. Using the SIFT1B exemplar data in Figure~\ref{ssd_comparison} (first column), \frameworkName{} delivers progressively higher query throughput at 98\% recall@10—achieving 1.4$\times$–7.0$\times$, 2.0$\times$–5.4$\times$, 2.7$\times$–5.9$\times$, and 3.9$\times$–5.7$\times$ QPS gains over baselines when scaling from 1 to 8 SSDs. 

\noindent \textbf{Comparison to the In-Memory Baseline. }Furthermore, we observe that FusionANNS underperformed relative to expectations in high I/O bandwidth scenarios. Our analysis suggests this stems from implicit inefficiencies in FusionANNS’ I/O submission queue scheduling mechanisms. To rigorously isolate algorithmic and architectural advantages from implementation-specific biases, we conduct a controlled comparison: we evaluate \frameworkName{} (using 4 SSDs) against an in-memory variant of FusionANNS (all indices loaded into DRAM), thereby removing any confounding effects of storage-layer optimizations. As shown in Figure~\ref{sift-inmemory-fusion}, \frameworkName{} achieves higher performance than the in-memory FusionANNS baseline. This result demonstrates that \frameworkName{}' SSD-based implementation maintains a performance advantage over even the in-memory indexed FusionANNS.

\subsection{Impact of Dependency-Relaxed Asynchronous I/O Pipeline}
\label{sec:pipe}

\begin{figure}
	\centering
	{\includegraphics[width=2.3in]{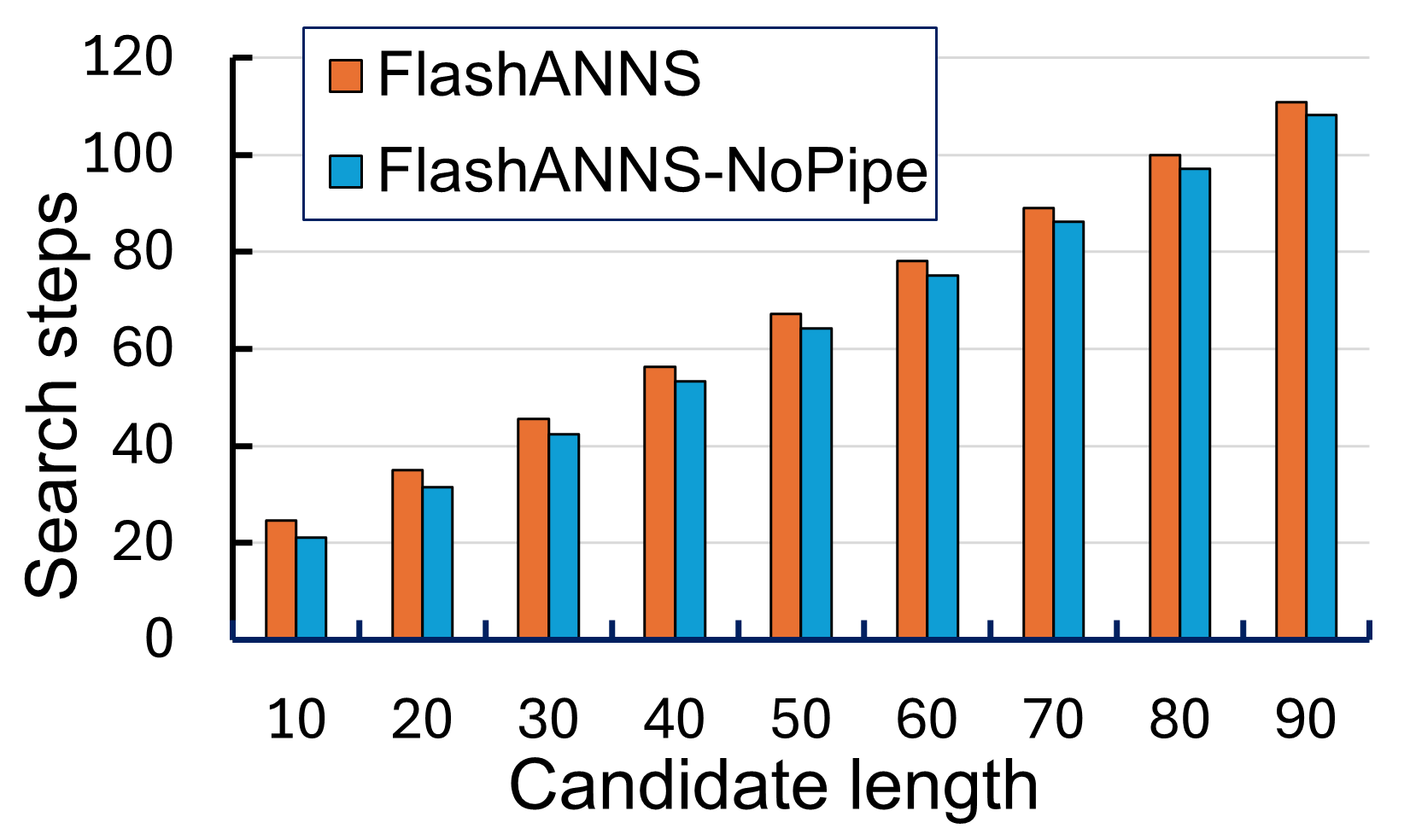} } 
	\caption{Query throughput of FlashANNS vs. FlashANNS-Nopipe under different candidate min-heap lengths} 
	\label{sift_250_4ssd_steps}
\end{figure}  

\begin{figure}
	\centering
	{\includegraphics[width=2.3in]{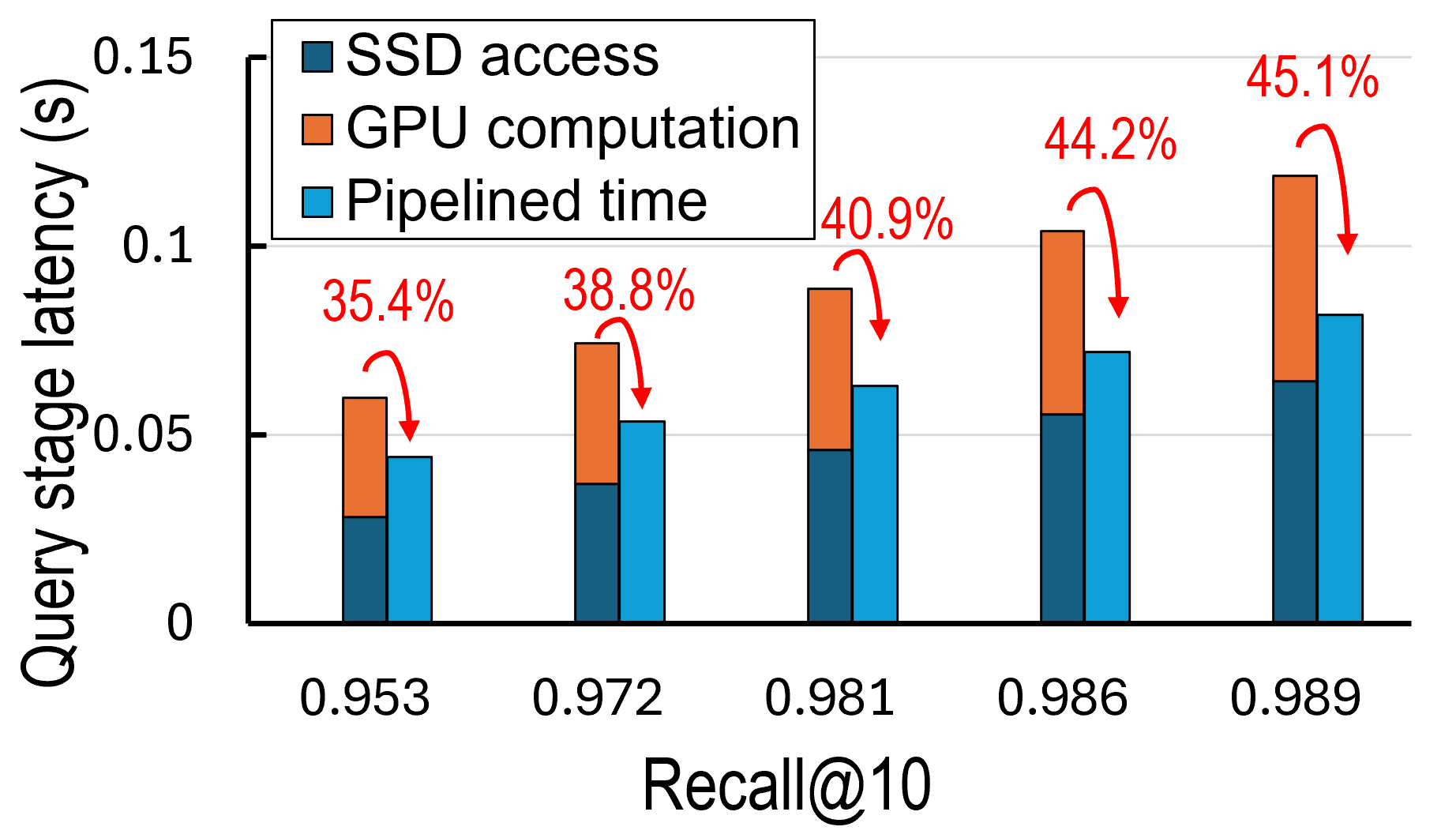} } 
	\caption{FlashANNS's query latency breakdown of SSD access, GPU computation and total pipelined time under different recall rates} 
	\label{sift_250_4ssd_section_delay}
\end{figure}  

\begin{figure}
	\centering
	{\includegraphics[width=2.3in]{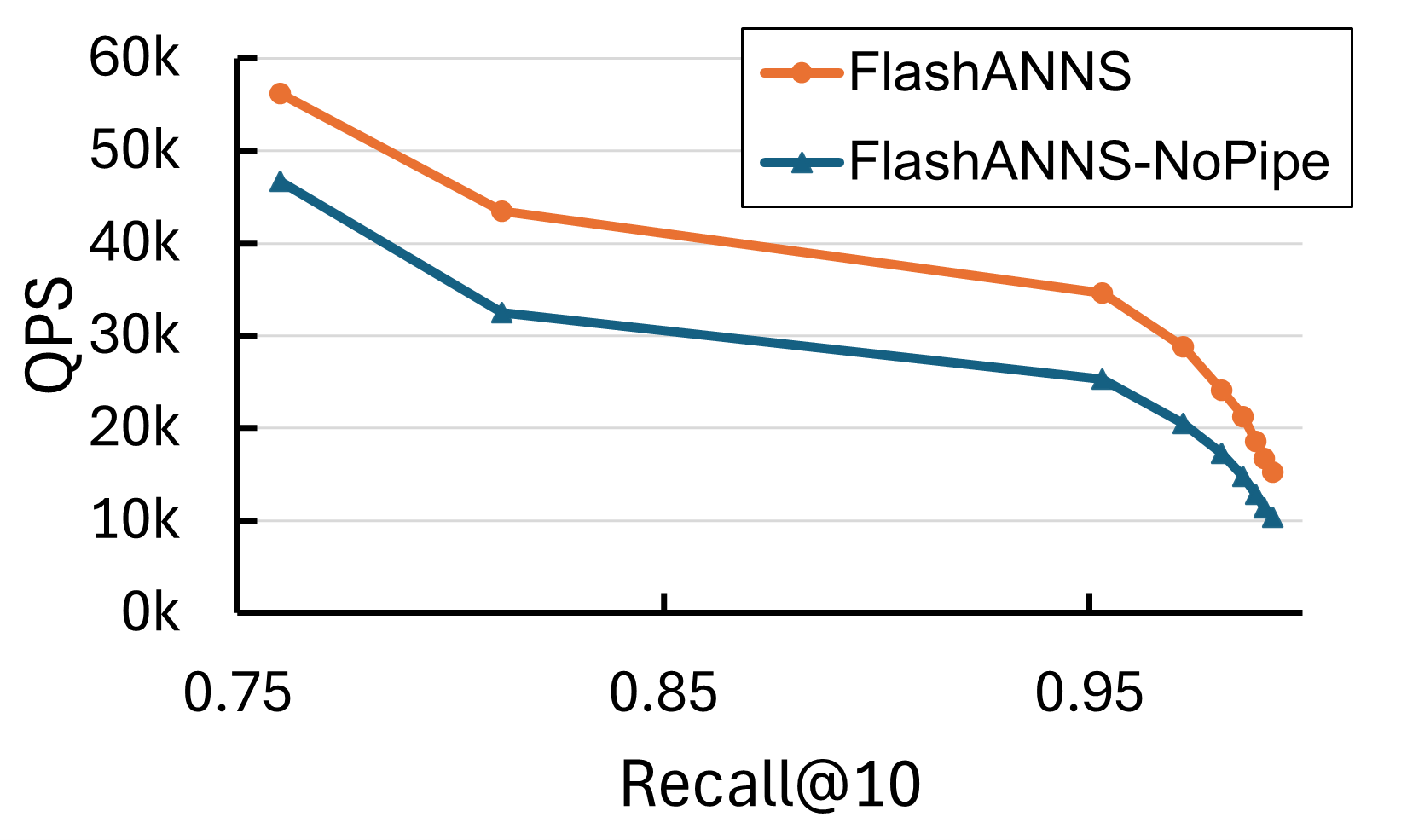} } 
	\caption{End-to-end QPS performance of \frameworkName{} and the \frameworkName{}-Nopipe} 
	\label{sift_150_4ssd_pipeline}
\end{figure}  

\begin{figure}
	\centering
	{\includegraphics[width=2.5in]{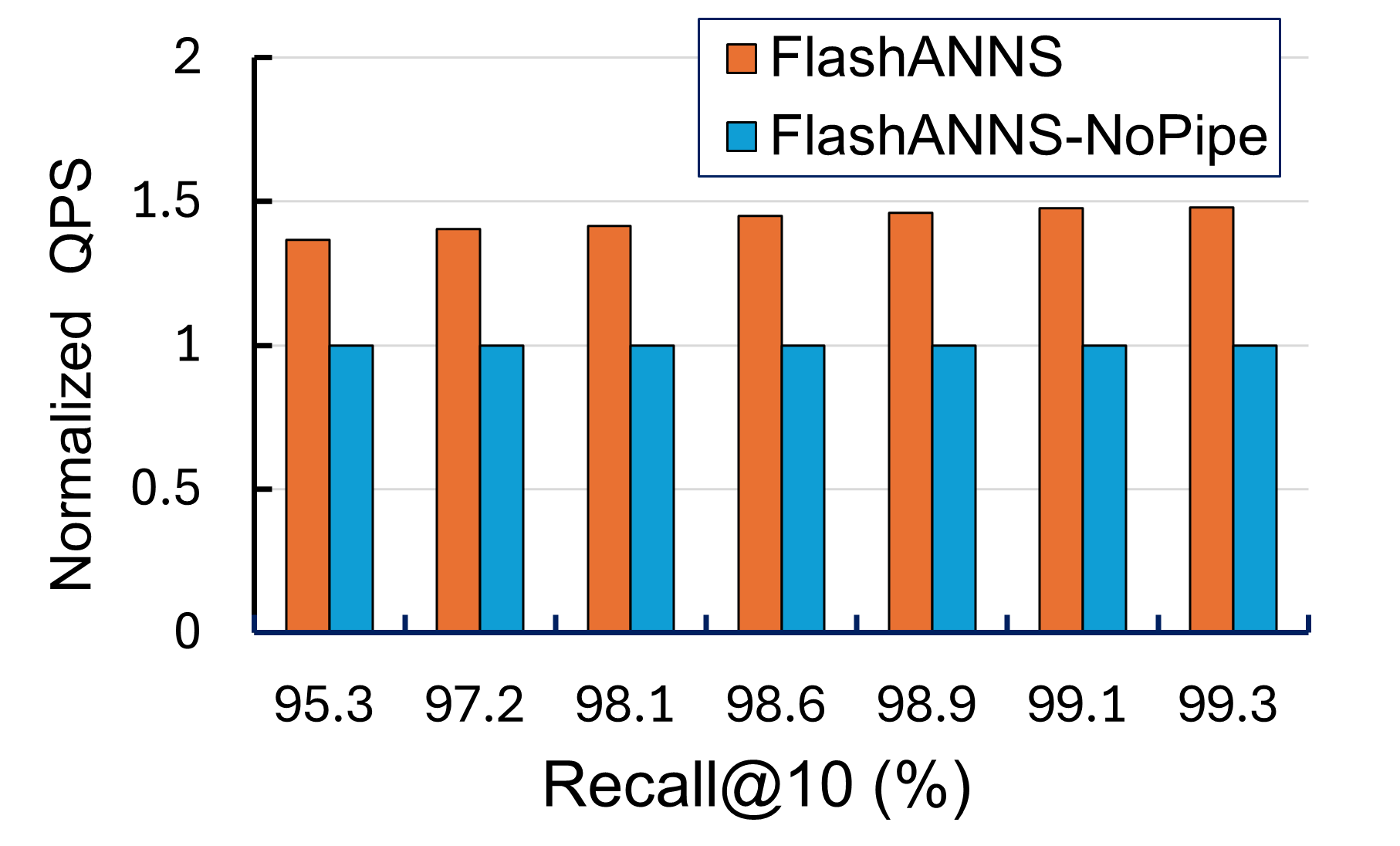} } 
	\caption{Normalized end-to-end QPS performance comparison of \frameworkName{} and the no pipeline version} 
	\label{sift_150_4ssd_pipeline_normalized}
\end{figure}  

In this subsection, we quantitatively evaluate how dependency-relaxed pipeline parallelism affects query throughput during execution. We separately evaluate: 1) the effect of dependency relaxation on step count in graph queries, 2) the latency reduction achieved through execution overlapping, and 3) the improvement in overall query throughput.

\textbf{Impact on Searching Steps. }
To elucidate the impact of dependency relaxation on search procedures, we present two implementations of \frameworkName{}: the pipeline version and the no-pipe variant. The pipeline implementation introduces dependency relaxation to enable concurrent execution between SSD read operations and data computation during graph traversal, whereas the no-pipe variant strictly adheres to the strong dependency constraints inherent in best-first search algorithms, enforcing serialized execution between SSD access and computational processing.

Figure \ref{sift_250_4ssd_steps} illustrates the impact of dependency relaxation on search path lengths, comparing the pipeline and best-first search implementations on the sift1B dataset under a 250-degree graph configuration. We observe that the relaxed-dependency scheme incurs only 2.8-3.2 additional steps, representing a mere 2.5\%-7\% overhead relative to the total query steps in the baseline search strategy. We conclude that dependency relaxation induces only minor variations in search step counts.

\textbf{Impact on Query Latency. }
This section quantifies pipeline overlap characteristics in terms of latency. Evaluated on the SIFT-1B dataset with a 250-degree graph and 4-SSD parallelism, Figure~\ref{sift_250_4ssd_section_delay} compares the average query latency of pipelined \frameworkName{} against the summed latency of its SSD access phase and GPU computation phase. The experimental results show that the overlapped execution time accounts for 35.4\% to 45.1\% of the total pipelined latency across different recall rates. This demonstrates that \frameworkName{} effectively overlaps the SSD accesses and GPU computation stages, thereby reducing search latency and improving QPS.

\textbf{Overall Pipeline Performance. } 
Collectively, the dependency-relaxed asynchronous I/O pipeline trades only 2.5\%-7\% additional computational steps for 36\%-47\% pipeline overlap ratio, thereby improving the overall query throughput. As Figures \ref{sift_150_4ssd_pipeline} and \ref{sift_150_4ssd_pipeline_normalized} demonstrate, this approach delivers 33.6\%–46.6\% higher QPS compared to the no-pipeline variant under the same recall.









\subsection{Impact of Query-grained Concurrent SSD Access}
\label{sec:warp-level}

\begin{figure}
	\centering
	{\includegraphics[width=2.3in]{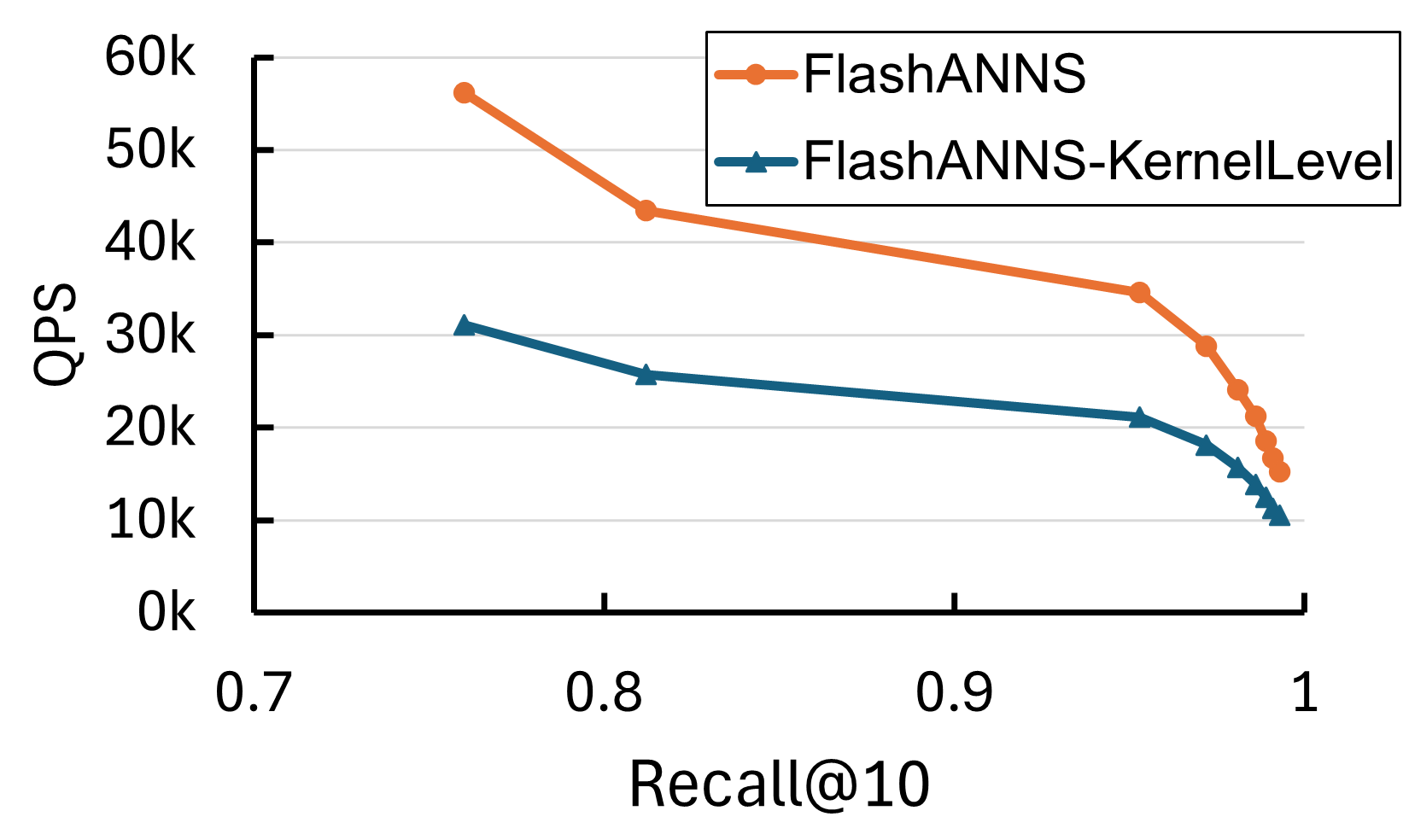} } 
	\caption{End-to-end QPS performance of \frameworkName{} and kernel-grained access version} 
	\label{sift_150_4ssd_warp_kernal}
\end{figure}  

\begin{figure}
	\centering
	{\includegraphics[width=2.5in]{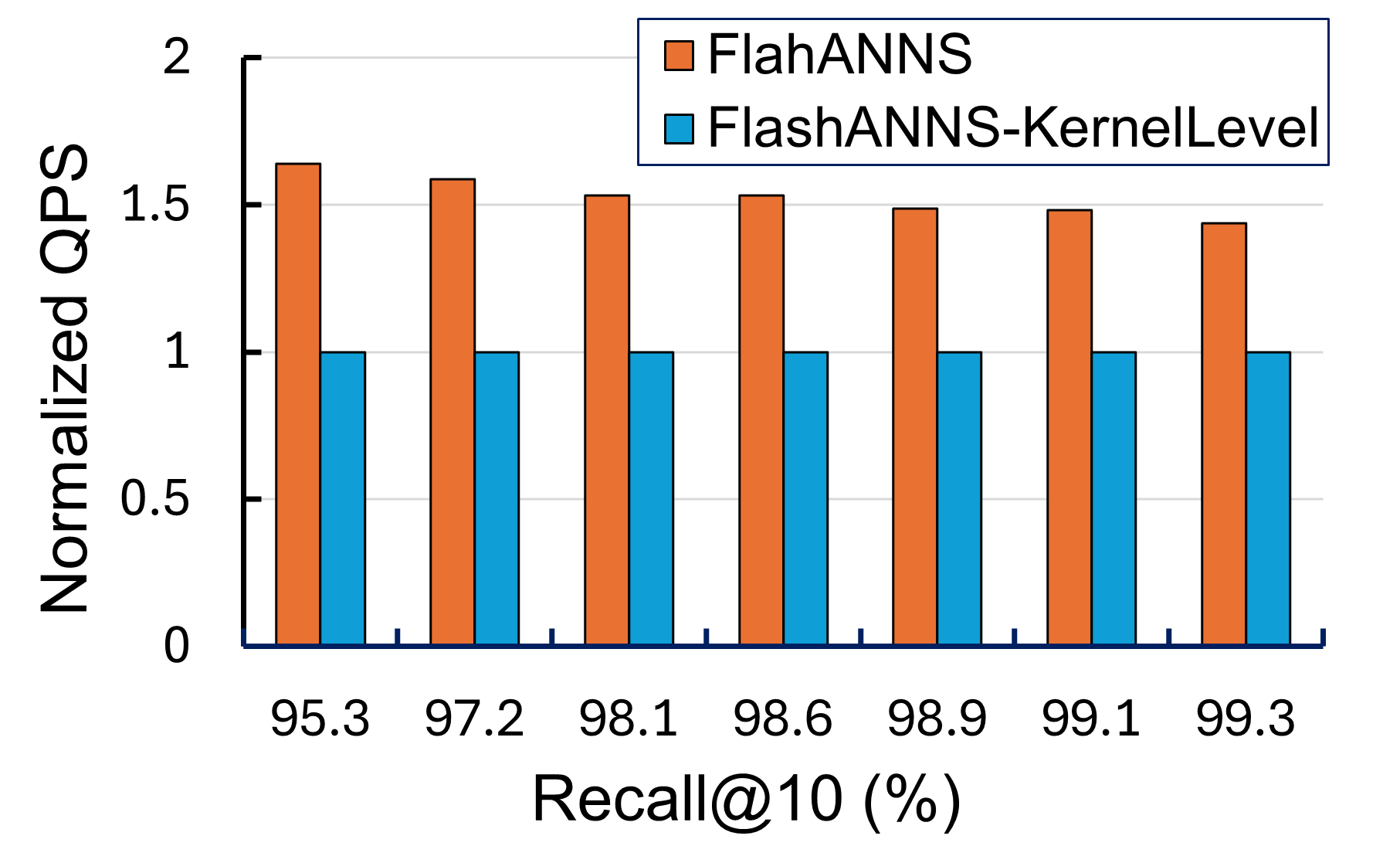} } 
	\caption{Normalized end-to-end QPS performance of \frameworkName{} and kernel-grained access version} 
	\label{sift_150_4ssd_warp_kernal_normalized}
\end{figure}  

To validate the effectiveness of query-grained concurrent SSD access, we employ an unoptimized kernel-grained I/O stack (CAM~\cite{CAM}) for comparison. Unlike query-grained SSD access that treats individual warp operations as autonomous units, the kernel-grained approach aggregates all SSD requests issued during a kernel function's execution phase into a batch process. This requires completing full data retrieval for an entire request group before initiating subsequent data processing and next-stage access operations. In contrast, query-grained SSD access enables a finer-grained synchronization mechanism for immediate processing of subsequent read requests upon completion of queries handled by individual warps. 

We evaluate our experiment on the sift1B dataset with a 4-SSD configuration under 250-degree graph parameters. As Figure~\ref{sift_150_4ssd_warp_kernal} and Figure~\ref{sift_150_4ssd_warp_kernal_normalized} show, the query-grained SSD access implementation achieves 43\%-68\%  query throughput improvement compared to its kernel-grained counterpart CAM. These performance gains are attributed primarily to the finer synchronization granularity of query-grained operations, which effectively amortize sporadic long-tail SSD read latencies from ANNS systems. 


\subsection{Scalability to Larger Return Sets}
\label{sec: return sets}

\begin{figure}
	\centering
	{\includegraphics[width=2.3in]{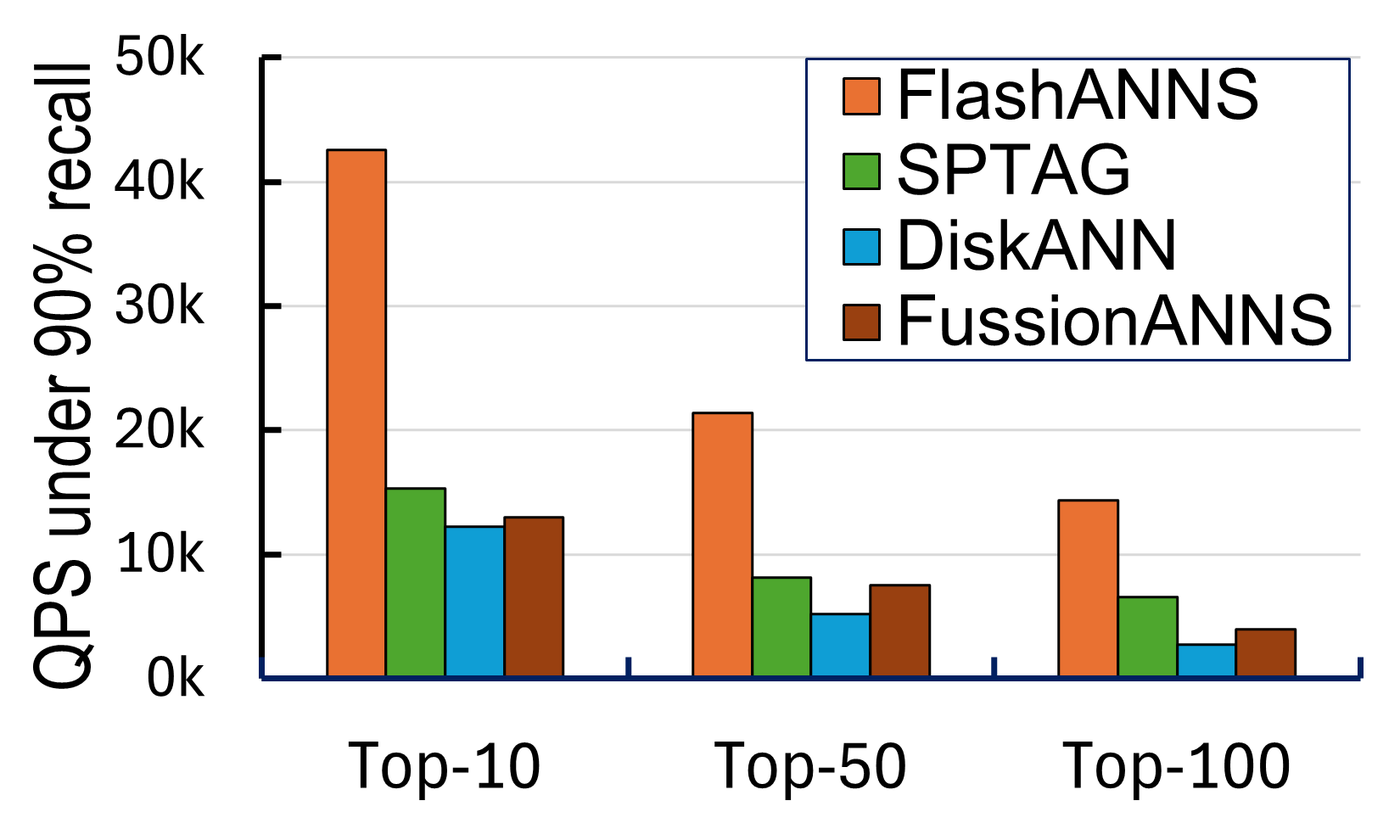} } 
	\caption{End-to-end QPS performance under different top-K sizes} 
	\label{recall@100_qps}
\end{figure}

{
To validate \frameworkName{}' performance with a larger returned set, we compare FlashANNS to three state-of-the-art baselines on SIFT-1B (4-SSD) under more top-K sizes. For each method and $K\!\in\!\{10,50,100\}$, we tune the parameters to measure the achievable QPS for different top-K sizes when the recall rate is higher than 90\%. 
As shown in Figure~\ref{recall@100_qps}, FlashANNS delivers \(\mathbf{2.2\text{--}5.2}\times\) higher QPS than the other three baselines, demonstrating that it can maintain high query throughput under different $K$ sizes.}

\begin{figure*}[htbp]
    \centering
    \begin{subfigure}[b]{0.24\textwidth}
        \includegraphics[width=\linewidth]{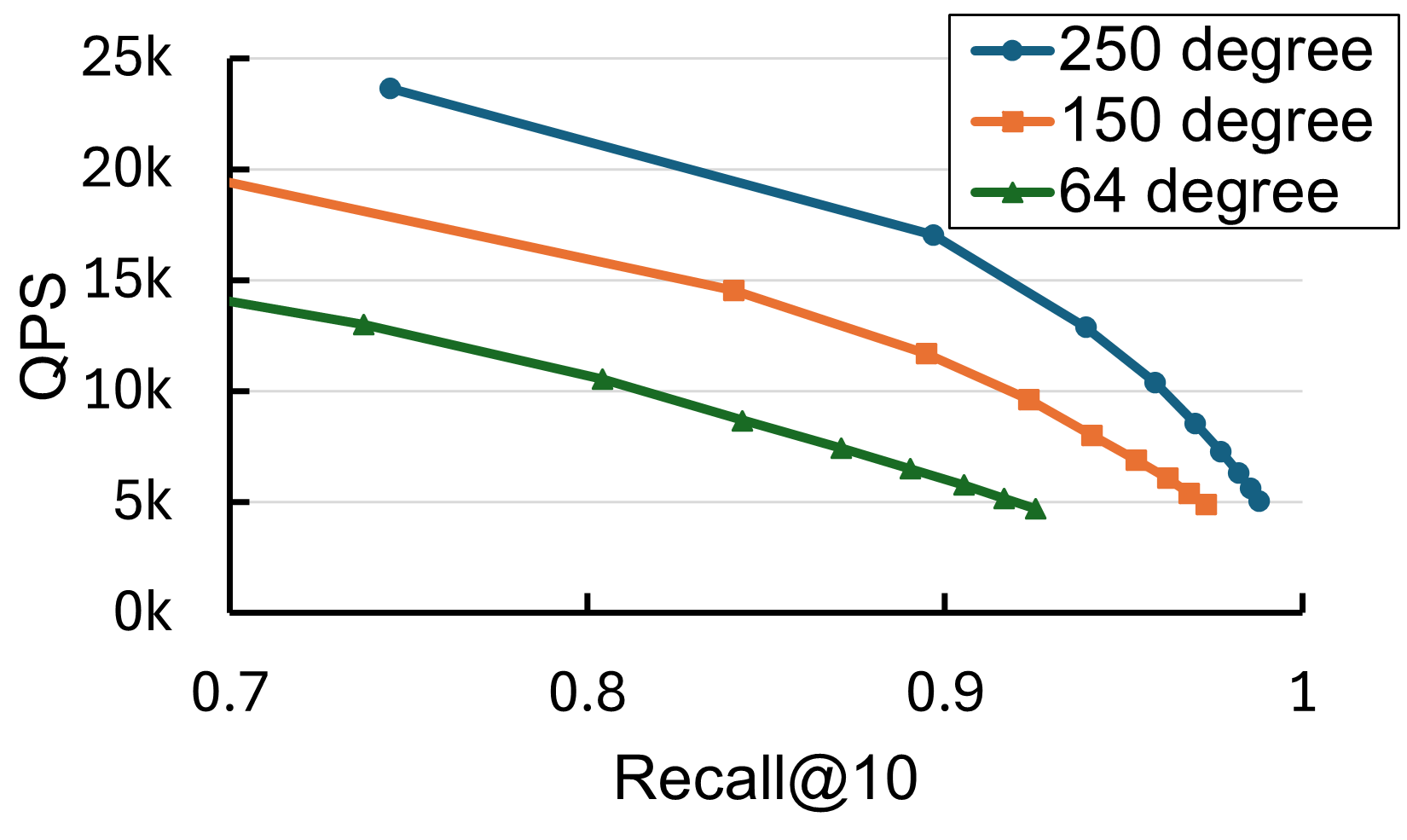}
        \caption{1 SSD}
        \label{fig:degree_1ssd}
    \end{subfigure}
    \begin{subfigure}[b]{0.25\textwidth}
        \includegraphics[width=\linewidth]{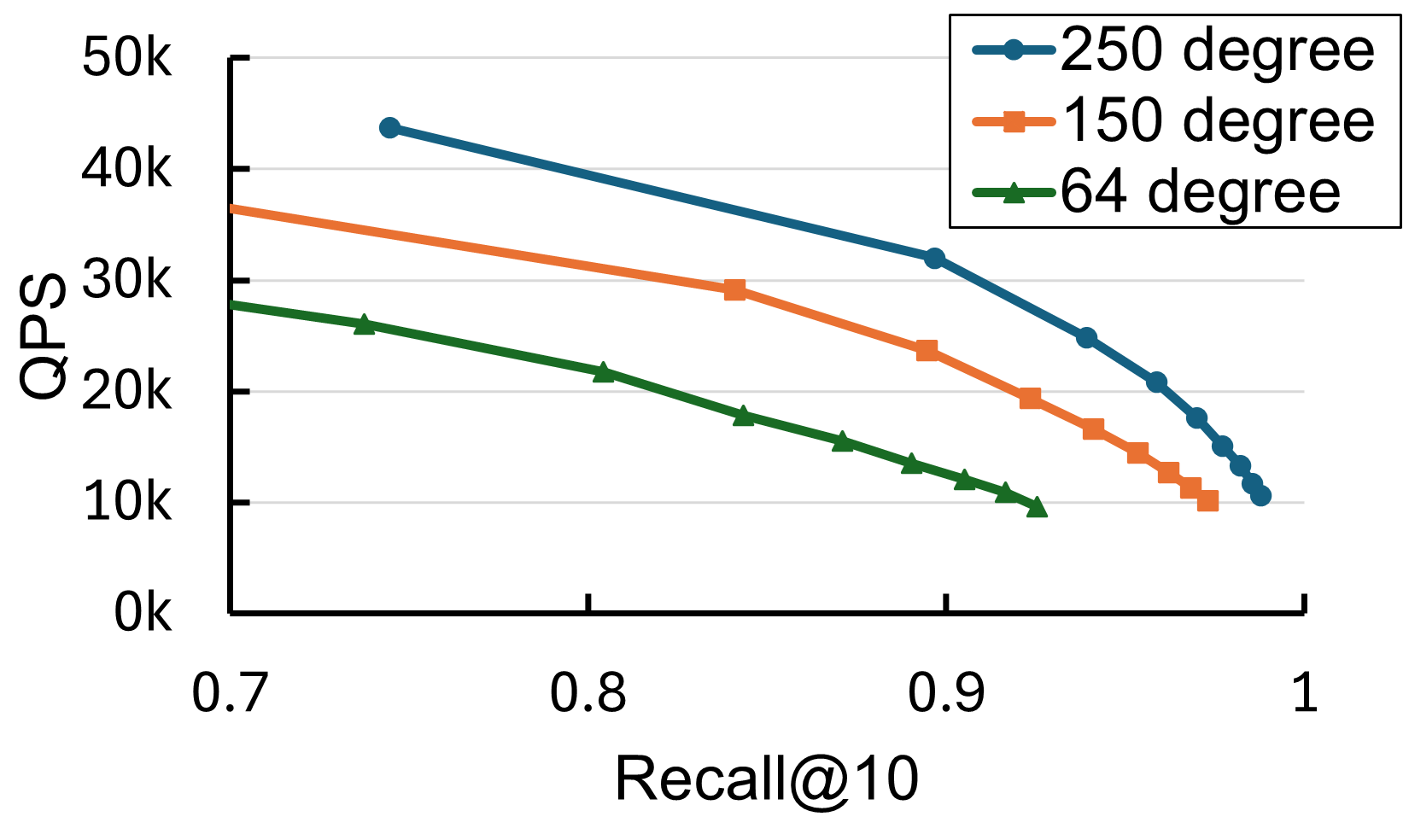}
        \caption{2 SSDs}
        \label{fig:degree_2ssd}
    \end{subfigure}
    \begin{subfigure}[b]{0.25\textwidth}
        \includegraphics[width=\linewidth]{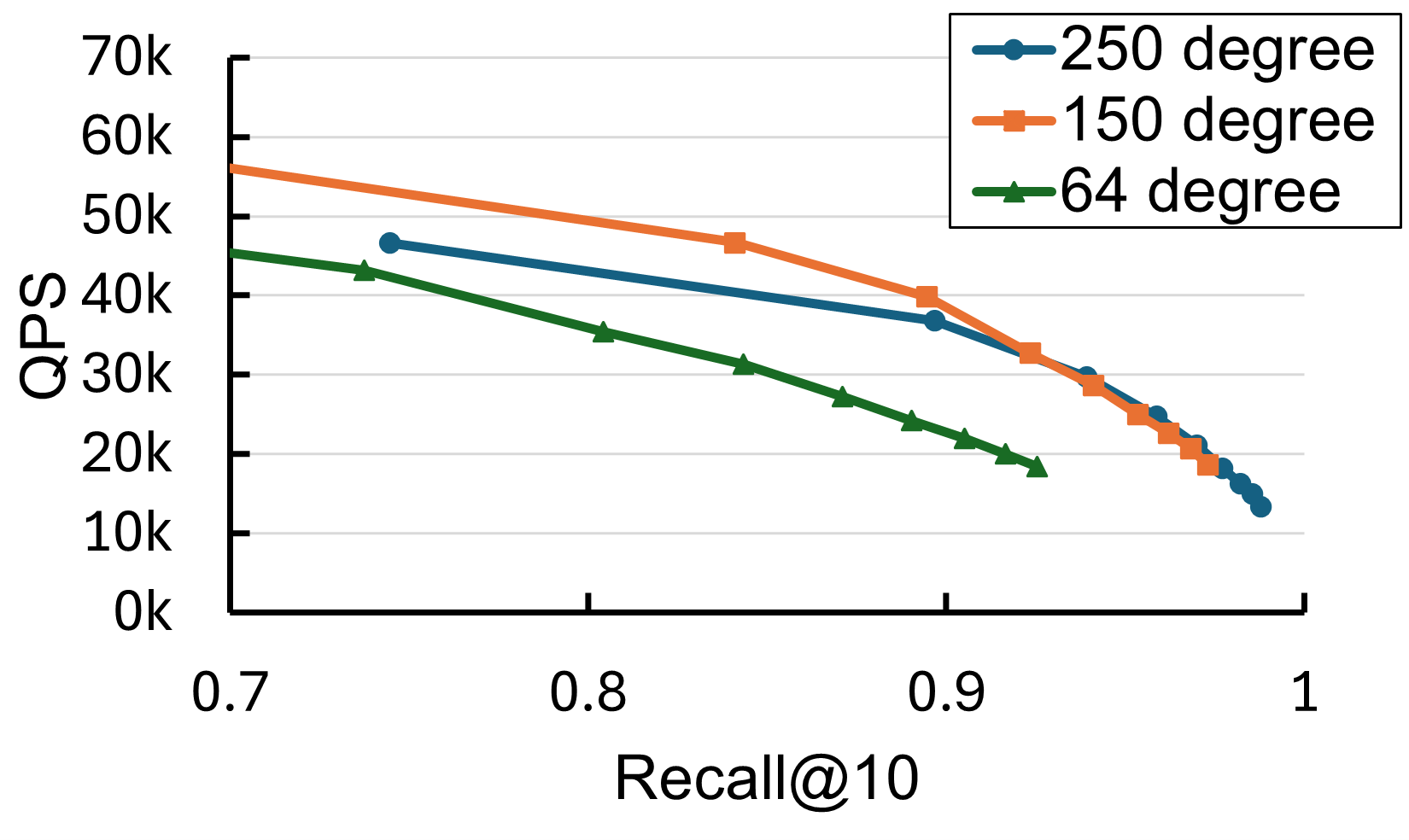}
        \caption{4 SSDs}
        \label{fig:degree_4ssd}
    \end{subfigure}
    \begin{subfigure}[b]{0.24\textwidth}
        \includegraphics[width=\linewidth]{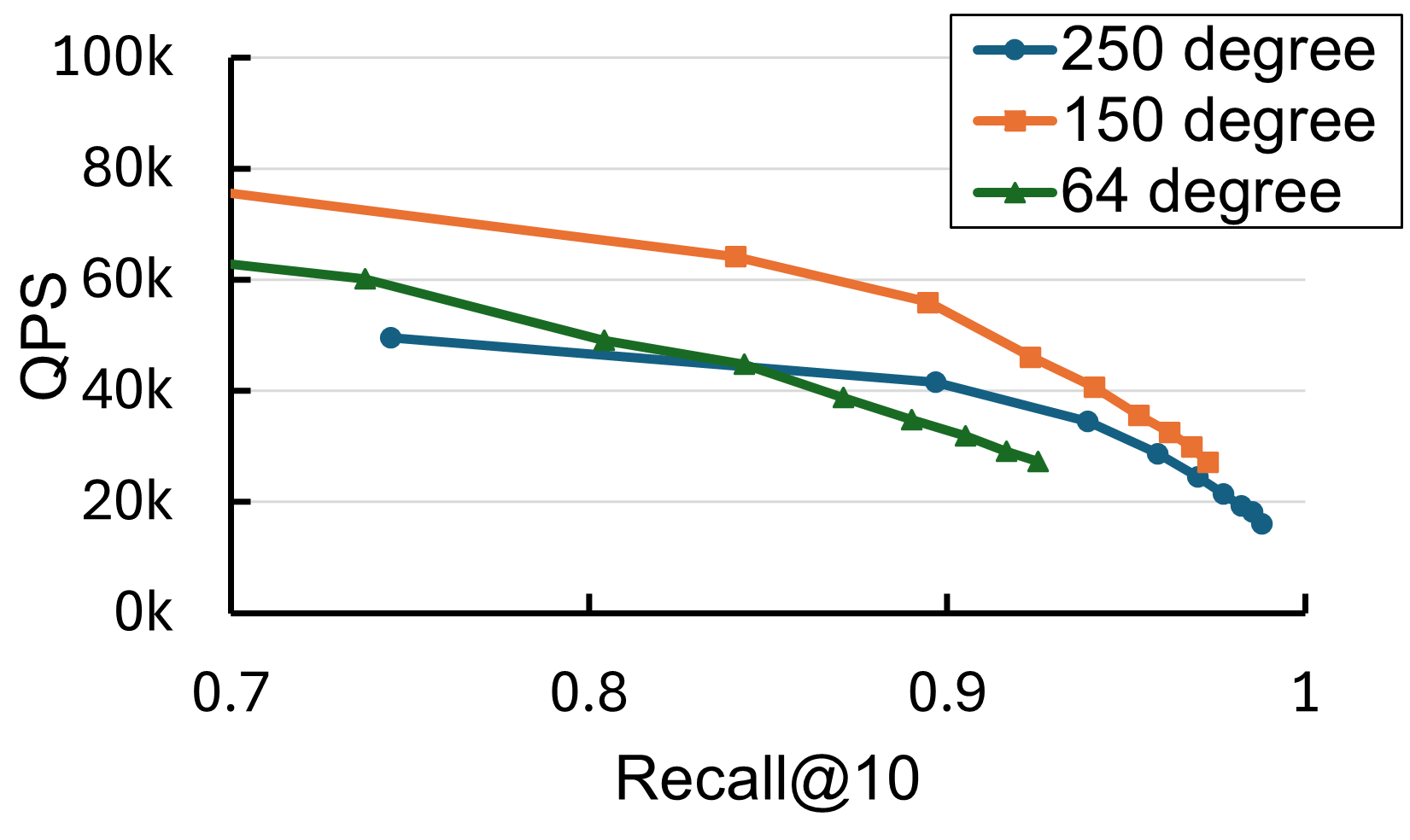}
        \caption{8 SSDs}
        \label{fig:degree_8ssd}
    \end{subfigure}
    \caption{FlashANNS QPS-recall performance on the dataset DEEP1B dataset under different SSD configurations }
    \label{fig:degree_comparison}
\end{figure*}

\begin{figure}
	\centering
	{\includegraphics[width=2.5in]{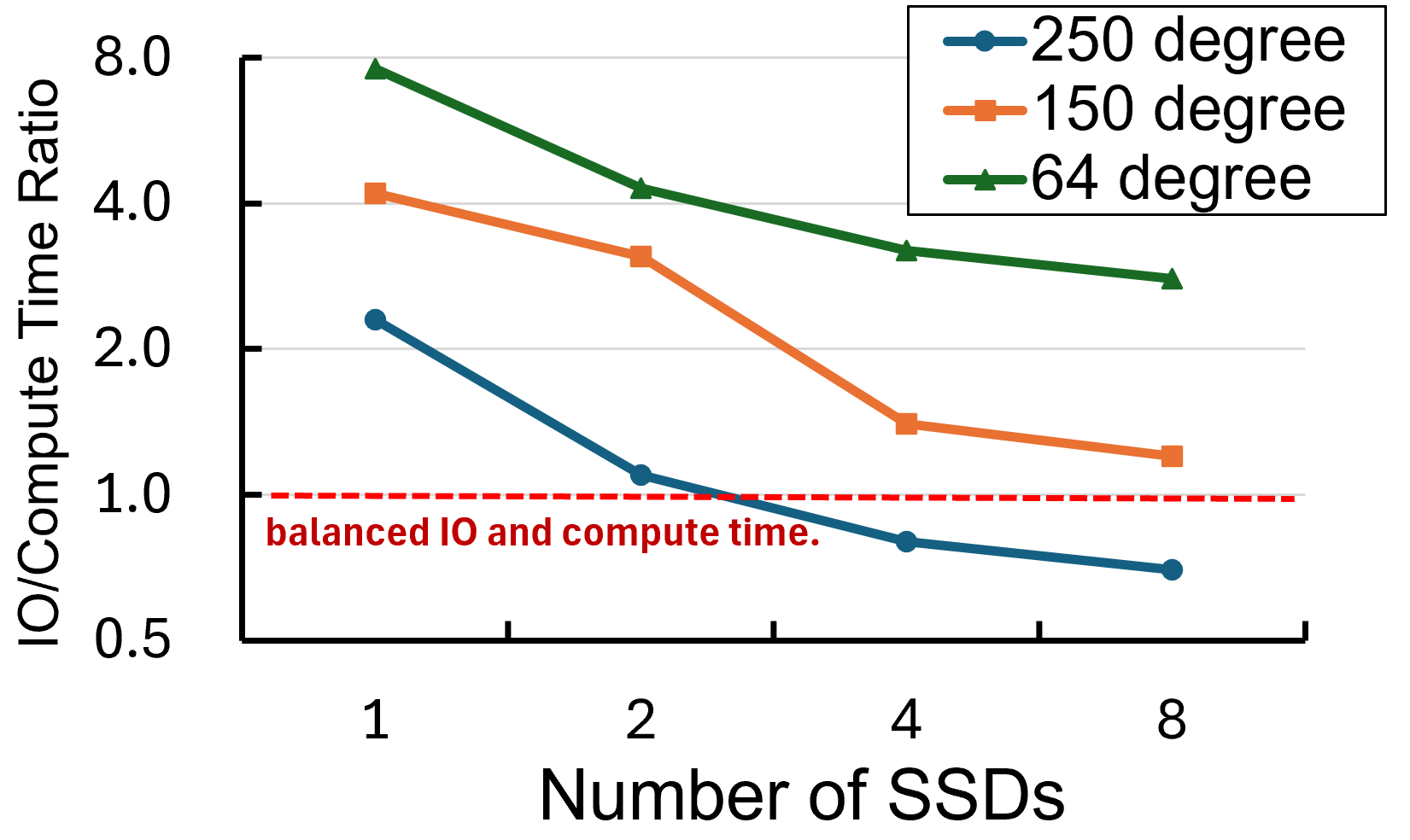} } 
	\caption{I/O-computation time ratio of sample indices in different SSD counts} 
	\label{IO_compute_ratio}
\end{figure}

\subsection{Process of Computation I/O-Balanced Graph Degree Selection}
\label{sec:degree_select}

We conduct index construction experiments across three graph degree configurations (64, 150, and 250 degrees) under varying SSD parallelism conditions (1, 2, 4, and 8 SSDs). Our evaluation systematically measures query throughput characteristics across different I/O bandwidth settings. Results demonstrate the effectiveness of selecting graph degrees via sample-indexed analysis.

We evaluate the QPS-recall tradeoffs of varying graph degrees on the DEEP1B dataset under different I/O bandwidth configurations. As demonstrated in Figure~\ref{fig:degree_comparison}, under low I/O bandwidth constraints (with indices stored on 1-2 SSDs), higher-degree graphs consistently achieve superior query throughput while satisfying equivalent recall targets compared to lower-degree graphs. This advantage stems from their enhanced ability to leverage GPU parallel computation units to mask I/O latency when SSD bandwidth is scarce.

However, as I/O bandwidth increases with higher SSD parallelism (up to 8 SSDs), the query throughput gap between high-degree and low-degree graphs narrows significantly. Notably, in the 8-SSD configuration, the 150-degree graph outperforms its 250-degree counterpart by 13\% in query throughput under 96\% recall@10. Beyond critical bandwidth thresholds, the higher computational demands inherent to high-degree graph processing (e.g., requiring access to more neighbors at each traversal step) may conversely create computational bottlenecks.

Before determining the optimal graph degree for index construction, we create sample indices (as described in Section~\ref{sec:degree_sel}) — which share the same data type as the vector index but exclude actual neighbor relationships — to pre-estimate pipeline stage latency. Experimental measurement shows the I/O-computation ratio characteristics for different graph degree sample indices in Figure~\ref{IO_compute_ratio}. These sample indices reflect the actual performance characteristics of full-scale index graphs at corresponding degree configurations.

With 1 SSD,
the I/O latency of the 150-degree graph is 4.2$\times$ its computational latency, while the 250-degree graph exhibits I/O latency at 2.3$\times$ computational latency, indicating that both configurations remain I/O-bound, thus higher-degree graphs retain their advantage.

With 2 SSDs,
the 150-degree graph’s I/O latency becomes 3.1$\times$ computational latency (still I/O-bound), whereas the 250-degree graph achieves near-balance at 1.1$\times$ I/O-to-compute latency ratio, enabling full pipeline utilization.

With 4 SSDs,
the 150-degree graph’s I/O latency reduces to 1.4$\times$ computational latency (marginally I/O-bound), while the 250-degree graph becomes compute-bound (0.7$\times$ I/O-to-compute ratio). This implies the optimal degree lies between 150 and 250 to avoid asymmetrical bottlenecks.

The performance trends predicted by our degree sampling via lightweight sample indices align closely with actual test results, empirically validating the effectiveness of graph degree pre-selection for workload-aware index optimization.

\subsection{Out-of-Core Efficiency on 30B-Vector Dataset}
\label{sec:30B}

\begin{figure}
	\centering
	{\includegraphics[width=2.5in]{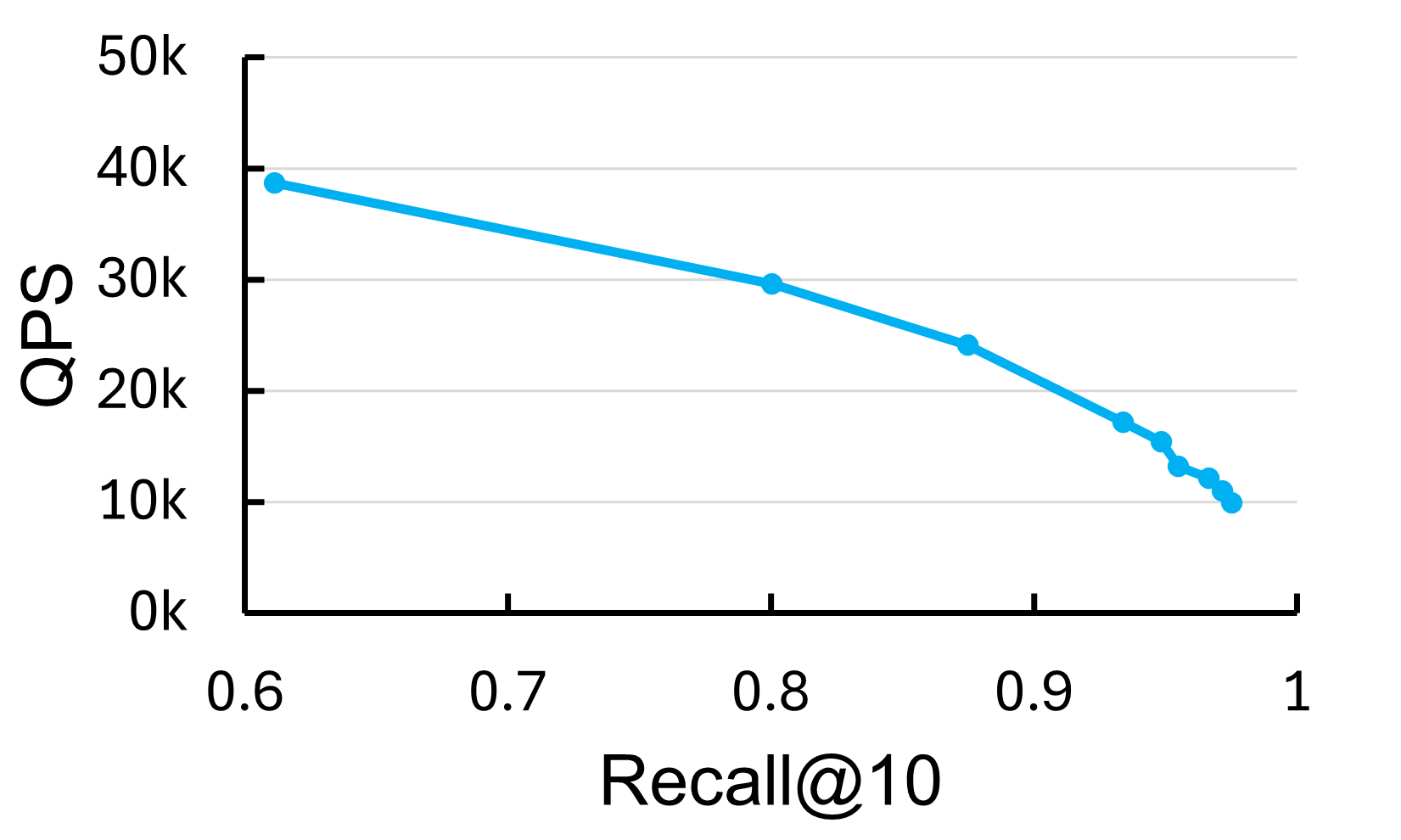} } 
	\caption{\frameworkName{}' QPS performance under TB-scale dataset} 
	\label{flash-1Tdataset}
\end{figure}  

To demonstrate \frameworkName{}' processing capability to out-of-core indices, we augment the DEEP1B dataset to a 30 billion-vector dataset (1,073 GB) by duplicating each vector twice with minor perturbations, thereby expanding its scale while preserving proximity relationships.
Figure~\ref{flash-1Tdataset} shows that \frameworkName{} achieves good QPS-recall trade-offs on this exabyte-scale dataset, validating \frameworkName{}’s capability to efficiently perform ANNS on huge datasets far exceeding DRAM capacity for RAG applications. 

\section{Related Works}
To our knowledge, this work presents the first GPU-accelerated, SSD-based graph ANNS framework. While Section~\ref{sec:experiments} provides comprehensive comparisons with state-of-the-art SSD-based ANNS systems (DiskANN, SPANN, FusionANNS), we review related work in two key areas: in-memory ANNS frameworks and  SSD I/O optimization implementations.

\noindent{\textbf{In-Memory ANNS Frameworks. }
In-memory ANNS systems are widely utilized, with their index structures primarily categorized into four types:
1. Tree-based indices~\cite{yianilos1993data,cayton2008fast,dasgupta2008random,image1,tree1,fukunaga1975branch,navarro2002searching,beygelzimer2006cover}: The core premise centers on constructing a tree-like data structure that hierarchically organizes vectors via partitioning criteria based on distance or density metrics. 
2. Hash-based~\cite{datar2004locality,andoni2008near,andoni2014beyond,terasawa2007spherical,lsh3,gionis1999similarity}: Locality-Sensitive Hashing (LSH) maps high-dimensional vectors into hash buckets while preserving similarity relationships, enabling efficient approximate nearest neighbor search. 
3. Quantization-based~\cite{jegou2010product,ge2013optimized,norouzi2013cartesian,babenko2015tree,xia2013joint,kalantidis2014locally,inverted}: This method divides high-dimensional vectors into subvectors, independently quantizes each subvector into compact codes. This method reduces the storage footprint and accelerates similarity computation through lookup tables. 
4. Graph-based~\cite{hajebi2011fast,dong2011high,jin2014fast,fu2016efanna,kleinberg2000navigation,graph1,graph2}: Graph-based indices demonstrate superior search performance in Euclidean spaces due to their explicit modeling of local neighbor proximity. In particular, the edges in the graph structure directly encode vector adjacency relationships, enabling greedy traversal toward nearest neighbors. In contrast, \frameworkName{} focuses on out-of-order ANNS systems while achieving slightly higher performance than the in-memory counterparts.

\noindent{\textbf{SSD I/O Optimizations. }
Recently, the SSD has been deployed in many applications for its massive storage volume while achieving high performance compared to traditional hard disk drives~\cite{haas2023modern, maschi2023difficult, alonso2023future, 10.5555/3650697.3650709}. There are many works proposed to exploit the potentialities of SSD~\cite{kroviakov2024heterogeneous,lerner2024data,maschi2023difficult, alonso2023future, von2022you, wei2022much, hyperion, liao2025ratel, papon2024enhancing, huang2024neos, wang2024boosting, wang2024leaderkv, do2021better, duffy2023dotori, chai2019ldc, ziegler2022scalestore, haubenschild2020rethinking, chu2020latte, wang2024bushstore, thonangi2017log, li2016hippogriffdb, lee2023lru}. Existing works~\cite{nvmmu} achieve direct GPU-SSD data transfer using the GPUDirect~\cite{gpudirect} technology. However, it relies on the CPU to initiate or trigger SSD access and fails to eliminate the OS kernel overhead entirely. 
Systems~\cite{silberstein2013gpufs} do not support GPUDirect~\cite{gpudirect}. These methods involve OS kernel overheads, especially making it hard to saturate SSD throughput for batching access patterns of ANNS workloads. 

BaM~\cite{qureshi2023gpu} proposes GPU-initiated on-demand direct SSD access without CPU involvement. However, BaM's design introduces new GPU core contention issues, which are intended to be used in computation tasks, while CAM~\cite{CAM} can achieve high throughput without GPU SM occupancy. However, its implementation enforces kernel-grained global synchronization: all warps within a kernel must wait for the slowest SSD I/O request to complete before proceeding. In contrast, \frameworkName{} enables query-grained concurrent SSD access to accelerate GPU-based ANNS.

\section{Conclusion}
In this work, we present \frameworkName{}, a GPU-accelerated, out-of-core graph-based ANNS system. Our approach achieves full I/O-compute overlap through three coordinated mechanisms: First, a dependency relaxation I/O pipeline to overlap SSD node retrieval with GPU computation while providing a rigorous theoretical convergence guarantee. Second, a query-grained concurrent SSD access that eliminates the GPU kernel-grained global synchronization inherent. Third, a hardware-aware graph degree selection mechanism maximizing pipeline stage overlap efficiency. Experimental results show that under the same recall accuracy, \frameworkName{} achieves 2.7–5.9$\times$ higher query throughput than existing SOTA methods with a single SSD configuration, and scales to 3.9–12.2$\times$ query throughput improvements in multi-SSD setups.

\begin{acks}
The work is supported by the following grants: 
National Science and Technology Major Project (2022ZD0117000), the Major Project of the Zhejiang Provincial Natural Science Foundation under Grant No. LD26F020002, 
the National Natural Science Foundation of China under the grant numbers (62472384, 62441236, U24A20326), 
Starry Night Science Fund of Zhejiang University Shanghai Institute for Advanced Study (SN-ZJU-SIAS-0010). Jie Zhang, Zeke Wang, and Fei Wu are the corresponding authors. 

\end{acks}

\bibliographystyle{ACM-Reference-Format}
\bibliography{sample-base}

\end{document}